\documentclass{osudissert96}
\input{epsf}

\def\BibTeX{{\rm B\kern-.05em{\sc i\kern-.025em b}\kern-.08em
    T\kern-.1667em\lower.7ex\hbox{E}\kern-.125emX}}

\renewcommand\typesetChapterTitle[1]{\uppercase{#1}}

\def\Dslash{\rlap{\,/}D}
\def\Re{\mbox{Re}}
\def\Im{\mbox{Im}}
\def\dione{$\Delta I=1/2$ }
\def\dithree{$\Delta I=3/2$ }
\def\epsp{$\varepsilon '/\varepsilon$ }
\def\openone{{\displaystyle\mathbf 1}}
\def\iden{\overline{(\openone\otimes \openone) } }
\def\gam#1{\overline{(\gamma_{#1}\otimes \openone)} }
\def\ggam#1{\overline{\overline{(\gamma_{#1}\otimes \openone)} } }
\def\sfno#1#2{\overline{(\gamma_{#1}\otimes\xi_{#2})}}
\def\ssfno#1#2{\overline{\overline{(\gamma_{#1}\otimes\xi_{#2})}}}
\def\sf#1#2#3#4{\overline{(\gamma_{#1}\otimes\xi_{#2})}_{#3#4}}
\def\ssf#1#2#3#4{\overline{\overline{(\gamma_{#1}\otimes\xi_{#2})}}_{#3#4}}

\begin{document}

\author{Dmitry Pekurovsky}
\title{Calculation Of Hadronic Matrix Elements Relevant For
\dione Rule And \epsp In Lattice QCD With Staggered Fermions}
\authordegrees{M.S.}  
\unit{Department of Physics}

\advisorname{Dr. G. Kilcup}
\coadvisorname{Dr. J. Shigemitsu}
\member{Dr. H. Kagan}
\member{Dr. R. Furnstahl}

\maketitle

\disscopyright

\begin{abstract}
  This thesis presents the setup and results of numerical 
calculation of hadronic matrix elements
of $\Delta S=1$ weak operators, with the aim of studying
the \dione  rule and direct CP violation. Such study
provides a useful comparison of the Standard Model 
predictions with experiment. We work within the framework of 
(partially quenched) Lattice QCD with the staggered 
(Kogut-Susskind) version of fermions. We have achieved a reasonable 
statistical accuracy in calculating all matrix elements in question.
Based on these results, we find that the \dione  channel in
$K \to \pi\pi$ decays is enhanced compared to the \dithree  
channel. This is consistent with experiment, although
the exact amount of enhancement is subject to considerable 
systematic errors. I also discuss the difficulties encountered
in attempting to calculate the direct CP violation parameter \epsp   
and our approximate solution leading to 
a crude estimate. In a related study we nonperturbatively compute
bilinear
renormalization factors $Z_S$ and $Z_P$, which are needed for
our estimate of \epsp  and also allow to calculate the light quark 
masses. For the strange quark mass in NDR $\overline{\rm MS}$ scheme 
at 2 GeV in the quenched approximation we find: $m_s = 102.9 \pm 5.1$ MeV.

\end{abstract}

%
% Uncomment the three lines below to generate the external abstract.  Two
% copies of this must be turned in to the graduate school.  These lines can
% be placed pretty much anywhere, since the page numbering should be
% independent of the rest of the thesis
%

\begin{externalabstract}
  This thesis presents the setup and results of numerical 
calculation of hadronic matrix elements
of $\Delta S=1$ weak operators, with the aim of studying
the \dione  rule and direct CP violation. Such study
provides a useful comparison of the Standard Model 
predictions with experiment. We work within the framework of 
(partially quenched) Lattice QCD with the staggered 
(Kogut-Susskind) version of fermions. We have achieved a reasonable 
statistical accuracy in calculating all matrix elements in question.
Based on these results, we find that the \dione  channel in
$K \to \pi\pi$ decays is enhanced compared to the \dithree  
channel. This is consistent with experiment, although
the exact amount of enhancement is subject to considerable 
systematic errors. I also discuss the difficulties encountered
in attempting to calculate the direct CP violation parameter \epsp   
and our approximate solution leading to 
a crude estimate. In a related study we nonperturbatively compute
bilinear
renormalization factors $Z_S$ and $Z_P$, which are needed for
our estimate of \epsp  and also allow to calculate the light quark 
masses. For the strange quark mass in NDR $\overline{\rm MS}$ scheme 
at 2 GeV in the quenched approximation we find: $m_s = 102.9 \pm 5.1$ MeV.

\end{externalabstract}

\dedication{To my parents}

%
% Bring in Acknowledgement and Vita from separate files named ``ack.tex''
% and ``vita.tex''.
%

\begin{acknowledgements}
I gratefully acknowledge the intellectual support and
guidance of my advisor, Prof. G. Kilcup, both in 
theoretical and computational aspects of this work.

The parallel FORTRAN code used for majority of the 
calculations in this thesis was written in collaboration
with Prof. Kilcup and Lakshmi Venkataraman, to both of whom 
I express my thankfullness. The large-scale computations were 
done on CRAY-T3E supercomputers at the Ohio Supercomputing
Center (OSC) and National Energy Research Scientific Computing Center
(NERSC). 
I am grateful to OSC and Prof. Kilcup for supercomputer
time allocation. The work was supported by Department of Energy grants.
I am grateful to Columbia University Lattice QCD
group for providing access to their dynamical configurations.

Finally, I would not survive through the years of Graduate School
without the love and encouragement of my partner, Natalie Prigozhina, 
as well as my family.

\end{acknowledgements}

\begin{vita}

\dateitem{November 29, 1971}{Born - Kiev, Ukraine, USSR}

\dateitem{1991}{B.S. equivalent, Physics, Moscow \\ Institute of Physics and Technology}

\dateitem{1993-1995}{Graduate Teaching Associate,\\
                         Department of Physics, \\
			 The Ohio State University.}
\dateitem{1995}{M.S., Physics, \\ The Ohio State Univerity} 
\dateitem{1995-present}{Graduate Research Associate,\\
                         Department of Physics, \\
			 The Ohio State University.}

\begin{publist}

\researchpubs

\pubitem{D. Pekurovsky, G. Kilcup
\newblock ``Lattice Calculation of Matrix Elements Relevant For
\mbox{$\Delta I=1/2$} Rule and $\varepsilon '/\varepsilon$''
\newblock Published electronically by the library of UCLA at: \\
\underline{http://www.dpf99.library.ucla.edu}}

\pubitem{D. Pekurovsky, G. Kilcup 
\newblock ``$\Delta I=1/2$ Rule from Staggered Fermions''
\newblock {\it Nucl. Phys.} B (Proc. Suppl.) 73, 309--311 (1999)}

\pubitem{D. Pekurovsky, G. Kilcup 
\newblock ``Weak Matrix Elements: On the Way
to $\Delta I=1/2$ rule and $\varepsilon '/\varepsilon$ with staggered 
fermions''
\newblock {\it Nucl. Phys.} B (Proc. Suppl.) 63, 293--295 (1998)}

\pubitem{G. Kilcup, D. Pekurovsky, L. Venkataraman
\newblock ``On the $N_f$ and $a$
dependence of $B_K$''
\newblock {\it Nucl. Phys.} B (Proc. Suppl.) 53, 345--348 (1997)}

\end{publist}

\begin{fieldsstudy}

\majorfield{Physics}

%%
%% Note:  If there were only one field of study, the following list 
%%        would best be done using the following command:
%%

\onestudy{Lattice QCD}{Prof. G. Kilcup}
\end{fieldsstudy}

\end{vita}

\tableofcontents
\listoftables
\listoffigures

\chapter{Introduction}
\label{intro}

%=============================================
\section{Overview}
%=============================================

One of the main directions of efforts in both experimental and
theoretical particle physics is obtaining reliable information about
the fundamental free parameters of the Standard Model (SM).
These parameters include the masses of quarks, leptons,
$W$ and $Z$ gauge bosons, and the Higgs particle; coupling
constants of electroweak and strong interactions; and
the elements of the Cabibbo-Kobayashi-Maskawa (CKM) matrix.
Without the exact knowledge of all of these parameters, the SM lacks
the full predictive power. Such knowledge would also allow
to perform various consistency checks based on results of
independent experiments. In addition, there is another important reason 
for knowing the SM parameters: many alternative theories of fundamental
particle interactions have been proposed. Among such theories are
various extensions of the SM, as well as Supersymmetric Grand Unified Theories
(SUSY GUTs), in which the SM itself is a low-energy limit
of more fundamental and unifying theories.
%; strings theories and M-theory,
%which operate on an even more fundamental level. 
In order to check if extensions of SM are really necessary and to 
check if any of the proposed alternatives to the SM are indeed
valid descriptions of physical reality, it is necessary to 
know the parameters of the SM. 

In many cases obtaining the SM parameters from experiment is 
difficult or impossible due to the presence of non-perturbative 
effects. For example, in order to extract certain CKM matrix elements,
very often it is necessary to calculate
matrix elements (MEs) of weak interaction operators between strongly
interacting states (hadrons). In these cases the weak and strong
interactions are tightly intertwined; hence there is
an obligatory non-perturbative component in computing MEs,
since the energy scales involved fall within the nonperturbative
region of Quantum Chromodynamics (QCD), the theory of strong interactions. 
In calculating this nonperturbative component, as in other
aspects of nonperturbative calculations in particle physics,
Lattice Gauge Theory (LGT) plays an increasingly significant role.
It is a first-principles method, and although several types of
errors are inevitably introduced by lattice computations, 
rapid advances in available
computational power as well as algorithmic techniques
are allowing for steadily better control of such errors.

This thesis deals with several cases of application of
Lattice QCD (LQCD) to calculating hadronic matrix elements
of weak operators with the purpose of bridging the gap between theory
and experiment and fixing the parameters of the SM. In addition, 
such calculations are aimed at testing nonperturbative QCD
calculations against experiment so that the developed machinery
of lattice calculations can be employed with confidence in
other useful applications. 

This work addresses the rich phenomenology of $K^0\to\pi\pi$ decays.
One of the long-standing puzzles is the ``$\Delta I=1/2$ rule'',
which is the observation that the transition channel with isospin changing
by 1/2 is enhanced 22 times with respect to transitions
with isospin changing by 3/2. Weak interactions and the short-distance
part of strong interactions are not enough to explain this phenomenon.
In order to study the long-distance part of strong interactions
(namely, to calculate hadronic matrix elements of the operators 
appearing in the effective weak Hamiltonian) a non-perturbative 
method has to be employed.
In the work of this thesis, LQCD is used to this purpose. 
The aim is to explain the $\Delta I=1/2$ rule and to check the validity
of several SM assertions used in obtaining the explanation,
as well as to test the technology of lattice calculations.

In addition, I address the somewhat related issue of $\varepsilon '$, the 
direct CP violation parameter in the neutral kaon system.
Experimentally, the value of \epsp is measured to be non-zero.
Resolving a 6 years old controversy, the Fermilab KTeV group~\cite{Fermilab} 
has recently announced their new
measurement of $\Re \,(\varepsilon '/\varepsilon)$ at 
$(28.0 \pm 4.1) \cdot 10^{-4}$. This is consistent with 
the latest reports from CERN group NA48~\cite{newCERN} of 
$\Re \, (\varepsilon '/\varepsilon) = 
(18.5 \pm 7.3) \cdot 10^{-4}$ as well as with the 
value of $\Re \,(\varepsilon '/\varepsilon) = (23.0 \pm 6.5) \cdot 10^{-4}$
measured by the CERN's previous experiment NA31~\cite{CERN}. 
Both Fermilab and CERN groups are currently
analyzing their data, so enhanced statistics will soon allow to 
decrease the above errors. In addition, an experiment
at Frascati is soon expected to measure the value of 
$\Re \,(\varepsilon '/\varepsilon)$ by a different technique.

It is clearly very interesting to compare the theoretical prediction
for \epsp with the experimental result. Such a comparison will 
directly show if the minimal version of the SM correctly accounts 
for direct CP violation.
Estimating \epsp is currently an area of considerable effort in 
the theoretical physics community~\cite{Efforts}. 
According to some current estimates~\cite{BurasNew}, it appears unlikely
that the SM description of direct CP violation is correct,
so that some extension or modification of the SM may be necessary.  
The weakest part of such estimates is the lack of knowledge of MEs of
certain weak operators~\cite{buras}, which require nonperturbative 
treatment. 
In this work, LQCD is used for calculation of the relevant MEs.

The core of our work is calculation of matrix elements 
$\langle\pi\pi|O_i|K^0\rangle$ for all  basis operators
entering the effective weak Hamiltonian 
(introduced in Sec.~\ref{framework}), within the framework
of LQCD with staggered (Kogut-Susskind) fermions. Then the theoretical
estimates for both $\Delta I=1/2$ rule and \epsp can be made
as mentioned above. 

Statistically significant
results have been obtained for MEs of the basis operators.
Our results for $\Delta I=1/2$ rule (described in Chapter.~\ref{chap:di12})  
confirm a significant enhancement of $\Delta I=1/2$ channel
compared to $\Delta I=3/2$ one.
The exact ratio of isospin amplitudes (Sec.~\ref{sec:ratio}) is subject
to considerable systematic uncertainties, but is consistent
with the experimental value. 

In principle, the MEs calculated in this work are sufficient to 
determine \epsp. However, currently the failure of lattice perturbation theory
in operator matching introduces a large error in the end results.
To reduce this error, nonperturbative operator matching 
has to be performed. Such matching would allow to use the
(statistically strong) results of this work to their fullest. 
To give a crude estimate, we have currently adopted a 
partial nonperturbative renormalization procedure (Sec.~\ref{sec:ansatz}), 
combining perturbative mixing with nonperturbative determination of 
bilinear renormalization constants $Z_S$ and $Z_P$.
This provides an admittedly rough way to do the operator matching. 
The estimates of \epsp obtained in this manner come as a surprise:
the sign of \epsp is negative (see Sec.~\ref{sec:epsp_res} for numbers). 
This result contradicts the experimental findings. Of course, 
the current systematic errors preclude any definite conclusion,
but if this situation persists in the future when such errors are reduced,
the Minimal SM description of CP violation might need a revision. 

Light quark masses are among the poorer known parameters of the SM.
Yet, knowing their values is important, for example for constraining
SUSY GUTs, where an independent prediction for these masses can be made.
LQCD is one of the best-suited, direct techniques for extracting the
light quark masses~\cite{GuptaQuarkMasses}. The major obstacle
of lattice quark mass estimation has been the poor suitability
of the perturbation theory for matching lattice and continuum
operators. In this work such matching is done nonperturbatively,
which provides better estimates of the light
quark masses than previously available. For the strange
quark mass in NDR $\overline{\rm MS}$ scheme at 2 GeV,
in quenched QCD we obtain: $$m_s = 102.9 \pm 5.1\; {\rm MeV}\ .$$

%=============================================
\section{Comparison To Previous Work}
%=============================================

A variety of phenomenological methods such as $1/N_c$ expansion~\cite{Nc} 
and chiral quark models~\cite{CQM} have been used   
for calculation of hadronic MEs relevant for both \epsp
and $\Delta I=1/2$ rule. Without underestimating the
strengths of such methods, it may be fair to say that in prospect
LQCD is one of the best choices for such calculations
(of course, assuming LQCD calculations are computationally
feasible). This is due to the fact that LQCD calculations 
are intrinsically reliable: they are based on first principles
in the sense that, by design, no unjustifiable approximations or
assumptions are introduced. At the same time, it is also true that 
in many cases this strength of the LQCD method lies mainly in the future, 
since, as will become clear from this thesis, sometimes the uncertainties 
associated with current lattice calculations are overwhelmingly large.
Most of these uncertainties will very likely be decreased in the future.  

%Calculation of $Z_S$ and $Z_P$ is itself a part of this thesis
%(Sec.~\ref{Zs}). Calculation of $Z_S$ also allows us to obtain
%the masses of the light quarks ($s$, $d$ and $u$). 

Concerning previous lattice calculations of the MEs discussed in
this thesis, there have been several attempts, but they fell short
of desired accuracy because of technical difficulties and/or insufficient
statistics (due to limited computational power). 
Works reported in Refs.~\cite{KilcupSharpe,BernardSoni,BernardSoniTwopi,MartinelliMaiani,MartinelliTwopi} and reviews in 
Refs.~\cite{BernardReview1,BernardReview2} discuss attempts of
several lattice groups to compute $\Delta I=1/2$ rule on the lattice,
and explain various encountered difficulties.
In addition, several groups~\cite{Twopi,BernardSoniTwopi,MartinelliTwopi} 
have studied MEs
$\langle \pi^+\pi^0|O_i|K^+\rangle$ with reasonable precision;
however, these MEs describe
only $\Delta I=3/2$, but not $\Delta I=1/2$ transition.
In the present simulation, the statistics is finally under control
for both $\Delta I=1/2$ and $\Delta I=3/2$ amplitudes. 

A previous attempt~\cite{KilcupSharpe} to compute MEs relevant
for \epsp on the lattice with staggered fermions did not take
into account operator matching. In this work we repeat this calculation
with better statistics and better investigation of systematic 
uncertainties. In particular, we take into account operator matching
by using a partially nonperturbative renormalization procedure 
(Sec.~\ref{sec:ansatz}). As already mentioned, and as will be discussed 
in further detail in Chapter~\ref{matching},
operator matching is currently the source of the biggest error
in computing $\varepsilon '/\varepsilon$. 

Recently, the RIKEN-BNL-Columbia group~\cite{blum} has reported the results
of their first study of \epsp on the lattice using domain wall fermions.
Compared to the staggered fermions, the domain wall fermion technique
has the advantage of preserving the chiral symmetry away from
continuum limit without introducing extra fermion flavors. The disadvantage
of this method is the need to simulate a lattice with an extra
(5th) dimension. 
The RIKEN-BNL-Columbia group is investigating whether simulations
with domain wall fermions are practically feasible, and whether
the advantages of domain wall fermion technique is worth the 
increased demand for computational power. In any case, it is 
interesting to compare results obtained with these different
methods. According to the latest work of the RIKEN-BNL-Columbia
group~\cite{blum}, the calculations with domain wall fermions 
also produce negative sign of \epsp. 
We note that our results
have higher statistical accuracy and come closer to the continuum
limit ($\beta =6.2$) than those of the RIKEN-BNL-Columbia group.
On the other hand, the RBC group uses fully nonperturbative matching, 
which is definitely an advantage compared to our partial scheme. 
The consistency of both groups' results at $\beta =6.0$ is very 
encouraging, and gives more credibility 
to our partially nonperturbative procedure. 

Concerning calculation of light quark masses,
only recently such calculations have started to use
nonperturbative matching. 
Recently, JLQCD collaboration~\cite{IshizukaNonpert} 
has done a study of light quark masses
with staggered fermions using nonperturbative matching.
The work in this thesis uses similar ensembles, but a 
different method:
we deduce $Z_S$ and $Z_P$ from considering the inverse averaged
propagator, while JLQCD group obtains these two factors from setting
renormalization conditions on appropriate amputated Green's functions.
There is also a small technical difference in the calculational
setup. Despite these differences, our results are consistent:
JLQCD obtains $106.0 \pm 7.1$ MeV, while we obtain
$102.9 \pm 5.1$ MeV for the strange quark mass in NDR $\overline{\rm MS}$
scheme at 2 GeV in quenched QCD.

In addition, some work has recenetly been reported~\cite{WilsonQuarkMasses}
on nonperturbative quark masses with Wilson fermions.
This is a completely independent calculation since Wilson
fermions are conceptually different from staggered fermions
in many respects. It is very encouraging to see that the
staggered and Wilson results are consistent.

Phenomenological methods, such as QCD sum rules,
can also be applied for deducing quark masses~\cite{SumRules}. 
However, they possess 
several major disadvantages when compared to lattice 
calculations, as explained in Ref.~\cite{GuptaSumRules}. 

%=============================================
\section{Organization Of This Thesis}
%=============================================

This thesis is structured as follows. In Chapter~\ref{framework}
we explain the framework in which we work: effective weak theory
obtained by Operator Product Expansion (OPE) and Renormalization Group (RG)
evolution. We also discuss the
exact connection between the MEs that we strive to compute and
the physical quantities of interest, as well as some other questions
related to continuum physics.

Chapter~\ref{sim} delves into the theory and practice of lattice
calculations. It
starts by an overview (Sec.~\ref{background}) of the generic setup 
of lattice calculations of hadron masses and MEs, including the most 
commonly introduced errors. This is meant to be not a systematic
introduction, but rather a quick refresher. Staggered fermions are 
discussed in a little more detail in Sec~\ref{sec:stag}, 
although the basic facts are 
only stated, not proven. The interested reader will readily find the
missing information in available texts and review articles. 
Then, concentrating on lattice calculations 
relevant for this thesis, Sec.~\ref{TheoryLattice} discusses some matters
of principle in such calculations. Sec.~\ref{sec:tricks} contains
detailed explanation of the techniques and tricks we have used
to implement the above strategies on the numerical level. 
Some of the techniques of dealing with staggered fermions are rather 
tedious, so I give explicit formulae for reference purposes. 
(Some other explicit expressions which 
may be useful to anyone attempting to perform similar calculations,
can be found in the Appendix). Sec.~\ref{sec:tricks} 
also contains the parameters of our simulations.
I close the chapter with a brief mention of some computational issues
(Sec.~\ref{sec:comp}).

Turning to results of our calculations, 
Chapter~\ref{chap:di12} presents the calculated 
$\Delta I=1/2$ and $\Delta I=3/2$ amplitudes and their ratio,
with a discussion of the systematic errors. 
Chapter~\ref{matching} is devoted to operator matching and \epsp.
Sec~\ref{sec:pert} explains how the operator matching problem 
together with other systematic errors precludes a reliable calculation of 
$\varepsilon '/\varepsilon$. Sec.~\ref{sec:bil} concerns calculation
of $Z_S$ and $Z_P$ bilinear renormalization factors nonperturbatively,
including the detailed setup of lattice calculations for this case
as well as lattice parameters used. Apart from results for $Z_S$ and $Z_P$,
given in Sec~\ref{sec:Zs}, I present our results for the light quark
masses obtained with nonperturbative matching in Sec~\ref{sec:QM}. 
Sec~\ref{sec:ansatz} describes the partially nonperturbative procedure 
for matching four-fermion operators relevant for estimating \epsp,
and Sec.~\ref{sec:epsp_res} contains the results for \epsp and a discussion 
of the errors in its determination. A brief summary of results for \epsp
is given in Sec.~\ref{epsp.sum}. 
Chapter~\ref{conclusion} contains conclusions.

\chapter{Theoretical Framework}
\label{framework}

%=============================================
\section{Effective Weak Theory}
%=============================================

The standard approach in applying the SM to topics mentioned in 
the introduction is to
use the Operator Product Expansion (OPE) at the $M_W$ scale,
that is to find
a suitable basis of four-fermion operators which give
effectively the same interactions as the charged-current
weak interactions for a given problem of interest (see Ref.~\cite{buras}
and references therein). This is done by matching the 
strengths of interaction vertices of the basis operators with
that obtained by a W exchange.
%For the next-to-leading order, at which we work,
%it is necessary to evaluate diagrams up to one loop. 
%The results of such matching are readily available (e.g., see~\cite{buras}). 
The effective weak Hamiltonian for nonleptonic kaon decays is given 
as a linear combination
of basis operators $O_1$ and $O_2$, defined as follows: 
\begin{eqnarray}
O_1 & = & (\bar{s}_\alpha \gamma_\mu (1-\gamma_5) u_\beta )
(\bar{u}_\beta \gamma^\mu (1-\gamma_5)d_\alpha )  \nonumber \\
O_2 & = & (\bar{s}_\alpha \gamma_\mu (1-\gamma_5)u_\alpha)
(\bar{u}_\beta \gamma^\mu (1-\gamma_5)d_\beta )  \nonumber
\end{eqnarray}
To compute physical amplitudes,
one needs to compute matrix elements of this Hamiltonian 
between the suitable hadronic states (in this work, kaon and two-pion
states). 

Note that the most natural energy scale $\mu$ for lattice calculations
is around or less than $1/a$, which is 2 -- 4 GeV in current simulations.
This is much less than the scale $M_W$ of the above 
effective Hamiltonian. It follows that if the QCD corrections 
are to be included in the effective theory description,
their contribution from one-loop diagrams (which are 
$O(\alpha_s \log (m_W^2/\mu^2))$) can be sizable.
Higher-order corrections are $O([\alpha_s \log (m_W^2/\mu^2)]^n)$, 
which should not be neglected.
A well-known method of resummation of these ``leading logarithm'' terms 
involves using Renormalization Group (RG) equations to evolve the theory down 
the energy scale to 
a region more accessible for current lattice calculations
(2 -- 4 GeV). Then the resulting theory is valid up to terms
$O(\alpha_s [\alpha_s \log (m_W^2/\mu^2)]^n)$, and the method is
called RG Improved Perturbative Expansion. 
To achieve an even higher
accuracy, as is quite necessary in the case of kaon nonleptonic decays,
next-to-leading order logarithms can be resummed by using 
one-loop matching at the $M_W$ threshold and using RG equations
with anomalous dimension matrix computed to two loops. 

In the process of RG evolution heavy quarks are integrated out
by matching the theories with and without
the given quark flavour at a threshold of the order of the
given quark's mass.  For detailed discussion of this procedure
refer to Ref.~\cite{buras}. Due to elimination of heavy quarks
from the effective theory and due to operator
mixing the basis of operators is enlarged to 10.
The effective Hamiltonian takes the following form: 
\begin{equation}
H_{\mathrm W}^{\mathrm eff} = 
\frac{G_F}{\sqrt{2}} V_{ud}\,V^*_{us} \sum_{i=1}^{10} \Bigl[
z_i(\mu) + \tau y_i(\mu) \Bigr] O_i (\mu) 
 \, , 
\end{equation}
where $G_F$ is the Fermi constant, 
$z_i$ and $y_i$ are Wilson coefficients (at two-loop order), 
$\tau \equiv - V_{td}V_{ts}^{*}/V_{ud} V_{us}^{*}$,
$V$ is the CKM mixing matrix,  
and $O_i$ is a basis of four-fermions operators defined as follows:
\begin{eqnarray}
\label{eq:ops1}
O_1 & = & (\bar{s}_\alpha \gamma_\mu (1-\gamma_5) u_\beta )
(\bar{u}_\beta \gamma^\mu (1-\gamma_5)d_\alpha )  \\
O_2 & = & (\bar{s}_\alpha \gamma_\mu (1-\gamma_5)u_\alpha)
(\bar{u}_\beta \gamma^\mu (1-\gamma_5)d_\beta )  \\
\label{eq:ops3}
O_3 & = & (\bar{s}_\alpha \gamma_\mu (1-\gamma_5)d_\alpha )
\sum_q(\bar{q}_\beta \gamma^\mu (1-\gamma_5)q_\beta ) \\
O_4 & = & (\bar{s}_\alpha \gamma_\mu (1-\gamma_5)d_\beta )
\sum_q(\bar{q}_\beta \gamma^\mu (1-\gamma_5)q_\alpha ) \\
O_5 & = & (\bar{s}_\alpha \gamma_\mu (1-\gamma_5)d_\alpha )
\sum_q(\bar{q}_\beta \gamma^\mu (1+\gamma_5)q_\beta )  \\
O_6 & = & (\bar{s}_\alpha \gamma_\mu (1-\gamma_5)d_\beta )
\sum_q(\bar{q}_\beta \gamma^\mu (1+\gamma_5)q_\alpha )  \\
O_7 & = & \frac{3}{2}(\bar{s}_\alpha \gamma_\mu (1-\gamma_5)d_\alpha )
\sum_q e_q (\bar{q}_\beta \gamma^\mu (1+\gamma_5)q_\beta ) \\
O_8 & = & \frac{3}{2}(\bar{s}_\alpha \gamma_\mu (1-\gamma_5)d_\beta )
\sum_q e_q (\bar{q}_\beta \gamma^\mu (1+\gamma_5)q_\alpha ) \\
O_9 & = & \frac{3}{2}(\bar{s}_\alpha \gamma_\mu (1-\gamma_5)d_\alpha )
\sum_q e_q (\bar{q}_\beta \gamma^\mu (1-\gamma_5)q_\beta ) \\ 
O_{10} & = & \frac{3}{2}(\bar{s}_\alpha \gamma_\mu (1-\gamma_5)d_\beta )
\sum_q e_q (\bar{q}_\beta \gamma^\mu (1-\gamma_5)q_\alpha ) 
\label{eq:ops10}
\end{eqnarray}
Here $\alpha$ and $\beta$ are color indices (summation over which 
is implied), $e_q$ are quark
electric charges, and summation over $q$ is done over all light quarks. 
This description is valid up to 
$O(\alpha_s^2 [\alpha_s \log (m_W^2/\mu^2)]^n)$ terms.

%------------------------------------------------------------
\subsection{Treatment Of Charm Quark}
%------------------------------------------------------------

A natural question at this point is whether one should consider
the charm quark as ``light'' or ``heavy''. In the first case, it is to be
included in the sum over $q$ in the above equations, and some
additional operators containing the charm quark should be considered; 
while in the case
it is considered ``heavy'' it is integrated out, and only
$u$, $d$ and $s$ quarks are left in the theory. Since the scales
we consider are of the same order as $m_c$, the choice
is ambiguous. The physical results are not expected to depend on this
choice, disregarding the $O(\alpha_s^2 [\alpha_s \log (m_W^2/\mu^2)]^n)$ 
terms and higher-dimensional operators (proportional to $1/m_c$).

In practice, when performing lattice calculations it is more 
convenient to work in the three-quark effective theory. In this
framework the set of operators is smaller than it would be
in the four-quark theory,
and the problems of discretizing the charm quark field are not present. 
Such approach is adopted in the present work. 

%=============================================
\section{Definitions}
%=============================================

Here we define some quantities of interest. The 
$\Delta I=1/2$ and $\Delta I=3/2$ transition amplitudes 
are defined as follows, according to the isospin of the
final two-pion state:
\begin{equation}
\label{amp}
A_{0,2} \equiv \langle (\pi\pi )_{I=0,2}|H_W|K^0\rangle \;e^{-i\delta_{0,2}},
\end{equation}
where $\delta_{0,2}$ are the final state interaction phases of the
two channels. Experimentally
%\footnote{The error is due to 
%interactions changing isospin by 5/2.},
\begin{equation}
\omega = \Re A_0 /\Re A_2 \simeq 22 \, .
\end{equation}
One goal of our work is to check if the theory gives the same ratio.
Also, we would like
to verify if predictions for $\Re A_0$ and $\Re A_2$ separately
agree with experiment.

Another direction of our work is calculating the size of direct 
CP violation. Direct CP violation refers to the fact that
the neutral kaon state $K_2$, defined as an odd CP eigenstate,
has a non-zero (though tiny) probability to decay
into two-pion state, which
is CP-even. This is to be contrasted with the indirect CP
violation, a bigger and better-studied phenomenon
arising from the fact that the eigenstates of the weak Hamiltonian
($K_S$ and $K_L$) are mixtures of CP eigenstates ($K_1$ and $K_2$).
Indirect CP violation is characterized by parameter $\varepsilon$,
which is defined as
\begin{equation}
\varepsilon = \frac{{\rm A}(K_L \to \pi\pi )}{{\rm A}(K_S \to\pi\pi )}
\end{equation}
and experimentally measured at $2.26 \cdot 10^{-3}$.
Direct CP violation is characterized by parameter $\varepsilon '$,
defined as
\begin{equation}
\varepsilon '= \frac{{\rm A}(K_2 \to \pi\pi )}{{\rm A}(K_1 \to\pi\pi )}\ .
\end{equation}
As mentioned in the introduction, two latest experiments report
the measured ratio {\mbox{Re (\epsp)} at $(28.0 \pm 4.1)\cdot 10^{-4}$ and
$(18.5\pm 7.3)\cdot 10^{-4}$. 

It can be shown~\cite{Georgi,Donoghue} that in the Standard Model 
$\varepsilon '$ can be computed in terms 
of imaginary parts of the isospin amplitudes defined above:
\begin{equation}
\varepsilon ' = -\frac{\Im A_0 - \omega \Im A_2}{\sqrt{2}\;\omega\;\Re A_0}
\; e^{i(\pi/2 + \delta_2 - \delta_0)}.
\end{equation}
Expressing the amplitudes in terms of the basis operators of
the effective Hamiltonian, we find that the experimentally measured ratio
of direct to indirect CP violation is given by  
\begin{equation}
\label{eq:epsp}
\Re \,(\frac{\varepsilon '}{\varepsilon}) =
\frac{G_F}{2\omega |\varepsilon |\Re{A_0}} \,
\mbox{Im}\, \lambda_t \, \,
 \left[ \Pi_0 - \omega \: \Pi_2 \right] ,
\end{equation}
where
\begin{eqnarray}
\label{P0}
 \Pi_0 & = &  \sum_i y_i \, \langle (\pi\pi )_{I=0}|O_i^{(0)}|K^0\rangle 
(1 - \Omega_{\eta +\eta '}) \\
\label{P2}
 \Pi_2 & = &  \sum_i y_i \, \langle (\pi\pi )_{I=2}|O_i^{(2)}|K^0\rangle  
\end{eqnarray}
with $\mbox{Im}\, \lambda_t \equiv \Im V_{td}V^*_{ts}$, and where
$\Omega_{\eta + \eta'} \sim 0.25\pm 0.05$ takes into account the effect
of isospin breaking (since $m_u \neq m_d$). $O_i^{(0)}$ and
$O_i^{(2)}$ are isospin 0 and 2 parts of the basis operators.
Their expressions are given in the Appendix for completeness.
%due to $\eta$ and $\eta '$ mixing. 
The matrix elements in the last two equations contain long-distance
dynamics of strong interactions, and therefore a non-perturbative
calculation, such as the one considered in this thesis, is necessary
to compute \epsp.  
\chapter{Lattice Simulations}
\label{sim}

As explained in the preceding chapter, crucial knowledge
about long-distance dynamics is contained in matrix elements
$\langle\pi\pi |O_i|K^0\rangle$ of weak operators from the 
basis in Eqs.~(\ref{eq:ops1})--(\ref{eq:ops10}). 
In this chapter we review the basics of lattice calculations,
discuss the setup and techniques of calculation of matrix elements
in question, and give detailed information about parameters
used in the simulation.

%======================================================
\section{Lattice Background}
%======================================================
\label{background}

Let me start with a quick overview of general facts about
Lattice Gauge Theory (LGT) calculations. More details
are available to the interested reader in numerous textbooks and 
review articles~\cite{SharpeTASI,montvay,GuptaReview,Kogut,creutz}.
LGT is a 
systematic, first-principles method for numerical solution
of various theories which are hard or impossible to solve
analytically. In particular, one of the most often mentioned
strengths of the LGT technique is its ability to deal with
non-perturbative aspects of theories such as QCD. 

LGT has been successfully applied and has made 
a number of important contributions in various areas
such as calculation of matrix elements, quark masses, glueballs, 
heavy flavour physics, finite temperature theories to mention 
a few. The impact of LGT on particle physics phenomenology
is likely to expand even further in the future, since most of the
uncertainties of LGT simulations can be improved without limit
with increasing computational resources.   
In this work I am going to concentrate on application of LGT
to calculating matrix elements of weak operators as well as
light quark masses in QCD.  

Computer simulations are made possible by discretizing 
Eucledian space-time
on a four-dimensional hypercubic lattice of finite extent
in all directions. This certainly introduces
various errors compared to the continuum description, for example,
violation of Eucledian invariance and introduction of ultraviolet
and infrared momentum cutoffs. In practice, the discretized version 
of the theory is
formulated in such a way as to make possible the extrapolation
to continuum limit by repeating identical calculations at 
progressively smaller lattice spacing values (denoted $a$). Since  
one wants to keep the physical size of the ``box'' fixed, the
continuum extrapolation means increasing the number of lattice sites,
and therefore making the computations progressively more 
computer-intensive as one approaches the limit of $a=0$. 
One also has to check that the physical size of the box is large
enough to contain the system in question. This can be established
by checking that results do not change in repeated calculations at 
bigger volumes. Most often periodic boundary conditions are used.

%----------------------------------------------------------
\subsection{Lattice QCD Lagrangian}
%----------------------------------------------------------

Lattice QCD simulations are done in the framework of Eucledian
functional integral formulation of the theory.  
As is well-known, the continuum QCD Lagrangian equals
\begin{equation}
{\cal L} = \frac{1}{4} \ F^a_{\mu\nu}F^a_{\mu\nu} + 
\sum_{q}\overline{\psi} (\Dslash +m_q)\psi
\end{equation}
The partition function can be written as
\begin{equation}
Z = \int {\cal D} U\; {\cal D} \psi {\cal D} \overline{\psi}\ 
e^{-S_g[U]-S_q[\psi ,\overline{\psi}, U]} = 
\int {\cal D}U \prod_q {\rm det} (M_q[U]) e^{-S_g[U]}\ ,
\end{equation}
where ${\rm det} (M_q[U]) \equiv {\rm det} (\Dslash +m_q)$. 

In the original formulation of Lattice QCD by K. Wilson~\cite{Wilson}
the quark fields are defined on each lattice site, while the gluon 
fields ``live'' on links connecting these sites. Namely, consider
a {\it gauge link}, an $SU(3)$ matrix variable corresponding to
the path-ordered line integral along one link in the continuum: 
\begin{equation}
U_\mu (x)= P\ \exp (ig\int^x_{x+a\hat{\mu}}dx' A^c_\mu (x')T^c)\ ,
\end{equation}
where $T^c$ are color group generators, $x$ is a lattice site,
$g$ is the QCD coupling constant. In the lowest-order in $a$,
these variables are associated with gauge fields as follows:
\begin{equation}
U_\mu (x)= \exp (-iagA^c_\mu (x+\hat{\mu}a/2)T^c)\ .
\end{equation}
The gauge links provide a convenient way to introduce gauge-invariance.
Basically, all gauge-invariant quantities are either products of 
gauge links along a closed path in space-time, or expressions
of the type $\overline{\psi} (x)U(x,y)\psi (y)$, where
$U(x,y)$ is the product of gauge links along a path from
$y$ to $x$. 
 
The construction of lattice QCD Lagrangian is a standard textbook
material. The gauge part of action equals
\begin{equation}
S_g[U] = \beta \sum_x\sum_{\mu < \nu} (1-\frac{1}{3} {\rm Re\ Tr} 
P_{\mu\nu}(x)) \ ,
\end{equation} 
where $\beta \equiv 6/g^2$ and $P_{\mu\nu}(x)$ is so-called {\it plaquette}, 
defined as the product of 4 gauge links around the minimal closed path:
\begin{equation}
P_{\mu\nu}(x) \equiv U_\nu (x) U_\mu (x+\hat{\nu}a) U_\nu^\dag (x+\hat{\nu}a+\hat{\mu}a) U_\mu^\dag (x+\hat{\mu}a)\ .
\end{equation}
The fermionic part of action is
\begin{equation}
S_q[\overline{\psi},\psi ,U] = \sum_x \overline{\psi} (\Dslash +m)\psi \ ,
\end{equation}
where $\Dslash$ is a suitably discretized, Euclidean covariant derivative
(to be defined later).
Quantities of interest in lattice simulations can be extracted from
correlation functions obtained by weighted average as follows:
\begin{equation}
\langle C\rangle = \frac{1}{Z}\int {\cal D}U \prod_q{\rm det}\;(M_q[U])\;
e^{-S_g[U]}\; C[U]\ .
\end{equation}
Taking such integrals is an enormous task, since it involves
integrating over $4L^4$ $SU(3)$ variables. In practice, Monte Carlo
with importance sampling is used. A number of 
efficient algorithms have been developed For this purpose, 
the simplest of them being
the Metropolis (heatbath) algorithm. We only mention the names of the 
algorithms used for this work in Section~\ref{sec:tricks}; however, 
the discussion of them can be 
found in many texts on lattice gauge theory and is beyond the scope
of this work. The bottom line is that any algorithm produces a set
(``ensemble'') of $N$ gauge configurations $U_i$ at given $\beta$, 
which are randomly sampled from probability distribution
\begin{equation}
\label{weight}
P \propto \prod_q{\rm det}(M_q[U])e^{-S_g[U]}\ .
\end{equation}
Correlation functions are then obtained by averaging over the ensemble:
\begin{equation}
\langle C\rangle = \frac{1}{N} \sum_{i=1}^N C[U_i]\ .
\end{equation}

%----------------------------------------------------------
\subsection{Quenched Approximation}
%----------------------------------------------------------

It should be mentioned that in practice calculation of fermionic 
determinants in Eq.~(\ref{weight}) is computationally very demanding.
Only several years ago realistic lattice calculations have started
to include it in simulations. The alternative is to replace it
by a constant. This is the so-called quenched
(or valence) approximation. It corresponds to a continuum theory
just like QCD but without any closed fermion loops. In other
words, it is assumed that the effects of the sea quarks can be absorbed
into renormalized charges. The error
introduced by such an approximation is not known exactly, but 
in most cases Quenched QCD (QQCD) is expected to produce results 
within 10--15\% of those obtained in QCD. This has been confirmed
numerically for a number of quantities. 
Thus, QQCD can be thought of as a convenient test platform for 
developing algorithms for many calculations, being very
similar to QCD. Moreover,
the results obtained in QQCD are sometimes valuable by themselves, 
especially in cases where
it is interesting to know the answers even at the order of magnitude.

In this work we have mostly used quenched approximation. To see if
quenching introduces any serious error in the quantities of interest,
we have also used one partially unquenched ensemble with 2 flavors
of sea quarks of mass $m=0.01$ at $\beta=5.7$. 

%----------------------------------------------------------
\subsection{Extraction Of Hadron Masses And Matrix Elements}
%----------------------------------------------------------

First consider {\it correlation function}
\begin{equation}
\label{Ct}
\langle C(t)\rangle = \langle 0|T[O^\dag (t)O (0)]|0\rangle\ ,
\end{equation}
where $O$ is an operator that has the suitable quantum numbers
to create a hadron. For example, in order to create $\pi^+$
at rest, the
operator $O(t) = \sum_{\mathbf x}\overline{u}(t,{\mathbf x})\gamma_5 
d(t,{\mathbf x})$ can be used. After insertion of complete set of states 
and assuming $t>0$ we obtain:
\begin{eqnarray}
\langle C(t)\rangle = \langle 0|O^\dag (0) e^{-Ht}O(0)|0\rangle = 
\sum_n |\langle n|O|0\rangle|^2e^{-E_n t}\frac{1}{2E_n V}\ .
\end{eqnarray}
If we consider operators of zero spatial momentum,
then the lightest particle for which $\langle n|O|0\rangle$ is
non-vanishing dominates the
latter expression for large $t$. In cases of our interest it is always 
the pion or kaon~\footnote{Note that in our simulations we have always taken 
$m_s=m_d$, so that the strange and down quarks are indistinguishable 
in terms of computations, and therefore kaons are indistinguishable
from pions.}. Making an exponential fit to the correlator $C(t)$
data, it is possible to extract the mass of the hadron. In addition,
it is possible to extract the matrix element 
$\langle \pi^+ |\overline{u}\gamma_5 d|0\rangle$. If we consider
operator $\overline{u}\gamma_\mu\gamma_5 d$, we can
extract the pseudoscalar decay constant $f_\pi$ (see Sec.~\ref{sec:fpi}).

In order to show how the calculation of $C(t)$ is done, let us rewrite
Eq.~(\ref{Ct}) as follows:
\begin{equation}
\langle C(t)\rangle = \sum_{\mathbf x,y}
\langle 0|T[\overline{d}_{\alpha i}({\mathbf x},t)
\;\Gamma_{ij}\; u_{\alpha j}({\mathbf x},t) \overline{u}_{\beta k}(
{\mathbf y},0)
\;\Gamma_{kl}\; d_{\beta l}({\mathbf y},0)|0\rangle\ ,
\end{equation}
where the summation is done both over color and spin indices
and $\Gamma$ are any matrices specifying the spin structure of the
operators. In terms of the propagator $G$(m;x,y) of a quark
of mass $m$ defined by  
\begin{equation}
G^{\alpha\beta}_{ij} (m;x,y) \equiv \langle 0|T[\psi^\alpha_i(x)
\overline{\psi}^\beta_j(y)]|0\rangle\ 
\end{equation}
the expression for $\langle C(t)\rangle$ can be rewritten 
as follows\footnote{The corresponding expressions for staggered fermions 
are more complicated, and they are given below in Sec.~\ref{sec:tricks},
along with detailed explanation of the meson sources and sinks
that we have used.}:
\begin{eqnarray}
  \langle C(t)\rangle & = &\sum_{\mathbf x,y}
\langle G^{\alpha\beta}_{jk}({\mathbf x},t;{\mathbf y},0)
\; \Gamma_{kl}\; G^{\beta\alpha}_{li}({\mathbf y},0;{\mathbf x},t)
\; \Gamma_{li}\;\rangle \nonumber \\
& = & \sum_{\mathbf x,y}
\langle {\rm Tr}[G({\mathbf x},t;{\mathbf y},0)\;
\Gamma \; G({\mathbf y},0;{\mathbf x},t)\; \Gamma ] \rangle \ ,
\label{prop1}
\end{eqnarray}
where the trace is taken over both color and spin indices and
the brackets imply averaging over gauge configurations.

\begin{figure}[htb]
\begin{center}
\leavevmode
\centerline{\epsfxsize=5.5in \epsfbox{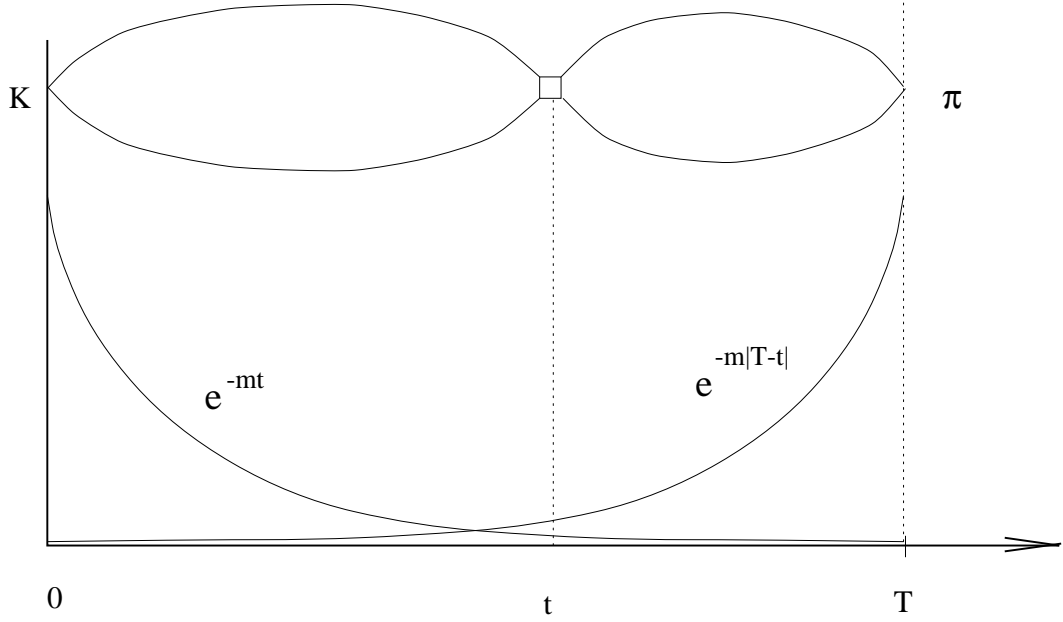}}
\end{center}
\caption{The general setup of the simulation. An ``eight'' 
contraction is shown for convenience. The kaon source is
at the timeslice 0, while the pion sink is at the timeslice $T$.
The operator is inserted at an adjustable time $t$. The result of this
contraction is proportional to the product of two exponentials
shown in the figure.}
\label{setup}
\end{figure} 

For our purposes it will be necessary to calculate three-point functions
for matrix elements of the type $\langle \pi|O_{K\pi}|K\rangle$, and here
I would like to mention the setup of such calculations. Suppose
the kaon source is at $t_0=0$, the operators are inserted at 
an adjustable time $t$, and the pion sink is located at the
time $T$ (see Fig.~\ref{setup}), where $T$ is sufficiently large.
Consider the correlation function
\begin{equation}
\langle C(t,T)\rangle = \langle 0|O_\pi (T) O_{K\pi}(t) O_K(0)|0\rangle\ .
\end{equation}
Inserting two complete sets of states as before, we obtain
\begin{equation}
\langle C(t,T)\rangle = \sum_{n_1,n_2}\langle 0|O_\pi (T)|n_1\rangle 
\frac{e^{-E_{n_1}(T-t)}}{2E_{n_1}V} 
\langle n_1|O_{K\pi}(t)|n_2\rangle \frac{e^{-E_{n_2}t}}{2E_{n_2}V} 
\langle n_2| O_K(0)|0\rangle\ .
\end{equation}
As before, the lightest states (in this case, kaon and pion) 
dominate the expression at large $t$, so we obtain:
\begin{equation}
\langle C(t,T)\rangle \to Z e^{-m_\pi T} \langle \pi|O_{K\pi}|K\rangle \ ,
\end{equation}
where $\sqrt{Z} = \langle 0|O_\pi |\pi\rangle \frac{1}{2m_\pi V}$, and
we have assumed that both kaon and pion are created at rest and have equal 
masses. Note that the last expression does not depend on $t$, so
we expect to see a plateau on the plot of $\langle C(t,T)\rangle$
vs. $t$. This plateau is our working region (see Fig.~\ref{plateau}). 
It is present only if $T$ is large enough. In order to know
exactly how large $T$ should be in order to be free from
excited states contamination and wraparound effects, we have done some 
checks (see discussion in Sec.~\ref{sec:B}). Explicit expressions
for $\langle C(t,T)\rangle$ in terms of propagators are 
given below in Section~\ref{sec:tricks}.

In addition, we will need to compute the ratio of matrix elements
\begin{equation}
\frac{\langle 0|O|K^0\rangle}{\langle 0|\overline{s}\gamma_5 d|K^0\rangle}\ , 
\end{equation}
in which the factors $\sqrt{Z}e^{-mt}$ cancel and one expects to see a 
plateau for large \mbox{enough $t$.}

Note that all quantities extracted from the lattice are in lattice units.
In order to convert to physical units, each quantity should be
multiplied by $a^{-p}$, where $p$ is the mass dimension of the quantity.
The lattice spacing uniquely corresponds to $\beta$. 
In particular, in the {\it perturbative
scaling} region, when the lattice spacing is small enough, the following
relationship exists:
\begin{equation}
a(\beta) = a_0 \biggl(\frac{\beta_0\;g_{\overline{\rm MS}}^2}{16\pi^2}\biggr)^
{-\beta_1/ 2\beta_0^2}
\exp\biggl(\frac{-1}{2\beta_0\; g_{\overline{\rm MS}}^2}\biggr)\ ,
\end{equation}
where $\beta_0 = 11-\frac{2}{3}N_f$, $\beta_1 = 102-\frac{38}{3}N_f$
and $N_f$ is the number of sea quark flavors. The coupling
constant $g_{\overline{\rm MS}}$ can be obtained from the average
plaquette value at a given $\beta$ 
(see Eqs.~(\ref{eq:coupling1}) and~(\ref{eq:coupling2})).
The overall scale $a_0$ can be obtained by dividing one chosen
quantity in lattice units by the experimental value in physical
units. For example, the mass of the $\rho$ meson can be 
calculated on the lattice, and the lattice spacing can be set by
$$a= \frac{(m_\rho a)_{\rm latt}}{m_\rho ({\rm GeV})}\ .$$
Exactly how the lattice scale is determined varies from computation
to computation.
In order to take the continuum limit, calculations should be repeated
with several values of $\beta$ in the region of small enough $a$.

%----------------------------------------------------------
\subsection{Discretizing Fermions}
%----------------------------------------------------------

So far we have not discussed the fermion part of the Lattice QCD action,
in particular the discretized version of the operator $\Dslash$.
One straightforward ({\it naive}) scheme is to write the fermion action as
\begin{equation}
S_N = \frac{1}{2}\sum_{x,\mu}\overline{\psi}(x)\gamma_\mu
[U_\mu (x)\psi (x+a\hat{\mu}) - U^\dag_\mu (x-a\hat{\mu})\psi (x-a\hat{\mu})]
+\sum_x m\overline{\psi}(x)\psi (x)\ ,
\end{equation}
where $\gamma_\mu$ are the Eucledian Dirac matrices, which are hermitian 
and satisfy $\{\gamma_\mu ,\gamma_\nu\} = 2\delta_{\mu\nu}$. 
This form of action
satisfies the important gauge invariance requirement and seems to correspond
to the QCD action in continuum limit. However, a well-known problem
with the above naive action lies in the following. 
According to this action, the free fermion
propagator is 
\begin{equation}
G(p) = \frac{1}{i\sum_\mu \sin (p_\mu a)\gamma_\mu +m} \ .
\end{equation}
This propagator implies 16 poles at $p\in \{0,\pi /a\}^4$
(in the limit of vanishing fermion mass $m$), which
includes not only the physical pole in the vicinity of 0, but
also 15 spurious poles corresponding to other species
of fermions (so-called {\it doublers}). Another important feature of 
the naive action is that the doublers come in pairs of opposite
chirality, so that the net chirality is always zero. In particular,
the chiral anomaly is completely absent~\cite{KarstenSmit}. 
Using the naive action, it is impossible to define any theory with 
chiral fermions, such as the electroweak theory. Moreover, the naive
action is unsuitable for describing QCD, since the 
chiral $SU(3)\times SU(3)$ symmetry, essential to the theory,
is reduced to vector $SU(3)$, and important axial Ward identities
are missing. 

Of course, the naive fermion action above is not the only possible action
with needed continuum limit.
Much work has been done in the direction of devising a fermion
discretization scheme that would be free from doublers and, more importantly,
have the correct chiral properties. In such work, one has to be aware of 
the Nielsen-Ninomiya theorem~\cite{Nielsen}, stating that the 
fermion doubling problem is unavoidable for any local, anti-hermitian 
discretization with periodic boundary conditions that preserves 
translational invariance.

%Several solutions to the ``doubling problem'' have been proposed. 
Currently most popular approaches which in some way deal with
the ``doubling problem'' include {\it Wilson
fermions}, {\it domain wall fermions} and {\it staggered fermions}. 
In the approach of 
Wilson fermions~\cite{WilsonFermions} an extra term is added
to the action which changes the propagator in such a way
as to make the effective physical mass of the 15 extra doublers 
infinite in the continuum limit; therefore they effectively decouple. 
Unfortunately this extra term completely breaks the chiral symmetry,
though it may be recovered in the continuum limit.
The approach of domain wall fermions is to overcome the
Ninomiya-Nielsen theorem by introducing the 5th dimension of
the lattice. This approach preserves chiral symmetry without
extra doublers. However, the computational cost of dealing
with the 5th dimension is high. The Wilson fermions and domain wall 
approaches have been described elsewhere in detail. 
In the following I concentrate
on the approach of staggered (Kogut-Susskind)
fermions~\cite{Susskind}, since it was assumed
throughout this work. 

%----------------------------------------------------------
\subsection{Basic Facts About Staggered Fermions}
%----------------------------------------------------------

\label{sec:stag}

In the staggered fermion approach, the extra doublers are kept.
However, their number is reduced from 16 to 4 (by using the
spin-diagonalization procedure, see below), and in such
a way as to obtain nice chiral properties even away from continuum.

The spin-diagonalization procedure~\cite{kluberg} is done by
changing variables from $\chi$ to $\psi$:
$$\psi (x) = \Gamma_x \chi (x)\ ,$$ 
where $4\times 4$ matrices $\Gamma_x$ 
satisfy the following property: 
$$\Gamma^\dag_x\gamma_\mu \Gamma_{x+a\hat{\mu}} = \eta_\mu (x)\ ,$$ 
with $\eta_\mu (x)$ being diagonal unitary matrices. One possible
realization is to use 
\begin{equation}
\Gamma_x = \gamma_1^{x_1} \gamma_2^{x_2} \gamma_3^{x_3} \gamma_4^{x_4} \ .
\end{equation}
With this choice the phases $\eta_\mu$ are equal to 
\begin{equation}
\eta_\mu (x) = (-1)^{x_1+\ldots +x_{\mu -1}}\ .
\end{equation}
It is easy to show that the naive action in terms of new variables
is
\begin{equation}
\label{eq:AcStag}
S = \frac{1}{2}\sum_{x,\mu} \eta_\mu (x)[\overline{\chi}(x)U_\mu (x)
\chi (x+a\hat{\mu}) - \overline{\chi}(x+a\hat{\mu})U^\dag_\mu (x)\chi (x)] + 
\sum_x m\overline{\chi}(x)\chi (x)\ .
\end{equation}
Since this action does not mix four components of $\chi$,
they contain redundant information, and therefore three of them 
may be dropped. From now on we will understand that in Eq.~(\ref{eq:AcStag})
$\chi(x)$ and $\overline{\chi}(x)$ are 
single-component Grassman variable fields. 
Note that in the kinetic term the fields at odd sites
are coupled  only to fields at even sites. Therefore,
in addition to the symmetries of charge conjugation,
parity, shifts and discrete rotations, the staggered
action preserves the $U_{\rm odd}(1)\times U_{\rm even}(1)$ 
symmetry in the massless limit, which corresponds to 
a combination of fermion 
number conservation and $U(1)$ chiral symmetry in the continuum.

The spin-diagonalization procedure allows one to decrease the number
of fermion doublers by 4 times. The remaining 4 species can be
identified as independent 4 flavors of physical fermions in 
the continuum as follows. In order to ``make contact'' with physical 
Dirac spinor fields, the space-time is divided into hypercubes,
each spanning 2 sites in each of the 4 dimensions.
The field $\chi$ at the 16 sites of each hypercube is interpreted
as various spinor components of the ``physical'' quark field. 
In this way, one introduces quark fields that ``live'' on a coarse 
lattice with spacing $2a$ and have 16 components, which
correspond to 4 Dirac (spin) indices and 4 indices corresponding
to independent, identical flavors of fermions in the continuum limit.
Mathematically, one defines 
\begin{eqnarray}
q^{\alpha a}(y) & = &\frac{1}{8}\sum_A \Gamma_{A;\alpha a}\;\chi (y,A) \ ,\\
\overline{q}^{\alpha a}(y) & = &\frac{1}{8}\sum_A \overline{\chi}(y,A)
\;\Gamma^\dag_{A;a\alpha } \ .
\end{eqnarray}
In this equation $\chi$ and $\overline{\chi}$ are labeled by
hypercube coordinate $y$ and coordinate inside the hypercube
$A\in\{0,1\}^4$, so that the total coordinate of a given hypercube
corner is $x=(2y+Aa)$. The sum runs over the 16 corners of a hypercube. 
Spin (flavour) indices are labeled by Greek (Roman) letters.
By using the properties 
$\frac{1}{4}{\rm Tr}[\Gamma_A^\dag\Gamma_B]=\delta_{AB}$ and
$\frac{1}{4}\sum_A \Gamma^\dag_{A;b\beta}\Gamma_{A;\alpha a}=\delta_{\alpha\beta}\delta_{ab}$, it is possible to show that the free action takes the form 
\begin{eqnarray}
S & = & 16\sum_y m(\overline{q}(y)\openone\otimes\openone q(y)) + 
16 \sum_y\sum_\mu \overline{q}(y)(\gamma_\mu\otimes\openone ) \Delta_\mu q(y) 
\nonumber \\ & +& 16 \;a \sum_y\sum_\mu \overline{q}(y)
(\gamma_5\otimes\xi_\mu\xi_5)\delta_\mu q(y)\ ,
\label{AcSF}
\end{eqnarray}
where $\xi_\mu\equiv\gamma_\mu^T$ and the direct product 
$\gamma\otimes\xi$ means that the $\gamma$ matrix acts on Dirac (spin) 
indices and $\xi$ acts on the flavour indices.
$\Delta_\mu$ and $\delta_\mu$
are the lattice versions of first and second derivative:
\begin{eqnarray}
\Delta_\mu q(y) & \equiv & \frac{1}{4a}(q(y+\hat{\mu})-q(y-\hat{\mu})) \longrightarrow \partial_\mu q(y)\ ,\\
\delta_\mu q(y) & \equiv & \frac{1}{4a^2}(q(y+\hat{\mu})+q(y-\hat{\mu})-2q(y))
\longrightarrow \partial^2_\mu q(y)\ .
\end{eqnarray}
As already mentioned, this action defines 4 flavors of
fermions which are intermixed by the last term of the action. However,
this term is proportional to $a$, and so in continuum limit 
the flavors completely decouple and become independent. 

To summarize, by introduction of quark fields on a coarse lattice, it
is possible to find a correspondence between the single-component
fermion field $\chi$ and spinor fields corresponding to 4
independent flavors in the continuum. The doubling problem of the
naive fermion action is overcome, since now the propagator
in momentum space has the form
\begin{equation}
G(p) \propto a\;\frac{\frac{i}{2}\sum_\mu \sin (2ap_\mu )(\gamma_\mu\otimes\openone ) + a\sum_\mu (\cos (2ap_\mu ) -1)(\gamma_5\otimes \xi_\mu\xi_5 )
-ma(\openone\otimes\openone )}{\sum_\mu\sin^2(ap_\mu ) +m^2a^2}\ ,
\end{equation}
and has only one pole (at $p=0$) within the allowed range of
momenta $-\pi /2a \le p_\mu < \pi/2a$, corresponding to the 
coarse lattice. The axial anomaly is non-vanishing  and has
the correct continuum limit. 
%This concludes the brief discussion
%of how staggered fermions allow to overcome the problems
%encountered with the naive fermion action. 

If it was not for the last term in Eq.~(\ref{AcSF}), the action would 
possess a $U(4)_{\rm vector}\times U(4)_{\rm axial}$ symmetry in the massless limit.
The last term breaks this symmetry down to $U(1)\times U(1)$, with
the generators of symmetry transformations being
$V=\overline{q}(\openone\otimes\openone )q$ and 
$A=\overline{q}(\gamma_5\otimes\xi_5) q$. 
If the axial symmetry is spontaneously
broken, the associated Goldstone boson is flavour non-singlet, with
the spin-flavour structure $\gamma_5\otimes\xi_5$. This will be important
in our calculations, wherein we evaluate matrix elements between 
states of Goldstone bosons associated with spontaneous breaking 
of the chiral symmetry. Thus, staggered fermions retain some basic
chiral properties of the continuum theory. In particular, the mass
is multiplicatively renormalized. A number of useful
Ward identities have been derived in Ref.~\cite{ToolKit} based on this
property, which greatly simplifies our calculations by 
predicting the chiral behaviour of matrix elements. The chiral
properties of staggered fermions are a significant advantage 
compared to the case of Wilson fermions, where one has to satisfy
the Ward identities by fine-tuning parameters and considering
an enlarged set of operators that can mix with the basic ones. 
This advantage arguably
outweighs the disadvantage of dealing with flavour in addition to spin
structure of operators, as well as intricate mixture of internal
and space-time indices.  

In numerical simulations we use the action in Eq.~(\ref{eq:AcStag}).
However, the interpretation of single-component fields $\chi$ in 
terms of spin-flavour is 
necessary if one wants to make contact with continuum physics.
In practice, this interpretation is important when one constructs
quark bilinears (or four-fermion operators) with given quantum
numbers $S$ (for spin) and $F$ (for flavour). Then it is necessary to 
be able to write down such
bilinear contractions in terms of the single-component field.
For such purposes, we use the operators
\begin{equation}
\label{eq:Oy}
O(y) = \frac{1}{16}\sum_{A,B} \overline{\chi}(y,B) \sf S F B A
\chi (y,A)\ , 
\end{equation}
where $\sf SFBA$ stands for the matrices $\gamma_S\otimes\xi_F$ in 
the following representation:
\begin{equation}
\sf SFBA \equiv \frac{1}{4} {\rm Tr}[\Gamma_B^\dag\Gamma_S\Gamma_A\Gamma_F^\dag ]\ .
\end{equation}
Note that the expression in Eq.~(\ref{eq:Oy}) is non-vanishing only 
for $A$ and $B$ such that $A_\mu +B_\mu +S_\mu +F_\mu = (0\ {\rm mod}\ 2)$. 
So one can rewrite it as 
\begin{equation}
\label{eq:Oy1}
O(y) = \frac{1}{16}\sum_{A,B} \overline{\chi}(y,B) \sf S F B A
\chi (y,A)\delta_{B,A+\Delta}\ ,
\end{equation}
where $\Delta_\mu =_2 S_\mu + F_\mu$. Note that all operations involving
$A$,$B$, $S$ and $F$ are understood modulo 2.
Four-fermion operators can be constructed by multiplication of two bilinears.
For explicit formulae used in calculations, see Sec.~\ref{sec:tricks}. 
Note that in general, the flavour structure of any physically
relevant operator does not matter since in the continuum all
flavors are equivalent. However, dealing with Goldstone bosons is special
since only the axial symmetry with quantum numbers $\gamma_5\otimes\xi_5$
is exact in chiral limit, and this dictates a choice for the flavor 
structure of 
operators in many cases. The rules for the flavour structure in such cases
were laid down in Ref.~\cite{WeakME}, and are summarized below in 
Sec.~\ref{sec:tricks}. 

So far, we have considered only the free fermion action. The description
of fermions interacting with gauge degrees of freedom can be 
obtained by inserting a factor $U(B,A)$ representing
a product of gauge links
along any path connecting the two sites.
This makes each such expression gauge-invariant.
The same is true for staggered fermion operators. 
(It is often advantageous, as in this work, 
to make all operators gauge-invariant. This restricts the number of operators
allowed to mix with the original ones.)
There is an ambiguity as to which path should be chosen.
All choices are equivalent since
they are supposed to give the same results in continuum limit.
For technical purposes, we have chosen to use the product of gauge
links along the shortest path, or, if this path is not unique, average
over all the shortest paths. See more discussion on this in Sec.~\ref{sec:tricks}.

%----------------------------------------------------------
\subsection{Summary: The Skeleton Of A Lattice Calculation}
%----------------------------------------------------------

I have overviewed the basics of generic LGT calculations.
Many important details regarding the calculations presented in this
thesis can be found below in Sec.~\ref{sec:tricks}.
Here I give a brief outline of a lattice calculation
from a procedural point of view.
\begin{enumerate}
\item
Discretize space-time on a 4D lattice of spacing $a$ and
a finite box size $L$. 
\item
Work in the framework of Euclidean functional integration,
with a suitably discretized QCD Lagrangian. 
\item
Generate an ensemble of gauge configurations using Monte Carlo
importance sampling. Each ensemble is characterized by a common
$\beta \equiv 6/g^2$. Often, it is necessary to quench
fermion loops in order to save computer time (quenched approximation). 
\item
Compute quark propagators from suitable sources,
on each gauge configuration of an ensemble.
\item
Construct correlators of quark operators with quantum numbers 
appropriate for given hadrons. Extract physical observables
(such as hadron masses and/or matrix elements) in lattice units.
\item
Average quantities over gauge configurations within a given ensemble.
\item
Set the lattice scale $a^{-1}$ by
using a suitable observable (for example, the mass of the $\rho$ meson).
Convert results to physical units by dividing by powers of $a$.  
\item
Sometimes (as in this work) it is necessary to perform chiral
extrapolation or interpolation: repeat calculations
at several quark masses, then extrapolate or interpolate
in quark mass to obtain results at quark masses corresponding
to physical hadron masses. Chiral perturbation theory
is useful in such extrapolations for predicting the chiral behaviour 
of quantities of interest.
\item
Lattice operators should be matched to continuum ones, as discussed
in Chapter~5. 
\item
Repeat steps 3 --- 9 for several values of $\beta$ (i.e. different
values of $a$), keeping the physical size of the box constant. 
Attempt to take continuum limit by extrapolating the quantities
of interest to $a=0$.
\end{enumerate}

%The error due to finite box size can be made small by
%repeating steps 3 --- 6 at larger sizes of the box until the
%results do not change with further increase of $L$.

%----------------------------------------------------------
\subsection{Sources Of Error In Lattice Simulations}
%----------------------------------------------------------

Although Lattice QCD is a systematic, first-principles approach,
certain statistical and systematic errors are present in current 
simulations. The hope of the future simulations is to decrease
these errors. Here I mention the most common of them.
\begin{enumerate}
\item
The {\it statistical error} comes from a finite number of 
gauge configurations (finite sample size). It is present
in virtually all lattice calculations.
It can be decreased by increasing the number of configurations
(the error behaves as $1/\sqrt{N}$). To estimate this error,
a number of intricate error analysis techniques can be employed,
taking into account correlations between various quantities.
\item
The error due to {\it finite cutoff} effects exists in current
simulations because the extrapolation to continuum limit is
not reliable in many cases. This has to do with the fact
that the calculations become increasingly computationally demanding
as one approaches the $a=0$ limit. Apart from just increasing the
computational power, several approaches exist in reducing
the finite cutoff effects, including improved and perfect actions.
\item
As mentioned, {\it finite box size} can introduce an error. Thus
it should be checked that the box is large enough.
\item
Often, {\it chiral errors} are present. Since the
inversion of the fermion matrix becomes computationally more
expensive due to the phenomenon of ``critical slowing down'',
one is forced to do the calculations at relatively high quark 
masses and extrapolate to lower ones. Such 
extrapolations introduce an error. In addition,
sometimes (as in this work) the quark masses in a given hadron
are taken equal, whereas they are not equal in the real world.
\item
{\it Quenched approximation} introduces an unknown error since
it completely removes the effects of sea quarks. Theoretical
as well as numerical estimates indicate that this error is not
more than 10--15\% for most quantities. However, one should
be careful since this error can vary from one quantity to another.
Ideally, one would like to perform full QCD calculations
(including the effects of the sea quarks), but such
calculations are presently 100 --- 1000 times more expensive than
quenched ones. 
\item
There is an error associated with {\it matching} of lattice
and continuum operators. I discuss it in detail in Chapter~5.
\end{enumerate}

This ends the overview of errors commonly encountered
in lattice calculations. Discussion
of these errors regarding the calculations presented in this thesis
can be found in Chapters~4 and~5.

%======================================================
\section{Theoretical Issues}
%======================================================

\label{TheoryLattice}

Before turning to detailed discussion of numerical simulation
techniques, let us consider
several issues of principle for our computations.

%------------------------------------------------------------
\subsection{Calculating $\langle \pi\pi|O_i|K^0\rangle$.}
%------------------------------------------------------------

As was shown by Martinelli and Testa~\cite{testa}, two-particle
hadronic
states are very difficult to directly construct on the lattice (as 
in any scheme using Euclidean space-time). We have
to use an alternative procedure to calculate the matrix elements 
involving two pions in the {\it out} state. 
We use the method due to Bernard {\it et al.}~\cite{bernard} which 
relies on chiral perturbation theory to relate 
$\langle \pi\pi |O_i|K^0\rangle$ to matrix elements involving 
one-particle states. That is, we use the following relationships
which are obtained in the lowest-order of chiral symmetry
breaking in QCD:
\begin{eqnarray}
\label{eq:chpt1}
\langle \pi^+\pi^-|O_i|K^0\rangle & = & \frac{m_K^2-m_\pi^2}{f}\gamma \\
\langle \pi^+|O_i|K^+\rangle & = & (p_\pi \cdot p_K)\gamma - 
                \frac{m_s+m_d}{f}\delta \\
\label{eq:chpt3}       
\langle 0|O_i|K^0\rangle & = & (m_s-m_d)\delta  ,
\end{eqnarray}
where $f$ is the lowest-order pseudoscalar decay constant.
The masses in the first of these formulae are the physical meson masses,
while the quark masses and the momenta in the second and third formulae
are meant to be from actual simulations on the lattice
(performed with unphysical masses). 

The above relationships ignore higher order terms in the chiral 
expansion, most importantly the interaction between the two pions 
in the final state. 
Therefore this method suffers from a significant uncertainty.
In principle, higher-order terms may be computed. For example, 
Golterman and Leung~\cite{golterman} have computed one-loop 
correction for $\Delta I=3/2$ matrix elements in a closed form. 
However, the values of several parameters (so-called contact terms 
and the momentum cut-off) are currently unknown and therefore subject 
to guessing. Varying the unknown parameters in a reasonable range, 
one obtains that the first-order corrections are about 30\% to 60\%. 

In our work we have chosen to use strictly lowest-order relationships, 
remembering that this method is subject to considerable systematic
error. (It is known that the \dione amplitude is 
underestimated while the \dithree amplitude is overestimated by this 
procedure). If certain progress is made in the future to 
improve the quantitative estimates of higher-order chiral
corrections, these developments can be easily incorporated into  
the present framework to produce more reliable answers.
In other words, it will not be necessary to repeat the extensive lattice 
calculations in order to reevaluate the main results obtained in this work.

%------------------------------------------------------------
\subsection{Mixing With Lower-dimensional Operators.}
%------------------------------------------------------------

Eqs.~(\ref{eq:chpt1}--\ref{eq:chpt3}) show that in order to recover 
$\langle\pi\pi|O_i|K^0\rangle$, it is necessary
to subtract from $\langle\pi^+|O_i|K^+\rangle$ the term proportional 
to $\delta$, which corresponds to unphysical $s \leftrightarrow d$ 
mixing. This term is proportional to $\langle 0|O_i|K^0\rangle$. 
There are several ways to implement this subtraction
in lattice simulations. We choose the method suggested in 
Refs.~\cite{WeakME}, which is to subtract the operator
\begin{equation}
O_{sub} \equiv (m_d+m_s)\bar{s}d + (m_d-m_s)\bar{s}\gamma_5d \,.
\label{eq:SubOp}
\end{equation}
This operator, having dimension 4, has to be subtracted 
non-perturbatively from the basis operators (all having dimension 6).
(A perturbative scheme would inevitably ignore some higher-order
terms, which may be proportional  to powers of $a^{-1}$ and therefore 
diverge in continuum limit.) 
Therefore, the mixing coefficients are determined non-perturbatively by 
requiring that the subtracted $\langle 0|O_i|K^0\rangle$ vanish. 
Thus we have: 
\begin{equation}
\label{eq:sub1}
\langle \pi^+\pi^-|O_i|K^0\rangle = 
\langle \pi^+|O_i - \alpha_i O_{sub}|K^+\rangle \cdot \frac{m_K^2-m_\pi^2}
{(p_\pi\cdot p_K)f} \, , 
\end{equation}
where $\alpha_i$ are found from 
\begin{equation}
0 = \langle 0|O_i - \alpha_i O_{sub}|K^0\rangle \, .
\label{eq:sub2}
\end{equation}
Kilcup and Sharpe~\cite{ToolKit} derived a number of Ward identities
which show that the lattice formulation of QCD with staggered fermions 
retains the essential chiral properties of the continuum theory. 
In consequence, as mentioned in Ref.~\cite{WeakME}, the procedure expressed 
in Eqs.~(\ref{eq:sub1},\ref{eq:sub2}) is a faithful lattice implementation
of the lowest-order chiral perturbation theory prescription, which
reproduces Eqs.~(\ref{eq:chpt1}--\ref{eq:chpt3}) in the continuum limit.
In particular, no other lower-dimensional operators except the 
one in Eq.~(\ref{eq:SubOp}) are to be subtracted non-perturbatively,
which is in striking contrast to the LQCD formulation with Wilson
fermions. Note that compared to other possible implementations
with staggered fermions, this present method has an advantage of 
the possibility of subtraction at each timeslice. 

Throughout the simulation we use only degenerate mesons, i.e. $m_s=m_d=m_u$.
Since only the negative parity part of $O_{sub}$ contributes in 
Eq.~(\ref{eq:sub2}), one naively expects infinity when calculating 
$\alpha_i$. However, the matrix elements 
$\langle 0|O_i|K^0\rangle$ of all basis operators 
vanish when $m_s=m_d$ due to invariance of both the Lagrangian
and all the operators in question under the CPS symmetry, which
is defined as the CP symmetry combined with interchange of $s$ and $d$ 
quarks~\cite{BernardReview2}. Thus calculation of $\alpha_i$ requires taking 
the first derivative 
of $\langle 0|O_i|K^0\rangle$ with respect to $(m_d-m_s)$. In order
to evaluate the first derivative numerically, we insert another
fermion matrix inversion in turn into all propagators involving
the strange quark, as explained below in Sec.\ref{sec:tricks}. 

\begin{figure}[p]
\begin{center}
\leavevmode
\centerline{\epsfxsize=5.5in \epsfbox{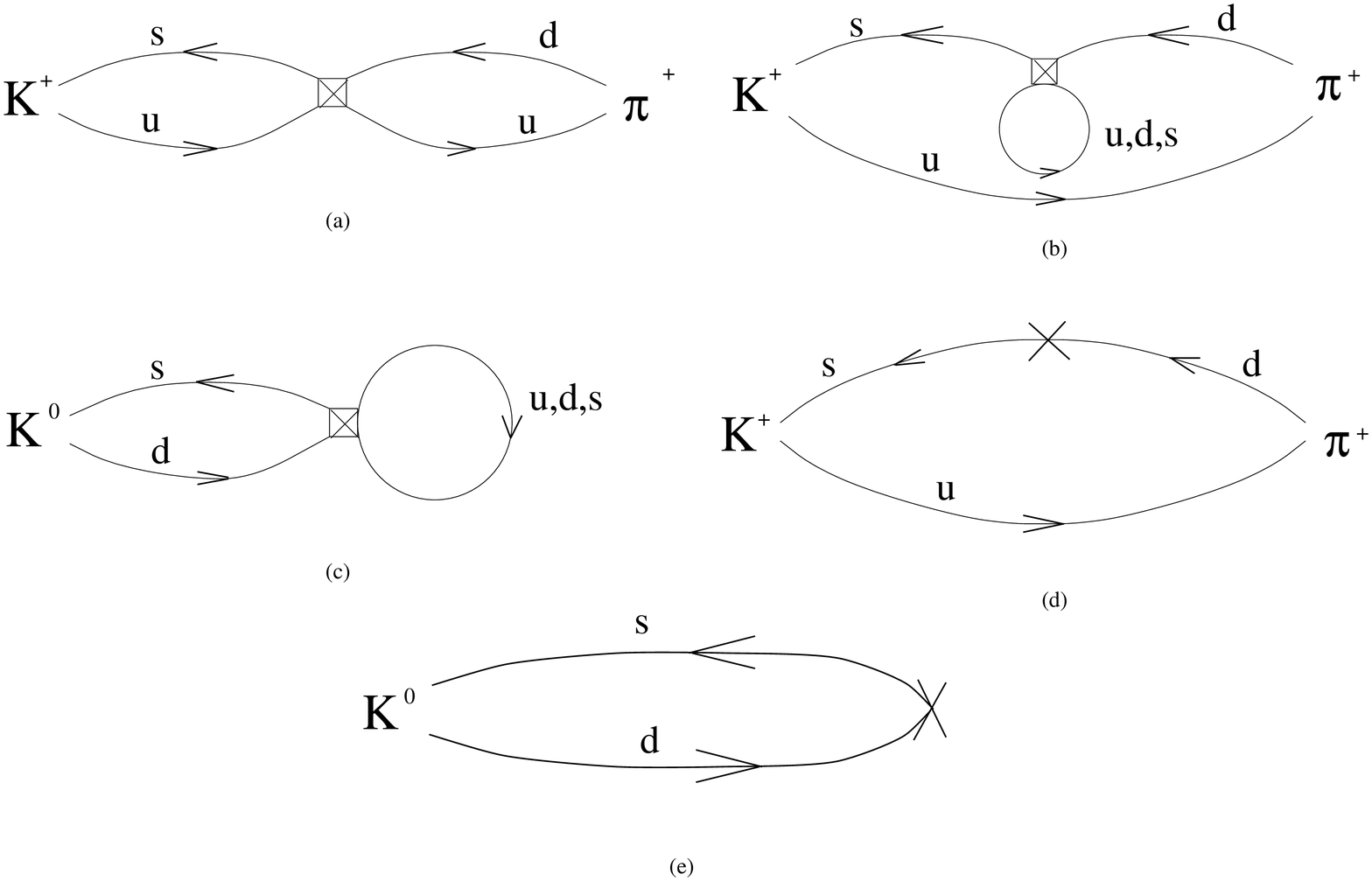}}
\end{center}
\caption{Five diagrams types needed to be computed: (a) ``Eight'';
(b) ``Eye''; (c) ``Annihilation''; (d) ``Subtraction''; (e)
two-point function.}
\label{diagrams}
\end{figure} 

%------------------------------------------------------------
\subsection{Contractions To Be Computed}
%------------------------------------------------------------

\label{sec:diag}

According to Eqs.~(\ref{eq:sub1}) and~(\ref{eq:sub2}), we need to compute three
diagrams involving four-fermion operators (shown in Fig.~\ref{diagrams})
and a couple of bilinear contractions. The exact expressions 
for these contractions are given in the following section.
Here I would like to note that the ``eight'' contraction type 
(Fig.~\ref{diagrams}a) is relatively cheap to compute. It is the only
contraction needed for the $\Delta I=3/2$ amplitude.
The ``eye'' and ``annihilation'' 
diagrams (Fig.~\ref{diagrams}b and~\ref{diagrams}c) are much more 
expensive since they involve calculation of propagators from
every point in space-time. Their calculation is essential for
estimation of the $\Delta I=1/2$ amplitude and \epsp.

%======================================================
\section{Numerical Techniques}
%======================================================

\label{sec:tricks}

In Sec.~\ref{sec:stag} I have briefly overviewed the conceptual basics
of staggered fermions. In this section I describe the multitude
of methods and tricks used for the purpose of calculating
the matrix elements of interest in the language of staggered fermions.

%------------------------------------------------------------
\subsection{Quark Operators}
%------------------------------------------------------------

There are a number of issues about quark operators that have
to be addressed. The subject is complicated since each operator
is characterized by a given physical flavour, staggered spin-flavour
and color flow. (Physical flavors should be distinguished from 
staggered flavors; the latter are 4 artificial independent species per 
each physical flavour). 

Firstly, there is a choice in constructing four-fermion operators
regarding the arrangement of physical flavour (such as $s$, $d$
and $u$) flow. It is related to the fact that there are
always two ways to arrange bilinears of any given operator, which are
related by the Fierz transformation. We choose to use such
form of the operators that all contractions can be represented
as a product of two flavour traces; in other words, that insertion of
the vacuum between the two bilinears would give non-vanishing results.
To be more precise, for ``eight'' contractions 
the operators  are rendered
in the form $(\overline{s}\;\Gamma \;u)(\overline{u}\;\Gamma\; d)$,
while for the ``eye'' and ``annihilation'' contractions
the appropriate form is $(\overline{s}\;\Gamma \;d)
(\overline{q}\;\Gamma\; q)$. This is done in the continuum, before
considering the staggered fermion transcription. 
%assigning the staggered fermion flavour. 
One should be careful in not forgetting that some operators
can have multiple non-vanishing contractions of a given type, 
in which case all of them should be cast in the two flavour trace
form. For example, the operator 
$(\overline{s}_\alpha\gamma_\mu (1-\gamma_5)d_\alpha)
(\overline{s}_\beta\gamma_\mu (1-\gamma_5)s_\beta)$ can have
non-vanishing two flavour trace ``eye'' type contractions between 
kaon and pion states in two forms: the present one as well as
$(\overline{s}_\alpha\gamma_\mu (1-\gamma_5)d_\beta)
(\overline{s}_\beta\gamma_\mu (1-\gamma_5)s_\alpha)$. The contributions 
of these two contractions should be added when calculating the
matrix element.

Secondly, the flavour structure of the operators should be assigned.
In general, flavour assignment is arbitrary, but in cases like
ours when one deals with matrix elements of Goldstone bosons,
there is no choice left~\cite{WeakME}.
As mentioned in Sec.~\ref{sec:stag}, the external Goldstone bosons 
states have spin-flavor
structure $\gamma_5 \otimes \xi_5$. The flavor structure of the operators
is defined by requiring non-vanishing of the staggered flavor traces,
and so it depends on the contraction type: the flavor structure is 
$\xi_5\otimes\xi_5$
in ``eights'', $\openone\otimes\openone$ in ``eyes''. 
In ``annihilation'' contractions the flavor structure is 
$\openone$ for the bilinear in the internal quark loop trace 
and $\xi_5$ for the one involved in the external trace.
The bilinears are transcribed in a similar manner: the 
Goldstone boson two-point function has the flavor $\xi_5$, 
while the $\overline{s}d$ operator in the ``subtraction''
contraction has flavor $\openone$.

Thirdly, either one or two color traces may be appropriate for a 
particular contraction with a given operator (see the next subsection
for details). In one trace contractions (type ``F'') the
color flow is exchanged between the bilinears, while in 
two trace contractions (type ``U'') the color
flow is contained within each bilinear so that the contraction
is the product of two color traces. In either contraction type,
when the distance between staggered fermion fields 
being color-connected is non-zero, a gauge connector is inserted
in the gauge-invariant fashion. The connector is computed as the average 
of products of gauge links along all shortest paths connecting the 
two sites. We also implement tadpole improvement
by dividing each link in every gauge connector by $u_0 = (\frac{1}{3}\; 
\mathrm{Tr}[U_P])^{1/4}$, where $U_P$ is the average plaquette value. 
This is done in order to improve the properties of lattice perturbation
theory by non-perturbatively removing ``tadpole diagrams''~\cite{Mackenzie}.
At the same time, in addition to rescaling the links, each bilinear 
should be multiplied by $u_0$ to account for quark field 
renormalization.

%------------------------------------------------------------
\subsection{Sources And Contractions}
%------------------------------------------------------------

\label{sources}

Here we show how sources and sinks are used to construct
the correlators, and give explicit expressions for 
various types of contractions which were described in words
in the previous subsection.

We use local $U(1)$ pseudofermion wall sources. That is, 
we set up a field of $U(1)$ phases $\xi_\alpha({\mathbf x};t_0)$
($|\xi_\alpha| = 1$) for each color and each site at a given timeslice 
$t_0$, which are chosen at random and independently, so they satisfy
\begin{equation}
\label{eq:noise}
\langle\xi_\alpha^*({\mathbf x};t_0)\; \xi_\beta ({\mathbf y};t_0)\rangle = 
\delta_{\alpha ,\beta }\; \delta_{{\mathbf x},{\mathbf y}}. 
\end{equation}
(Boldface characters designate spatial parts of the 4-vector
with the same name.) 
We proceed to explain how this setup works in the case of the
two-point function calculation, with trivial generalization to ``eight''
and ``annihilation'' contractions.

Consider the propagator from a wall source at $t_0=0$ in a given background
gauge configuration, computed by
solving the linear system of equations
\begin{equation}
\label{eq:prop}
\sum_y(\rlap{\,/}D +m)^{\alpha\beta}_{x\;y}\; \chi_{\beta}(y) = 
\xi_\alpha ({\mathbf x};0)\delta_{x_4,0}.
\end{equation}
In other words, we need to find $(\Dslash +m)^{-1}\;b$ for 
an appropriate source $b$. 
Note that it is equivalent to computing
\begin{equation}
\chi_\beta (y) = \sum_{\mathbf x} \, \xi_\alpha ({\mathbf x};0)
G_{\beta\alpha}(y;{\mathbf x},0)\, ,
\end{equation}
where $G(y;x)$ is the propagator from 4-point $x$ to 4-point $y$. 
For staggered fermions description we label
the fields by hypercube 
index $h$ and the hypercube corner indices $A_\mu \in \{0,1\}^4$,
as mentioned in Sec.~\ref{sec:stag}. 

The two-point function is constructed as follows:
\begin{equation}
\label{eq:2p-function}
{\rm TP} = \frac{1}{16}\sum_{h,A} \chi^*_{\alpha} (h,A) 
U_{\alpha\beta}(h,A,A+\Delta ) \chi_{\beta}(h,A+\Delta ) \;
\phi (A)\;(-1)^A ,
\end{equation}
where phases $\phi (A) = \frac{1}{4} {\mathrm Tr} [\Gamma_A^\dag\,\Gamma_S\,
\Gamma_{A+\Delta}\,\Gamma_F^\dag ]$  and distances 
$\Delta_\mu =_2 S_\mu +F_\mu$ are defined for a given bilinear operator with 
spin-flavour quantum numbers $S\otimes F$; $U(h,A,A+\Delta )$ is
%
%\footnote{For a given bilinear with spin-flavor structure 
%  $\Gamma_S\otimes\Gamma_F$, these are determined as follows: 
%\mbox{$\Delta_\mu = |S_\mu-F_\mu|^2$} and )$, 
%where $S_\mu$ and $F_\mu$ are spin and flavor vectors
%such that $\Gamma_S = 
%\gamma_1^{S_1}\gamma_2^{S_2}\gamma_3^{S_3}\gamma_4^{S_4}$
%and $\Gamma_F = \gamma_1^{F_1}\gamma_2^{F_2}\gamma_3^{F_3}\gamma_4^{F_4}$,
%and $\Gamma_A = \gamma_1^{A_1}\gamma_2^{A_2}\gamma_3^{A_3}\gamma_4^{A_4}$.}
%
the appropriate gauge connector (see below), modulo 2 summation is implied
for hypercube indices $A$, and $h$ runs over all hypercubes
in a given timeslice $t$ where the operator is inserted. The factor $(-1)^A$ 
takes into account that for staggered fermions 
$G(x;y) = G^\dag (y;x) (-1)^x(-1)^y$.
Equation~(\ref{eq:2p-function}) corresponds to 
\begin{equation}
{\mathrm TP} \propto \sum_{{\mathbf x,\mathbf z},y} G_{\alpha\beta}(z,y) \;\Gamma \;G_{\beta\gamma}(y,x)\; (-1)^z 
\;\xi^*_\alpha (z)\xi_\gamma(x) \, ,
\end{equation}
where $\Gamma$ is used for simplicity to show the appropriate operator
structure. The summation over $y$ is done over the entire space-time.
The summation over ${\mathbf x}$ and ${\mathbf z}$ 
over the spatial volume at the timeslice $t_0$ averages over the noise, 
so the last equation is equivalent to
\begin{equation}
\label{eq:123}
{\mathrm TP} \propto \sum_{\mathbf x,y}{\mathrm Tr}\; [G(x,y) \,\Gamma\, G(y,x)\; (-1)^x]
\end{equation}
(compare to Eq.~(\ref{prop1})).
Therefore, using the pseudofermion wall source is equivalent to 
summation of contractions obtained with independent local delta-function 
sources.
Note that the factor $(-1)^x$ and zero distance in the staggered fermions 
language are equivalent to spin-flavor structure $\gamma_5 \otimes \xi_5$. 
This means the source creates pseudoscalar mesons at rest, which 
includes Goldstone bosons.
Strictly speaking, this source also creates mesons with spin-flavor
structure $\gamma_5\gamma_4 \otimes \xi_5\xi_4$, since it is defined 
only on one timeslice instead of two. 
However, as explained in the first footnote
in Section~2.3 of Ref.~\cite{WeakME}, these states do not contribute. 
We have used one copy of pseudofermion sources per configuration.

Numerically, the inversion of the $(\Dslash +m)$ matrix is done by the 
conjugate gradient algorithm. For details, the interested reader may 
consult Ref.~\cite{lakshmi}.

Analogously to the kaon source, the pion sink at time $T$ is constructed
by using another set of $U(1)$ random noise 
($\langle\xi_\alpha^*({\mathbf x};T)\; \xi_\beta ({\mathbf y};T)
\rangle = \delta_{\alpha ,\beta }\; \delta_{{\mathbf x},{\mathbf y}}$, 
$|\xi |=1$). The propagator $\Phi$ is computed by solving
\begin{equation}
\sum_y(\rlap{\,/}D +m)^{\alpha\beta}_{x\;y}\; \Phi_{\beta}(y) = 
\xi_\alpha ({\mathbf x};T)\delta_{x_4,T}. 
\end{equation}

Suppose $\Delta_1$, $\Delta_2 \in \{0,1\}^4$ and $\phi_1 (A)$, $\phi_2 (A)$
are distances and phases of the two staggered fermion bilinears making
up a given four-fermion operator. 
The expression for the ``eight'' contraction (Fig.~\ref{diagrams}a) with two
color traces (``U'' type) is given by
\begin{eqnarray}
\label{eq:eight}
{\mathrm E_U} = \frac{1}{16^2} \sum_{h,A,B} & 
\chi^*_{\alpha} (h,A) U_{\alpha\beta}(h,A,A+\Delta_1 ) 
\chi_{\beta }(h,A+\Delta_1 ) \;\phi_1 (A)\;(-1)^A \nonumber \\
& \times\; \Phi^*_{\rho} (h,B) U_{\rho\sigma}(h,B,B+\Delta_2 ) 
\Phi_{\sigma}(h,B+\Delta_2 ) \;\phi_2 (B)\;(-1)^B \ .
\end{eqnarray}
%up to various normalization factors which cancel in the $B$ ratio. 
In this expression $A$, $B \in \{0,1\}^4$ run over 16 corners 
of a given hypercube (modulo 2 summation is implied for these indices). 
The hypercube
index $h$, as before, runs over the entire spatial volume of the
timeslice $t$ of the operator insertion. The gauge connector
$U (h,A,B )$ is the identity matrix when $A = B$, otherwise it is the 
average of products of gauge links in the given configuration along all 
shortest paths from $A$ to $B$ in a given hypercube $h$. The 
expression~(\ref{eq:eight}), as well as all other contractions,
is computed for each background gauge configuration 
(generated beforehand by one of the algorithms of Monte Carlo
importance sampling) and
is subject to ensemble averaging.
Whenever several contractions are combined in a single
quantity, such as a $B$ ratio, we use jackknife to estimate
the statistical error.

The expression for one color trace (``F'' type) contraction is similar:
\begin{eqnarray}
{\mathrm E_F} = \frac{1}{16^2} \sum_{h,A,B}  &
\chi^*_{\alpha} (h,A) U_{\alpha\beta}(h,A,B+\Delta_2 ) 
\chi_{\sigma}(h,A+\Delta_1 ) \;\phi_1 (A)\;(-1)^A \nonumber \\
& \times \Phi^*_{\rho} (h,B) U_{\rho\sigma}(h,B,A+\Delta_1 ) 
\Phi_{\beta }(h,B+\Delta_2 ) \;\phi_2 (B)\;(-1)^B ,
\end{eqnarray}

For ``eye'' and ``subtraction'' diagrams (Fig.~\ref{diagrams}b
and~\ref{diagrams}d) the source setup is a little more
involved, since the kaon and pion are directly connected
by a propagator. In order to construct a wall source we need to compute 
the product
$$
\psi (y) = \sum_{\mathbf{x}} G({\mathbf y},t; {\mathbf x},T) \cdot 
G({\mathbf x},T;{\mathbf x_0},t_0) (-1)^x.
$$
In order to avoid computing propagators from every point $\mathbf{x}$ 
at the timeslice $T$, we first compute propagator
$G({\mathbf x},T;{\mathbf x_0},t_0)$, 
cut out the timeslice $T$ and use it as the 
source for calculating the propagator to $({\mathbf y},t)$. This amounts
to inverting the system
\begin{equation}
\sum_y(\rlap{\,/}D +m)_{x\;y}^{\alpha\beta}\; \psi_{\beta}(y) = 
\chi_{\alpha} (x) \; \delta_{(x_4,T)} (-1)^x\, ,
\end{equation}
where $\chi_\alpha (x)$ is the propagator from the wall source 
at $t_0$ defined in 
Eq.~(\ref{eq:prop}). We use the following expression for evaluating the
``subtraction'' diagram:
\begin{equation}
\label{eq:sub_diag}
{\mathrm S} = \frac{1}{16} \sum_{h,A} \chi^*_{\alpha} (h,A) 
U_{\alpha\beta}(h,A,A+\Delta )\psi_{\beta} (h,A+\Delta ) \;
\phi (A)\;(-1)^A ,
\end{equation}
Again, averaging over the noise leaves only expressions local 
in both source and sink, so in the continuum language we obtain:
\begin{equation}
{\mathrm S} \propto \sum_{\mathbf x,y,z} {\mathrm tr} \;
G({\mathbf x},0;{\mathbf z},t) \;
\Gamma \; G({\mathbf z},t;{\mathbf y},T) \;
G({\mathbf y},T;{\mathbf x},0)\; (-1)^x (-1)^y\, .
\end{equation}
In this work we are mostly interested in subtracting the operator
$\overline{s} (\openone\otimes\openone )d$,
so in Eq.~(\ref{eq:sub_diag}) $\Delta = (0,0,0,0)$ and $\phi (A)=1$.

In order to efficiently compute fermion loops for ``eye'' and 
``annihilation'' diagrams (Fig.~\ref{diagrams}b and~\ref{diagrams}c), 
we use $U(1)$ noise copies $\zeta^{(i)}$, $i=1,\dots ,N$, 
at every point in space-time. We compute $\eta^{(i)}$ by inverting
$(\rlap{\,/}D +m)\eta^{(i)} = \zeta^{(i)}$. 
It is easy to convince oneself that the propagator from $y$ to $x$ equals
\begin{equation}
G(x;y) = \langle \eta_x \zeta_y^*\rangle.
\end{equation}
The efficiency of this method is crucial for
obtaining good statistical precision: in order to obtain accurate
results $N$ should be sufficiently large. 
In practice we average over $N=10$ noise samples per site per color.
Note that, since our lattice is copied 2 or 4 times
in the time direction (see next Subsection), independent noise samples
are used for each of the lattice copies to increase the efficiency.  

The expression for 
``U'' and ``F'' type ``eye'' diagrams are as follows:
\begin{eqnarray}
{\mathrm I_U} & =\frac{1}{16^2}  \displaystyle\sum_{h,A,B} &
\chi^*_{\alpha} (h,A) U_{\alpha\beta}(h,A,A+\Delta_1 ) 
\psi_{\beta}(h,A+\Delta_1 ) \;\phi_1 (A)\;(-1)^A \nonumber \\
\times & \displaystyle\frac{1}{N}\sum_{i=1}^N & 
\zeta^{(i)*}_\rho (h,B)U_{\rho\sigma}(h,B,B+\Delta_2) 
\eta^{(i)}_\sigma (h,B+\Delta_2 ) \;\phi_2 (B)\;(-1)^B  \\
{\mathrm I_F} & = \frac{1}{16^2}  \displaystyle\sum_{h,A,B} &
\chi^*_{\alpha} (h,A) U_{\alpha\sigma}(h,A,B+\Delta_2 ) 
\psi_{\beta}(h,A+\Delta_1 ) \;\phi_1 (A)\;(-1)^A \nonumber \\
\times & \displaystyle\frac{1}{N}\sum_{i=1}^N & 
\zeta^{(i)*}_\rho (h,B)U_{\rho\beta}(h,B,A+\Delta_1) 
\eta^{(i)}_\sigma (h,B+\Delta_2 ) \;\phi_2 (B)\;(-1)^B 
\end{eqnarray}

The computation of ``annihilation'' diagrams (Fig.~\ref{diagrams}c)
is similar to the two-point function, except the fermion loop is 
added and the derivative with respect to the quark mass 
difference $m_d-m_s$ 
is inserted in turn in every strange quark propagator. 
Derivatives of the propagators with respect to quark mass can be
obtained by inverting equations
\begin{eqnarray}
(\rlap{\,/}D +m)\chi ' & = & -\chi\  , \\
(\rlap{\,/}D +m)\eta^{'(i)} & = & -\eta^{(i)}\  .
\end{eqnarray}
Indeed, 
$$\frac{\partial \chi}{\partial m} = \frac{\partial}{\partial m}
\frac{1}{\Dslash +m}\xi = -\frac{1}{(\Dslash +m)^2}\xi = 
-\frac{1}{\Dslash +m}\chi \ .$$
We have, therefore, four kinds of ``annihilation'' contractions.
\begin{eqnarray}
  {\mathrm A_{1U}} & = \frac{1}{16^2} \displaystyle\sum_{h,A,B} &
\chi^{'*}_{\alpha} (h,A) U_{\alpha\beta}(h,A,A+\Delta_1 ) 
\chi_{\beta }(h,A+\Delta_1 ) \;\phi_1 (A)\;(-1)^A \nonumber \\
\times & \displaystyle\frac{1}{N}\sum_{i=1}^N &
\zeta^{(i)*}_\rho (h,B)U_{\rho\sigma}(h,B,B+\Delta_2) 
\eta^{(i)}_\sigma (h,B+\Delta_2 ) \;\phi_2 (B)\;(-1)^B  \\
  {\mathrm A_{1F}} & =\frac{1}{16^2}\displaystyle\sum_{h,A,B} &
\chi^{'*}_{\alpha} (h,A) U_{\alpha\sigma}(h,A,B+\Delta_2 ) 
\chi_{\beta}(h,A+\Delta_1 ) \;\phi_1 (A)\;(-1)^A 
\nonumber \nolinebreak\\
\times & \displaystyle\frac{1}{N}\sum_{i=1}^N &
\zeta^{(i)*}_\rho (h,B)U_{\rho\beta}(h,B,A+\Delta_1) 
\eta^{(i)}_\sigma (h,B+\Delta_2 ) \;\phi_2 (B)\;(-1)^B  \\
  {\mathrm A_{2U}} & =\frac{1}{16^2} \displaystyle\sum_{h,A,B} &
\chi^{*}_{\alpha} (h,A) U_{\alpha\beta}(h,A,A+\Delta_1 ) 
\chi_{\beta }(h,A+\Delta_1 ) \;\phi_1 (A)\;(-1)^A \nonumber \\
\times & \displaystyle\frac{1}{N}\sum_{i=1}^N &
\zeta^{(i)*}_\rho (h,B)U_{\rho\sigma}(h,B,B+\Delta_2) 
\eta^{'(i)}_\sigma (h,B+\Delta_2 ) \;\phi_2 (B)\;(-1)^B  \\
  {\mathrm A_{2F}} & =\frac{1}{16^2} \displaystyle\sum_{h,A,B} &
\chi^{*}_{\alpha} (h,A) U_{\alpha\sigma}(h,A,B+\Delta_2 ) 
\chi_{\beta}(h,A+\Delta_1 ) \;\phi_1 (A)\;(-1)^A 
\nonumber \nolinebreak\\
\times & \displaystyle\frac{1}{N}\sum_{i=1}^N &
\zeta^{(i)*}_\rho (h,B)U_{\rho\beta}(h,B,A+\Delta_1) 
\eta^{'(i)}_\sigma (h,B+\Delta_2 ) \;\phi_2 (B)\;(-1)^B 
\end{eqnarray}

In this subsection I have given explicit expressions
for calculating several types of contractions. 
These contractions should be combined in an
appropriate way for each given operator. This is spelled out
in the Appendix.

%------------------------------------------------------------
\subsection{Simulation Parameters}
%------------------------------------------------------------

The parameters of simulation are listed in the Tables~\ref{tab:param1}
and~\ref{tab:param2}.
We use periodic boundary conditions in both space and time.
Our main ``reference'' ensemble is a set of quenched configurations
at $\beta \equiv 6/g^2 =6.0$ ($Q_1$). In addition, we use an
ensemble with the same $\beta$ but a larger volume 
($Q_2$), an ensemble
with $\beta =6.2$ ($Q_3$) for checking the lattice spacing dependence,
and an  ensemble with 2 dynamical flavors ($m=0.01$) generated by the 
Columbia group, used for checking the impact of quenching. 
In addition, a number of quenched ensembles (Table~\ref{tab:param2}) 
with various $\beta$ are used for taking the continuum limit of $B_K$
and $f_K$.
The quenched ensembles were obtained using 4-to-1 ratio of 3-hit $SU(2)$
overrelaxed and heatbath algorithms.  The configurations were separated by
1000 sweeps. The dynamical configurations were obtained by 
the R-algorithm~\cite{Columbia1}.

\begin{table}[tbh]
\begin{tabular}{cccccccc}
\hline\hline
Ensemble & $N_f$ & $\beta $ & Size & $a^{-1}$, & L, & Number of & 
Quark masses\\ name & & & & GeV & fm & conf. & considered \\
\hline
$Q_1$ & 0 & 6.0 & $16^3\times (32\times 4)$ & 2.07 & 1.5 & 216 & 
0.01 --- 0.05 \\
$Q_2$ & 0 & 6.0 & $32^3\times (64\times 2)$ & 2.07 & 3.1 & 26 & 
0.01 --- 0.05 \\
$Q_3$ & 0 & 6.2 & $24^3\times (48\times 4)$ & 2.77 & 1.7 & 93 & 
0.005 --- 0.030 \\
$D$   & 2 & 5.7 & $16^3\times (32\times 4)$ & 2.0  & 1.6 & 83 & 
0.01 --- 0.05 \\
\hline\hline
\end{tabular}
\caption{Parameters of ensembles used for calculation of $\langle \pi\pi|O_{1-10}|K\rangle$.}
\label{tab:param1}
\end{table}

\begin{table}[tbh]
\begin{tabular}{ccccccc}
\hline\hline
$N_f$ & $\beta $ & Size & $a^{-1}$, GeV & L, fm & Number of & 
Quark masses\\  & & & & & configurations & considered \\
\hline
2 & 5.7 & $16^3\times (32\times 4)$ & 2.0  & 1.6 & 83 & 0.01 --- 0.05 \\
0 & 5.7 & $16^3\times (32\times 4)$ & 1.18 & 2.7 & 259 & 0.01 --- 0.08 \\
0 & 5.8 & $16^3\times (32\times 4)$ & 1.48 & 2.2 & 200 & 0.01 --- 0.04 \\
0 & 5.9 & $16^3\times (32\times 4)$ & 1.77 & 1.8 & 200 & 0.01 --- 0.04 \\
0 & 6.0 & $16^3\times (32\times 4)$ & 2.07 & 1.5 & 221 & 0.01 --- 0.04 \\
0 & 6.05 &$16^3\times (32\times 4)$ & 2.26 & 1.4 & 306 & 0.01 --- 0.05 \\
0 & 6.1 & $24^3\times (48\times 4)$ & 2.44 & 2.0 & 100 & 0.01 --- 0.04 \\
0 & 6.2 & $24^3\times (48\times 4)$ & 2.77 & 1.7 & 121 & 0.005 --- 0.035 \\
0 & 6.4 & $32^3\times (64\times 4)$ & 3.64 & 1.8 & 50  & 0.005 --- 0.030 \\
\hline\hline
\end{tabular}
\caption{Parameters of ensembles used for calculation of $B_K$ and $f_K$.}
\label{tab:param2}
\end{table}

We use renormalized coupling constant $g_{\overline{\rm MS}}$
which is found
as follows. We calculate the Lepage-Mackenzie coupling $\alpha_V$ from
\begin{equation}
\label{eq:coupling1}
-\ln \langle \frac{1}{3}{\rm Tr} U_P\rangle = \frac{4\pi}{3}
\alpha_V(3.41/a)\; [1-(1.185-0.070N_f)\alpha_V]\ ,
\end{equation}
where $N_f$ is the number of sea quark flavors.
The coupling in $\overline{\rm MS}$ scheme can be obtained as
\begin{equation}
\label{eq:coupling2}
 \alpha_{\overline{\rm MS}}(3.41/a) = \alpha_V(\frac{3.41\;e^{5/6}}{a})\;
[1+\frac{2}{\pi}\alpha_V]\ .
\end{equation}
Then using the two-loop running we can obtain $\alpha_{\overline{\rm MS}}$
at any scale $q^*$.

As mentioned in Sec.~\ref{background}, the lattice scale $a$
can be determined by comparing a suitable quantity computed
on the lattice with the experimental value. In this work we use
the mass of the $\rho$ meson for such purposes.

The quenched lattice scale was set as in Ref.~\cite{PK1},
i.e. we demand perturbative scaling of the form
\begin{equation}
a(\beta) = a_0 \biggl({16 \pi^2 \over 11 g_{\overline{\rm MS}}^2}\biggr)^{51\over121}
\exp\biggl({-8\pi^2\over 11 g_{\overline{\rm MS}}^2}\biggr),
\end{equation}
normalizing so that the world data for the $\rho$ mass~\cite{spectrum} 
is well fit by 
\begin{equation}
m_\rho(a) = (770 \mbox{ MeV })\cdot(1 + \Lambda^2 a^2(\beta))\ .
\end{equation}
This procedure takes into account the fact that the $\rho$ mass
itself receives corrections at finite $a$. Thus
cutoff effects are reduced at any given $a$.

The lattice spacing for the dynamical 
ensemble was also set by the $\rho$ mass~\cite{Columbia}. 

%------------------------------------------------------------
\subsection{$B$ Ratios}
%------------------------------------------------------------

\label{sec:B}

It has become conventional
%is sometimes convenient 
to express the results for matrix elements
in terms of so-called {\it $B$ ratios}, which are the ratios of desired 
four-fermion matrix elements to their
values obtained by {\it vacuum saturation approximation} (VSA).
VSA is a crude method for obtaining matrix elements of four-fermion
operators based on inserting a complete set of states 
between the bilinears and keeping only the vacuum terms.
For example, the kaon matrix element of the $\Delta S=2$ operator
can be obtained as 
$$\langle\overline{K^0}|(\overline{s}\gamma^\mu (1-\gamma_5)d)\;(\overline{s}\gamma_\mu (1-\gamma_5)d)\;|K^0\rangle 
= \frac{8}{3} \langle\overline{K^0}|(\overline{s}\gamma^\mu (1-\gamma_5)d)
|0\rangle\; \langle 0|(\overline{s}\gamma_\mu (1-\gamma_5)d)\;|K^0\rangle$$
VSA provides a convenient scale, so that the answers can be
quoted as the VSA result times a dimensionless quantity called $B$ ratio. 

Sometimes it is convenient to compute $B$ ratios directly on
the lattice.
For example, the $B$ ratios of operators $O_2$ and $O_4$ are formed by
dividing the full matrix element by the product of axial-current
two-point functions (Fig.~\ref{ratio}).
We expect the denominator to form a plateau 
in the middle of the lattice, equal to
$Z e^{-m_\pi T} \, \langle\pi|A_\mu|0\rangle \,\cdot \,
\langle 0|A^\mu|K\rangle$,
where $A^\mu$ are the axial vector currents with appropriate flavor quantum
numbers for kaon and pion. The
factor $Z e^{-m_\pi T}$ cancels, leaving the desirable ratio
$\langle\pi|O|K\rangle \, / \,
(\langle\pi|A_\mu|0\rangle\, \cdot \, \langle 0|A^\mu|K\rangle)$. 
%This is convenient since $Z_\pi$ contains various factors
%(including color and staggered flavors factors) omitted 
%in the equations of Sec.~\ref{sources} and the Appendix.  
Apart from common normalization factors, 
a number of systematic errors also tend to cancel in this ratio,
including the uncertainty in the lattice spacing, quenching and
in some cases perturbative correction uncertainty. 
Therefore, it is sometimes reasonable to give lattice answers in terms
of the $B$ ratios. 

\begin{figure}[htb]
\begin{center}
\leavevmode
\centerline{\epsfxsize=5.5in \epsfbox{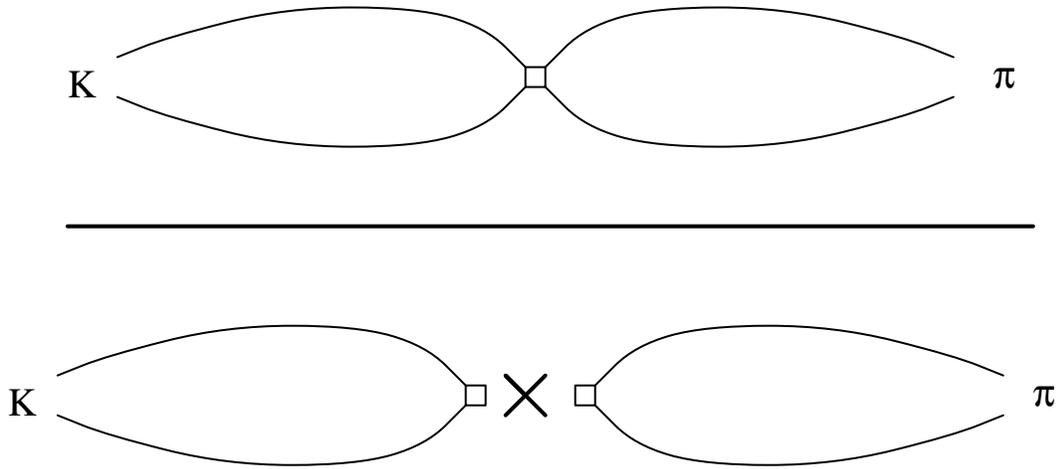}}
\end{center}
\caption{$B$ ratio is formed by dividing the four-fermion matrix element
by the product of two-point functions, typically involving $A_\mu$
or $P$ bilinears. All the operators involved are inserted at the same
timeslice $t$, and the external meson sources are also located at the same
timeslices for numerator and denominator. 
This enables cancellation of various common factors.}
\label{ratio}
\end{figure} 

However, eventually the physical matrix element
needs to be reconstructed by using the known experimental parameters 
(namely $f_K$) to compute VSA. In some cases, for example in
calculation of $B_K$ (see below), this is easy. 
In some other cases, such as for operators
$O_5$ --- $O_8$, the VSA itself is given in terms of $\langle 0|P|K\rangle$,
which is known very imprecisely due to the
failure of perturbative matching (see Sec.~\ref{sec:pert}).
Then it is more reasonable to give answers in terms of matrix elements
in physical units. We have adopted the strategy of expressing all matrix 
elements in units of $\langle\pi|A_\mu|0\rangle \, \langle 0|A^\mu|K\rangle
= (f_K^{latt})^2 m_M^2$ at an intermediate stage, and using 
pre-computed $f_K^{latt}$ at the given meson mass to convert to physical 
units. This method is sensitive to the choice of the lattice spacing. 

\begin{figure}[htb]
\begin{center}
\leavevmode
\centerline{\epsfysize=8cm \epsfbox{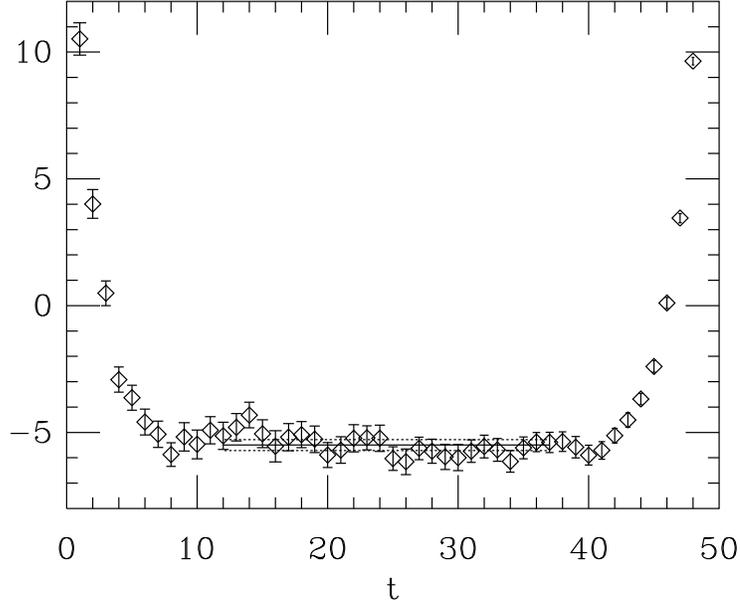}}
\end{center}
\caption{An example of the signal we get for one of the $B$ ratios
(in this case, for the ``eye'' part of the $O_2$ operator on $Q_1$ ensemble,
using quark mass 0.01). 
The wall sources are at $t=1$ and $t=49$. We see that
the excited states quickly disappear and a stable, well-distinguished
plateau is observed. We perform jackknife averaging in the range of $t$
from 12 to 37 (shown with the horizontal lines). }
\label{plateau}
\end{figure} 

It is very important to check that the time distance between the
kaon and pion sources $T$ is large enough so 
that the excited states do not contribute. That is, the plateau
in the middle of the lattice should be sufficiently flat,
and the $B$ ratios should not depend on $T$. We have found that 
in order to satisfy this requirement the lattice has to be
artificially extended in time direction by copying the gauge links 
(4 times in the case of the smaller volume lattices, 2 times 
otherwise). Thus we
get rid of excited states contamination and wrap-around effects. 

\begin{figure}[htb]
\begin{center}
\leavevmode
\centerline{\epsfysize=8cm \epsfbox{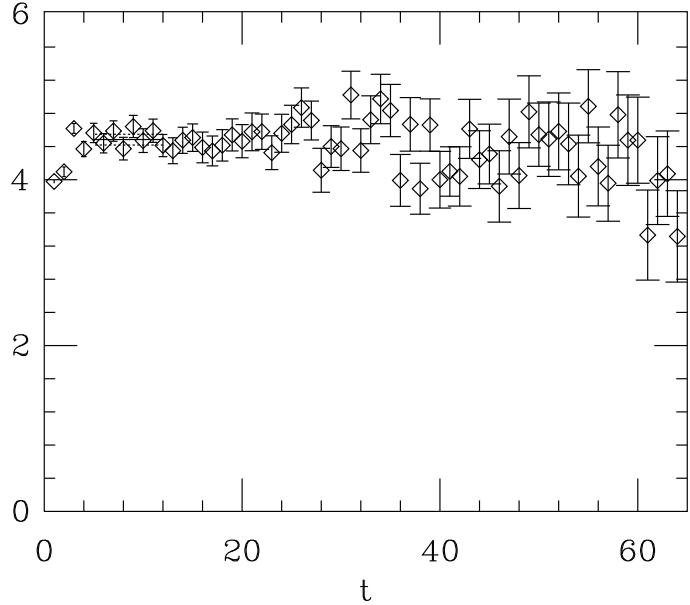}}
\end{center}
\caption{An example of the signal for 
$\langle 0|O_2|K^0\rangle\,/\,
[(m_d-m_s)\,\langle 0|\overline{s}\gamma_5 d|K^0\rangle]$ on $Q_1$
ensemble with quark mass 0.01. The kaon source is at $t=1$. We average 
over the range of $t$ from 5 to 12 (shown with horizontal lines).}
\label{ann}
\end{figure} 

We are using $T=72$ for $Q_3$ 
($\beta =6.2$) ensemble, and $T=48$ for the rest of the ensembles.
An example of a plateau that we obtain
with this choice of $T$ is shown in Figs.~\ref{plateau} 
and~\ref{ann}.
To read off the result, we average over the whole extension
of the plateau, and use jackknife to estimate the statistical
error in this average. 

%======================================================
\section{Computational Summary}
%======================================================

\label{sec:comp}

The bulk of the calculations in this thesis consisted of
computing fermion propagators on a number of gauge configurations
and combining the results in ways described above in Sec.~\ref{sec:tricks}.
The first task by far consumes most of computer resources,
although the second one is no less consuming of human labour.  
Calculations were done on high-performance parallel computers
CRAY-T3E at Ohio Supercomputing Center (OSC) and NERSC. 
The architecture of CRAY-T3E is very fitting for lattice
computations since it allows to assign a subregion of the lattice
to each processor and run the
program in parallel, with some exchange of 
information between the processors when necessary. 
A parallel \mbox{FORTRAN} code was developed by a collaboration of 
Dr. G. Kilcup, Dr. L. Venkataraman
and the author. The basic task of inverting the fermion matrix
(Eq.~(\ref{eq:prop})) is done by the conjugate gradient algorithm,
adapted to the staggered fermion case (discussed in detail 
in the Ph.D. thesis of L. Venkataraman~\cite{lakshmi}). 
The computer at OSC is capable of providing a peak performance
of 600 MFLOPS (millions of floating  point operations per second) 
per processor. Up to 128 processors have been used. As is well known, 
the performance sustained in realistic computations is significantly
less that the theoretical peak. Our program has been able to 
sustain performance of 100 MFLOPS per processor. The total running
time taken to collect the data in this thesis is approximately
one year. 

In addition to the large-scale calculations on the supercomputer, 
the raw data obtained thereby
were analyzed and combined into meaningful results on a UNIX workstation. 
This included statistical analysis, linear and non-linear fitting,
solving ODE etc.

\chapter{\dione Rule On The Lattice}
\label{chap:di12}

Using the data obtained for matrix elements of basis operators,
in this section I report numerical results for $\Re A_0$
and $\Re A_2$ amplitudes as well as their ratio. These
amplitudes are discussed separately since the statistics for 
$\Re A_2$ is much better and the continuum limit extrapolation 
is more reliable. I also present our results for $B_K$ and $f_K$.

%======================================================
\section{\dithree Amplitude}
%======================================================
\label{sec:A2}

The expression for $\Re A_2$ can be written as
\begin{equation}
\Re A_2 = \frac{G_F}{\sqrt{2}}\, V_{ud}V_{us}^*\, z_+(\mu )
\langle O_2(\mu )\rangle _2,
\end{equation}
where $z_+ (\mu )$ is a Wilson coefficient (we use $\mu =2$ GeV) and
\begin{equation}
\langle O_2\rangle _2 \equiv \langle (\pi\pi)_{I=2}|O_2^{(2)}|K\rangle .
\end{equation}
Here
\begin{eqnarray}
O_2^{(2)} & = & O_1^{(2)}  =  \frac{1}{3} 
[ (\overline{s}\gamma_\mu(1-\gamma_5)u)(\overline{u}\gamma^\mu(1-\gamma_5)d) 
\nonumber \\ & &
+(\overline{s}\gamma_\mu(1-\gamma_5)d)(\overline{u}\gamma^\mu(1-\gamma_5)u)
-(\overline{s}\gamma_\mu(1-\gamma_5)d)(\overline{d}\gamma^\mu(1-\gamma_5)d)].
\end{eqnarray}
In lowest-order chiral perturbation theory the matrix element
$\langle O_2\rangle_2$ can be expressed as
\begin{equation}
\langle O_2\rangle _2 
= \sqrt{2} \,\frac{m_K^2-m_\pi^2}{f} 
\,\frac{\langle\pi^+|O_2^{(2)}|K^+\rangle}{m_M^2},
\end{equation}
so that
\begin{equation}
\Re A_2 = G_F\, V_{ud}V_{us}^*\, \frac{m_K^2-m_\pi^2}{f} \, R_2\, ,
\end{equation}
where $R_2$ is calculated on the lattice and is defined as
\begin{equation}
R_2 \equiv z_+\, \frac{\langle \pi^+|O_2^{(2)}|K^+\rangle}{m_M^2}\, .
\end{equation}

It can be shown that this quantity involves only ``eight'' diagrams
and there are no subtractions to be made. This follows from the fact
that $O_2^{(2)}$ operator belongs to the (27,1) representation
of the $SU(3)_L \times SU(3)_R$ symmetry group, while any contractions
of the type ``eye'' or ``subtraction'' can give rise only to
a term belonging to the (8,1) representation. Moreover,
in the limit of exact $SU(3)_{\mathrm flavor}$ symmetry
$R_2$ is directly related~\cite{donoghue} to parameter $B_K$ (which is the 
$B$ ratio of the neutral kaon mixing operator, see next subsection),  
so that
\begin{equation}
R_2 = \frac{4}{9} \, z_+(\mu ) \, B_K(\mu ) \, f_K^2 \, .
\label{R2}
\end{equation}

%------------------------------------------------------------
\subsection{Calculating $B_K$}
%------------------------------------------------------------

The parameter $B_K$ is defined as follows:
\begin{equation}
\label{eq:Bk}
B_K = \frac{3}{8} \frac{\langle \overline{K^0}|(\overline{s}\gamma_L d) 
\;(\overline{s}\gamma_L d) | K^0\rangle}
{\langle \overline{K^0}|(\overline{s}\gamma_L d)|0\rangle \;
\cdot \;\langle 0|(\overline{s}\gamma_L d) | K^0\rangle} =
\frac{3}{8} \frac{\langle \overline{K^0}|(\overline{s}\gamma_L d) 
\;(\overline{s}\gamma_L d) | K^0\rangle}{m_K^2f_K^2} 
\end{equation}

\begin{figure}[htb]
\begin{center}
\leavevmode
\centerline{\epsfysize=8cm \epsfbox{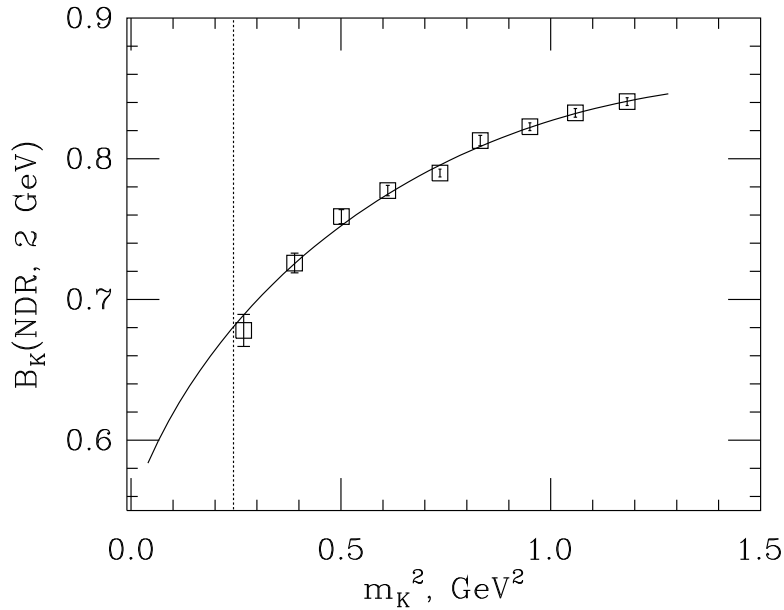}}
\end{center}
\caption{Parameter $B_K$ in NDR $\overline{\mathrm MS}$ scheme at 2 GeV
on the ensemble $D$ vs. the meson mass squared. The fit
is of the form \mbox{$B_K=a+bm_K^2+c\;m_K^2\log{m_K^2}$.} The vertical line
here and in the other plots below marks the physical kaon mass.}
\label{Bk}
\end{figure} 

Its calculation is relatively straightforward, based on the
techniques presented in Chapter~\ref{sim} (detailed descriptions
of the method are also readily available in the literature,
for example in~\cite{SharpeTASI}). It is also relatively
cheap to compute since, as already mentioned, it involves only
``eight'' diagrams''. We have computed $B_K$ on all ensembles
listed in Table~\ref{tab:param2}. For each ensemble, we 
study $B_K$ at a variety of quark masses, and find the mass
dependence well-fitted by the form predicted by the 
chiral perturbation theory~\cite{sharpe1}:
$$B_K=a+bm_K^2+c\;m_K^2\log{m_K^2}$$ 
(for example, see Fig.~\ref{Bk} for $D$ ensemble). 
Here we are mostly interested in the value of $B_K$ at kaon mass,
so there is almost no chiral extrapolation involved. 

\begin{figure}[htb]
\begin{center}
\leavevmode
\centerline{\epsfysize=8cm \epsfbox{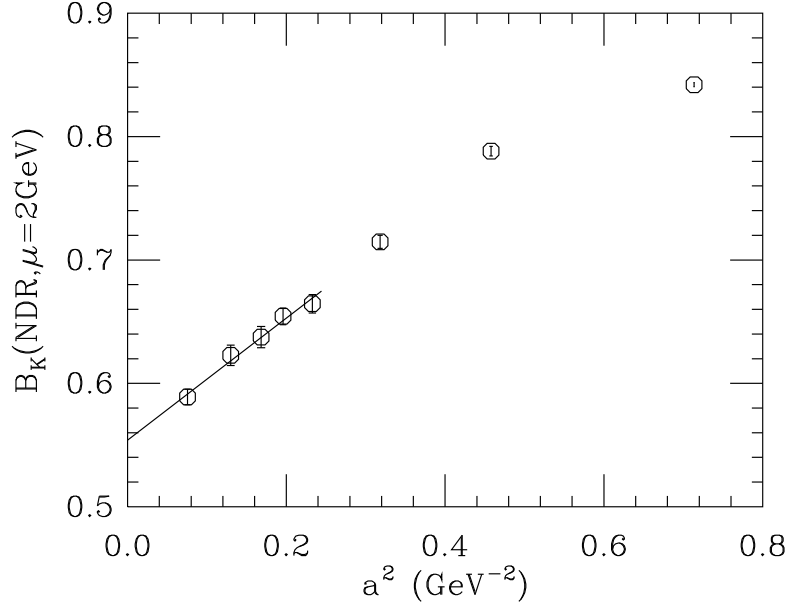}}
\end{center}
\caption{Parameter $B_K$ in NDR $\overline{\mathrm MS}$ scheme at 2 GeV
in quenched QCD vs. $a^2$. Extrapolation to continuum limit is linear,
according to Sharpe's prediction.}
\label{Bkvsa2}
\end{figure} 

Sharpe~\cite{sharpe2} showed that calculation of $B_K$ with 
staggered fermions introduces scaling violations at worst
at the order of $a^2$. Our data confirm this prediction
(see Fig.~\ref{Bkvsa2}). Extrapolating to the continuum limit,
we obtain in quenched QCD:
\begin{equation}
B_K ({\rm NDR},\mu = 2\;\mbox{GeV})= 0.554 \pm 0.009.
\end{equation}
We also found the effect
of quenching to be very small, not more than a few percent.

%------------------------------------------------------------
\subsection{Calculating $f_K$}
%------------------------------------------------------------
\label{sec:fpi}

The pseudoscalar decay constant is obtained from the amplitude
$C_A$ multiplying the exponential in the expression for meson 
correlator with the axial current operator (see Eq.~(\ref{Ct})). 
The absolute normalization is given by the pion field
renormalization factor $C_P$ obtained in the analogous manner from 
the pion wall correlator with the pseudoscalar density operator. 
Taking various factors into account, one obtains~\cite{ToolKit}:
\begin{equation}
f_K = \frac{C_A}{\sqrt{C_P}} \; \frac{\sqrt{\sinh{m_K}}}
{2\sinh{\frac{m_K}{2}}\sqrt{N_f V}} e^{m_K/2}
\end{equation}  

\begin{figure}[phtb]
\begin{center}
\leavevmode
\centerline{\epsfysize=8cm \epsfbox{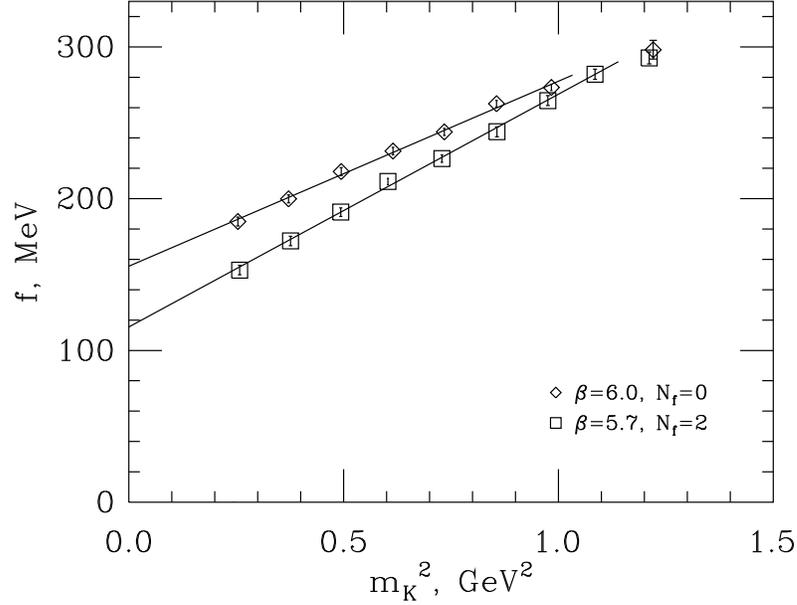}}
\end{center}
\caption{Pseudoscalar decay constant on the dynamical $\beta=5.7$ 
(squares) and quenched $\beta=6.0$ (diamonds) ensembles vs. 
meson mass squared.
The experimental number is shown with the horizontal line.}
\label{fk}
\end{figure} 

Fig.~\ref{fk} shows our results for the pseudoscalar decay constant
plotted against the mass of the meson squared. Extrapolation to
the chiral limit produces the value for $f_\pi$. Thus computing
$f_\pi$ across all quenched ensembles, the continuum limit value
of $134.5 \pm 3.6$~MeV can be obtained (see Fig.~\ref{fpivsa2}). 
Note that this value is allowed to be different from the experiment 
since we consider QCD in the quenched approximation.

\begin{figure}[phtb]
\begin{center}
\leavevmode
\centerline{\epsfysize=8cm \epsfbox{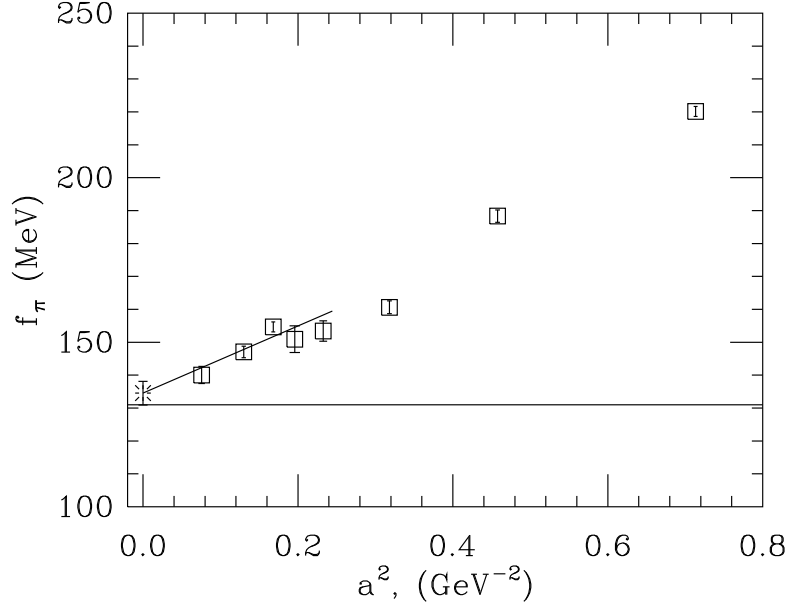}}
\end{center}
\caption{Continuum limit of $f_\pi$ in quenched QCD.}
\label{fpivsa2}
\end{figure} 

%------------------------------------------------------------
\subsection{$\Re A_2$ Results}
%------------------------------------------------------------

Using our data for $B_K$ and $f_K$, we found that the ratio $R_2$ 
shows a large dependence on the meson mass used in the simulation 
(Fig.~\ref{A2}). This is not surprising since both $B_K$ and $f_K$ 
depend on this mass quite significantly.
Which meson mass should be used to read off the $R_2$ value for 
estimation of $\Re A_2$ becomes an open question. 
If known, the higher order chiral terms would remove this ambiguity.
Forced to make a choice, we extrapolate to 
\mbox{$M^2=(m_K^2+m_\pi^2)/2$}.

\begin{figure}[phtb]
\begin{center}
\leavevmode
\centerline{\epsfxsize=5.5in \epsfbox{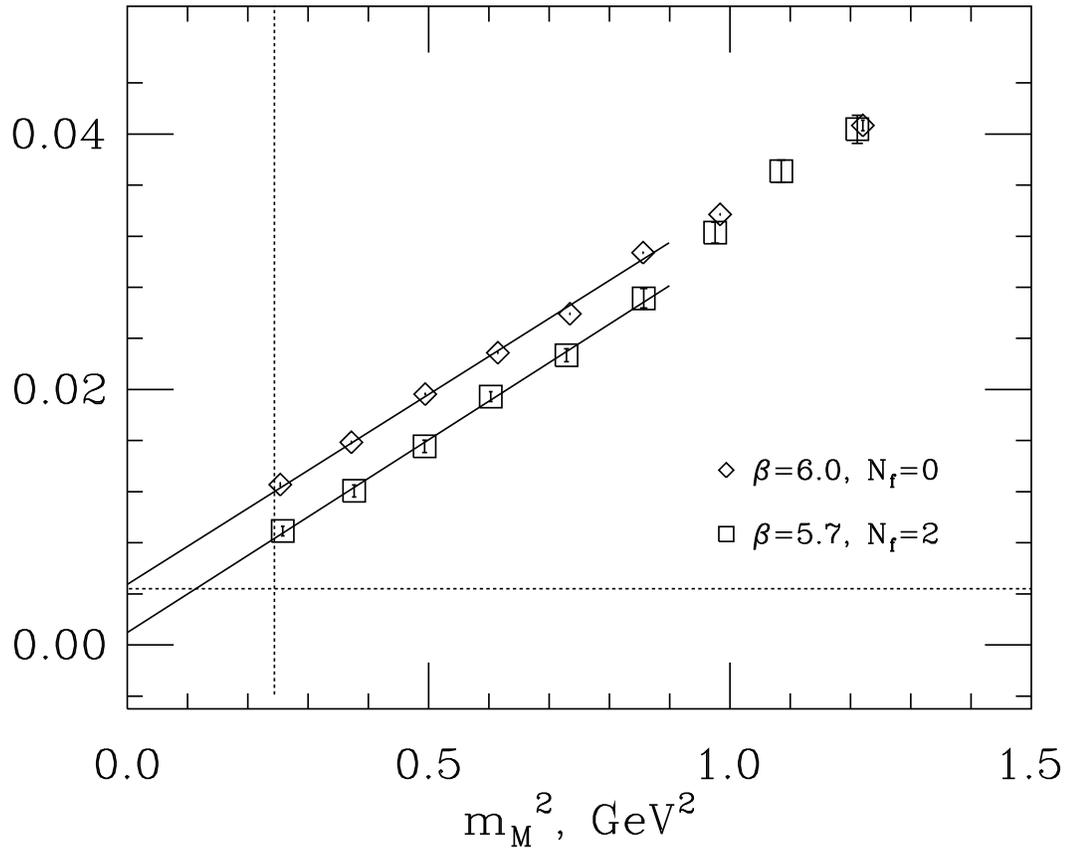}}
\end{center}
\caption{Matrix element $R_2$ computed on
the dynamical $D$ (squares) and quenched $Q_1$ (diamonds) ensembles. 
$\Re A_2$ is proportional to this quantity at the lowest order in 
chiral perturbation theory. The horizontal line shows
the value corresponding to the experiment.}
\label{A2}
\end{figure} 

Taking the continuum limit we obtain in quenched QCD:
$$\Re A_2 = (1.77 \pm 0.07)\cdot 10^{-8}\;\mbox{GeV},$$
where the error is only statistical,
to be compared with the experimental result
$\;\Re A_2 = 1.23 \cdot 10^{-8}\;\mbox{GeV}$. 

Higher order chiral terms (including the meson mass dependence)
are the largest systematic error in this determination.
According to Golterman and Leung~\cite{golterman}, one-loop corrections
in (quenched) chiral perturbation theory are expected to be
as large as 30~---~60\%. Other uncertainties (from lattice scale
determination, from perturbative operator matching and
from using finite lattice volume) are much smaller. 

%------------------------------------------------------------
\section{\dione Amplitude}
%------------------------------------------------------------
\label{sec:dione}

%We work in the normalization in which $\Re A_0 = 27.2 \cdot 10^{-8}$ 
%GeV, which is consistent with the normalization used for $\Re A_2$. 
Using Eqs.~(\ref{eq:sub1},\ref{eq:sub2}), $\Re A_0$ can be expressed 
as
%\footnote{In our normalization $\Re A_0 = 27.2 \cdot 10^{-8}$.}
\begin{equation}
\Re A_0 = \frac{G_F}{\sqrt{2}}V_{ud}V_{us}^* \frac{m_K^2-m_\pi^2}{f}
 R_0 ,
\end{equation}
where 
$$
R_0 \equiv \sum_i\, z_i\,\frac{\langle \pi^+|O_i^{(0)}|K^+\rangle_s}{m^2}.
$$
Here $z_i$ are Wilson coefficients, and
the subscript '$s$' indicates that these matrix elements already
include subtraction of $\langle \pi^+|O_{sub}|K^+\rangle$. 
$O_i^{(0)}$ are isospin 0 parts of operators 
$O_i$ (given in the Appendix~\ref{sec:app} for completeness). For example,
\begin{eqnarray}
O_1^{(0)} & = & \frac{2}{3} 
(\overline{s}\gamma_\mu (1-\gamma_5)d)(\overline{u}\gamma^\mu (1-\gamma_5)u)
-\frac{1}{3}(\overline{s}\gamma_\mu (1-\gamma_5)u)(\overline{u}\gamma^\mu 
(1-\gamma_5)d)  \nonumber \\
& + & \frac{1}{3}(\overline{s}\gamma_\mu (1-\gamma_5)d)
(\overline{d}\gamma^\mu (1-\gamma_5)d) \\
O_2^{(0)} & = & \frac{2}{3} 
(\overline{s}\gamma_\mu (1-\gamma_5)u)(\overline{u}\gamma^\mu (1-\gamma_5)d)
-\frac{1}{3}(\overline{s}\gamma_\mu (1-\gamma_5)d)(\overline{u}\gamma^\mu 
(1-\gamma_5)u) \nonumber \\
& + & \frac{1}{3}(\overline{s}\gamma_\mu (1-\gamma_5)d)
(\overline{d}\gamma^\mu (1-\gamma_5)d) 
\end{eqnarray}

\begin{figure}[phtb]
\begin{center}
\leavevmode
\centerline{\epsfxsize=5.5in \epsfbox{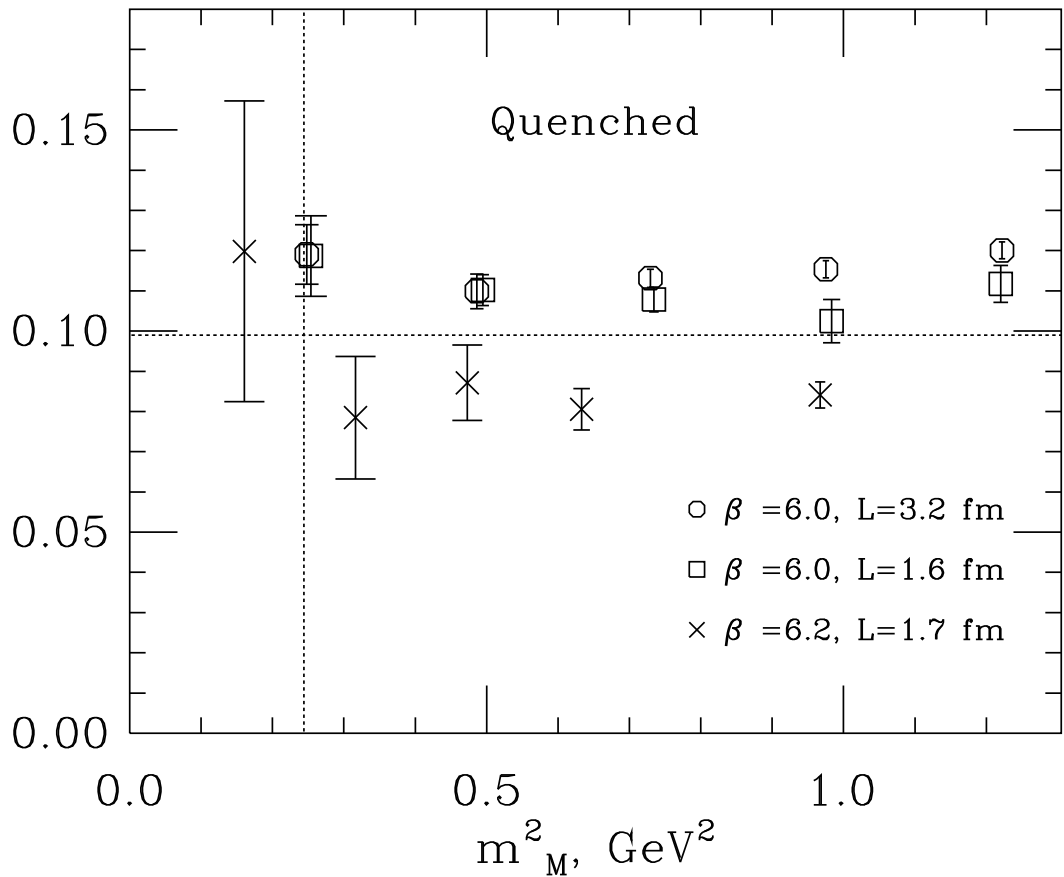}}
\end{center}
\caption{Matrix element $R_0$
for quenched ensembles plotted against the meson mass squared. 
$\Re A_0$ is proportional to this quantity in the lowest order in
chiral perturbation theory. The upper group of points
is for ensembles $Q_1$ and $Q_2$, while the lower group is for $Q_3$. 
Only statistical errors are shown. The horizontal line shows the
value corresponding to the experiment. }
\label{A0}
\end{figure} 

\begin{figure}[tbh]
\begin{center}
\leavevmode
\centerline{\epsfysize=8cm \epsfbox{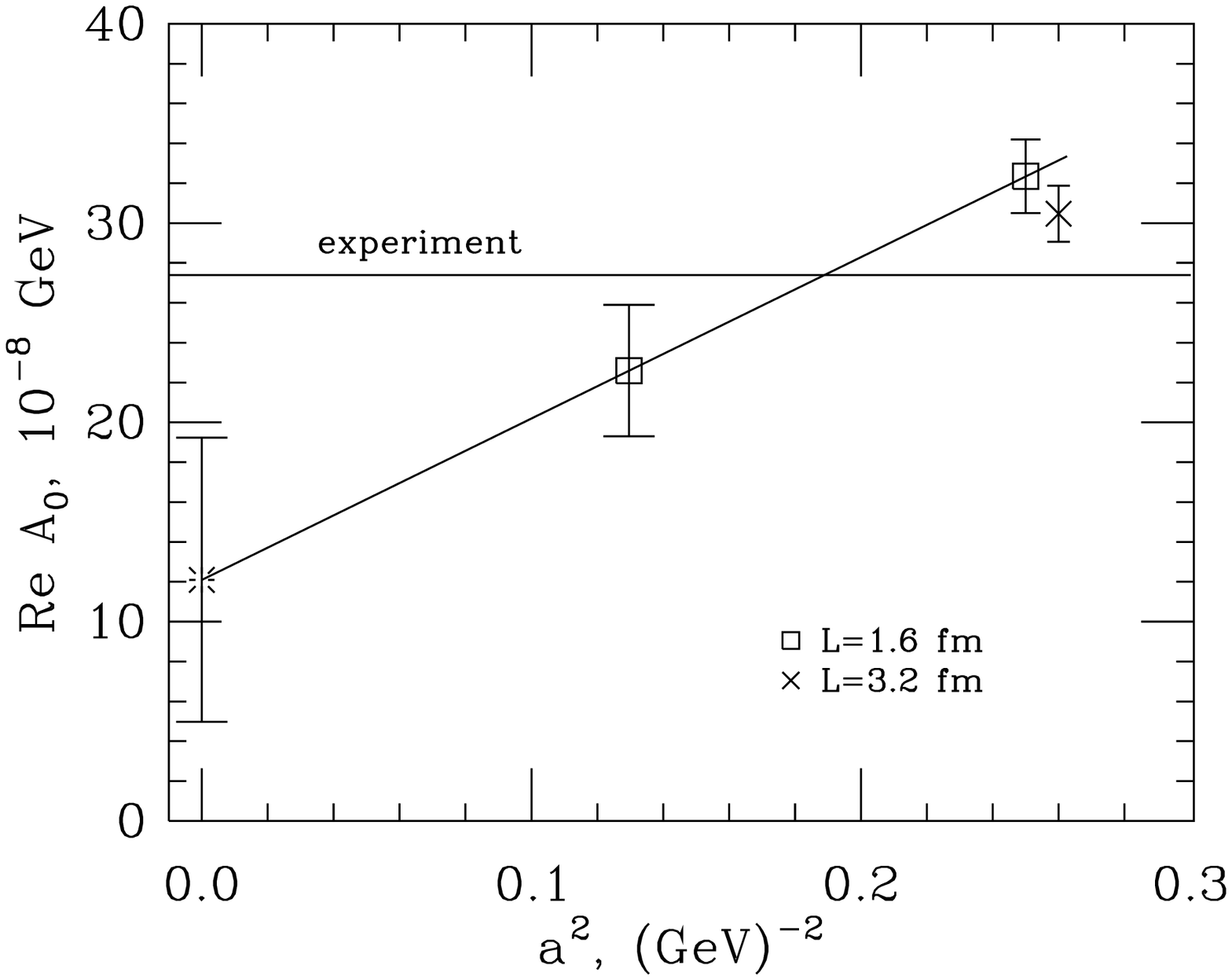}}
\end{center}
\caption{$\Re A_0$ for quenched ensembles plotted against lattice spacing
squared. A naive extrapolation to the continuum limit is made. 
The horizontal line corresponds to the experimental result of 
$27.4\cdot 10^{-8}$ GeV. Only statistical errors are shown.}
\label{A0cont}
\end{figure} 

All contraction types are needed in order to calculate 
$R_0$ (as opposed to the case of calculation of $\Re A_2$), 
including the expensive ``eyes'' and ``annihilations''.
%Our ``reference'' volume is $(1.6\; {\mathrm fm})^3$. 
The results for quenched $\beta =6.0$ and $\beta =6.2$ ensembles
are shown in Fig.~\ref{A0}. Dependence of $R_0$ on the 
meson mass is small, so there is no big ambiguity about the mass
prescription as in the $R_2$ case. 

It is apparent from Fig.~\ref{A0cont} that considerable cutoff 
dependence is present. The continuum extrapolation shown in the
figure is, naturally, not as reliable as one would desire.
In the future, data can be collected on a few more ensembles
to allow a more reliable extrapolation. Note that I have plotted
$\Re A_0$ vs. $a^2$. Strictly speaking, it is not known
whether the leading order cutoff effects are $O(a)$ or $O(a^2)$.
For the case of $B_K$ it was shown~\cite{sharpe2} that these effects
are $O(a^2)$, and we have assumed it is still the case for $\Re A_0$.
In any case, with present data the precision of the extrapolation
is far from the level where the form of the fit matters
significantly.

The effect of
the final state interactions (contained in the higher order
terms of the chiral perturbation theory) is likely to be large.
%(and positive), so that +50\% to 100\% correction is not unlikely. 
This is the biggest and most poorly estimated uncertainty. 

\begin{figure}[hbt]
\begin{center}
\leavevmode
\centerline{\epsfysize=8cm \epsfbox{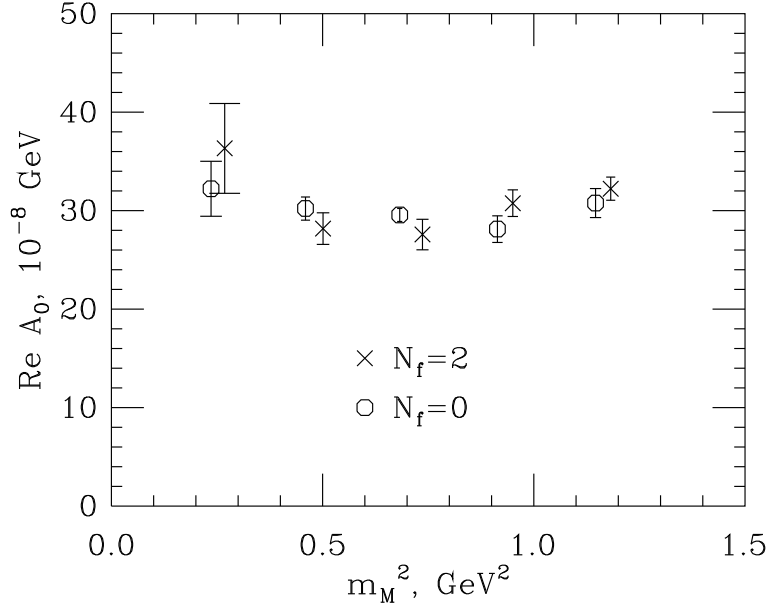}}
\end{center}
\caption{Comparison of quenched ($Q_1$) and dynamical results for $\Re A_0$
at comparable lattice spacings.}
\label{A0quench}
\end{figure} 

In addition, there is an operator matching uncertainty coming from mixing 
of $O_2$ with $O_6$ 
operator through penguin diagrams in lattice perturbation
theory. This is explained in the Section~\ref{sec:pert}. We estimate
that this uncertainty does not exceed 20\% for all ensembles.  

As for other uncertainties, we have checked the lattice volume
dependence by comparing ensembles $Q_1$ and $Q_2$ (1.6 and 3.2 fm
at $\beta =6.0$).
The dependence was found to be small, so we consider $(1.6 \;\mbox{fm})^3$ 
as a volume large enough to contain the system. We have also checked the effect
of quenching and found it to be small compared to noise
(see Fig.~\ref{A0quench}). 

\begin{figure}[htb]
\begin{center}
\leavevmode
\centerline{\epsfysize=8cm \epsfbox{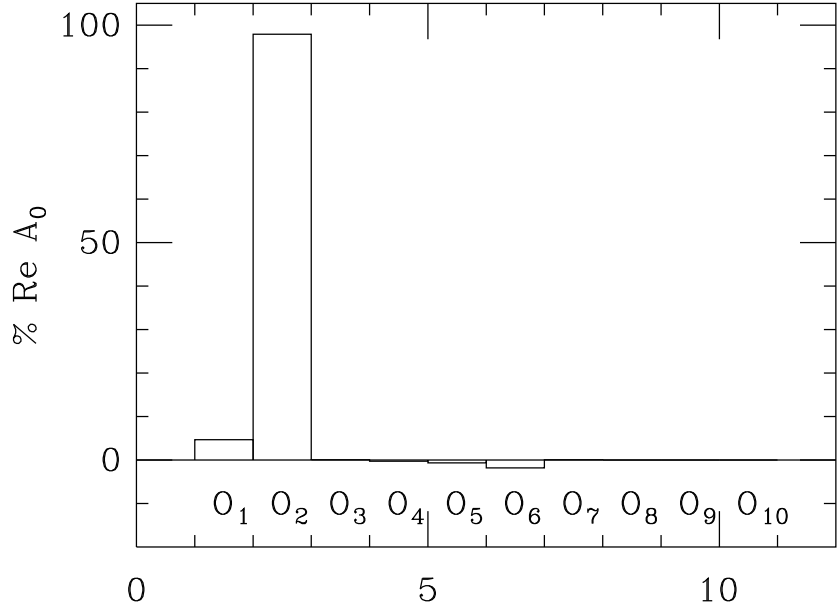}}
\end{center}
\caption{Contribution of various operators to $\Re A_0$.}
\label{A0hist}
\end{figure} 

The breakdown of contributions of various basis
operators to $\Re A_0$ is shown in Fig.~\ref{A0hist}.
By far, $O_2$ plays the most important role, whereas penguins
have only a small influence.

%------------------------------------------------------------
\section{Amplitude Ratio}
%------------------------------------------------------------
\label{sec:ratio}

Shown in Fig.~\ref{omega} is the ratio $\Re A_2/\Re A_0$ as directly 
computed on the lattice for quenched and dynamical data sets.
The data exhibit strong dependence on
the meson mass, primarily due to the chiral behaviour of $\Re A_2$ 
(compare with Fig.~\ref{A2}). 

\begin{figure}[htbp]
\begin{center}
\leavevmode
\centerline{\epsfxsize=5.5in \epsfbox{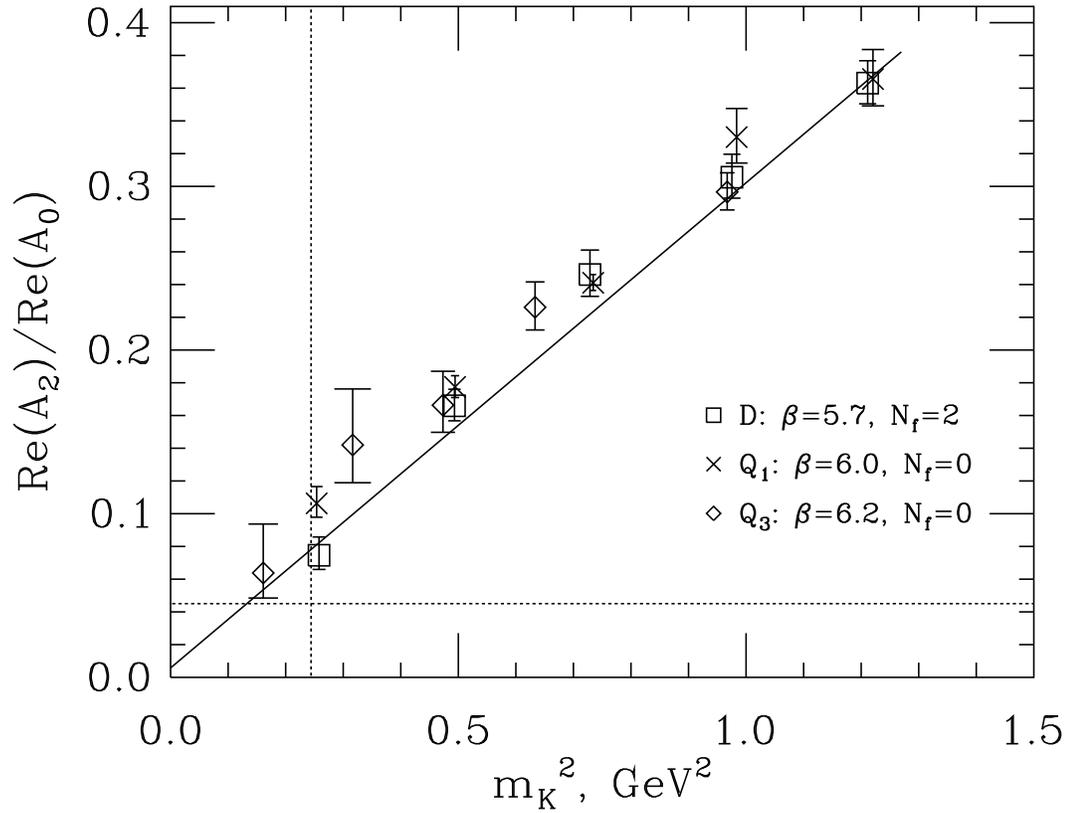}}
\end{center}
\caption{$\Re A_2/\Re A_0$ versus the meson mass squared
for quenched and dynamical ensembles.
Ensembles $Q_1$ and $D$ have comparable lattice spacings. 
The dynamical ensemble data were used for the fit. 
The horizontal line shows the experimental
value of $1/22$. The error bars show only the statistical errors 
obtained by jackknife.}
\label{omega}
\end{figure} 

\section{Summary}

Within our errors the results seem to confirm, indeed, the 
common belief that most of the $\Delta I=1/2$ enhancement comes 
from strong interactions, in particular from the ``eye'' and 
``annihilation'' diagrams. The exact amount
of enhancement is broadly consistent with experiment while being
subject to considerable uncertainty due to higher-order chiral terms.
Other systematic errors are the same as those described 
in Section~\ref{sec:dione}.
%With current level of knowledge we can only conclude that 
%QCD effects are in principle capable of bringing about $\Delta I=1/2$ 
%channel enhancement consistent with experiment.

\chapter{Operator Matching and \epsp}
\label{matching}

As mentioned before, we have computed the matrix elements of all 
relevant operators with an acceptable statistical accuracy.
These are regulated in the lattice renormalization 
scheme. To get physical results, 
operators need to be matched to the same scheme in which the Wilson
coefficients were computed in the continuum, namely $\overline{\mathrm MS}$
NDR. While perturbative matching works quite well for
$\Re A_0$ and $\Re A_2$, it seems to break down severely for
matching operators relevant for $\varepsilon '/\varepsilon$.

%=====================================================
\section{Perturbative Matching}
%=====================================================

\label{sec:pert}

Conventionally, lattice and continuum operators are matched using
lattice perturbation theory:
\begin{equation}
\displaystyle
\label{eq:matching}
O_i^{\it cont}(q^*) =  O_i^{\it lat} + \displaystyle\frac{g^2(q^* a)}{16\pi^2}\displaystyle\sum_j(\gamma_{ij}\ln (q^* a)
 + C_{ij})O_j^{\it lat} + O(g^4) + O(a^n) ,
\end{equation}
where $\gamma_{ij}$ is the one-loop anomalous dimension matrix 
(the same in the continuum 
and on the lattice), and $C_{ij}$ are finite coefficients calculated
in one-loop lattice perturbation theory~\cite{Ishizuka,PatelSharpe}. 
We use the ``horizontal matching'' 
procedure~\cite{horizontal}, whereby the same coupling constant
as in the continuum ($g_{\overline{\rm MS}}$) is used.
The operators are matched at an intermediate scale 
$q^*$ and evolved using the continuum renormalization
group equations to the reference scale $\mu$, which we take 
to be 2 GeV.
%By varying $q^*$ we can check how well the perturbation theory works
%because $q^*$ dependence is a next-order effect. 

\begin{figure}[htbp]
\begin{center}
\leavevmode
\centerline{\epsfxsize=5.5in \epsfbox{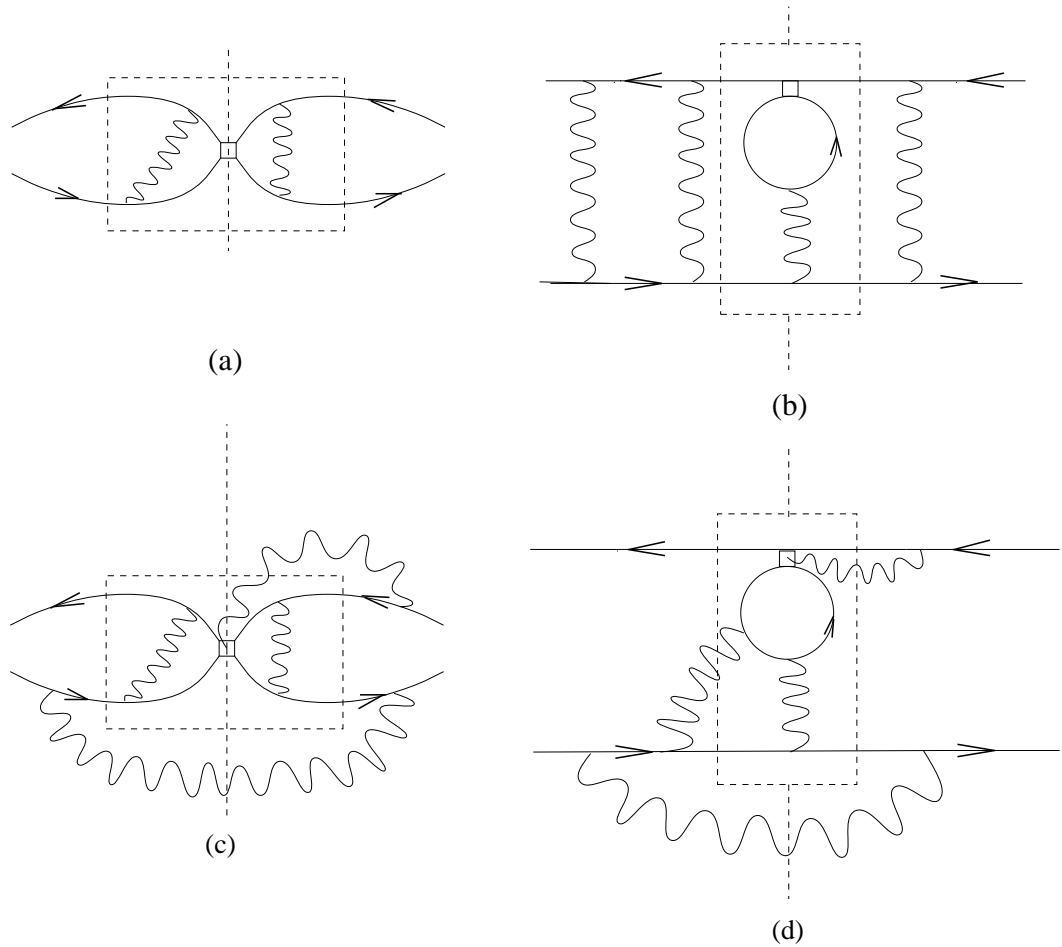}}
\end{center}
\caption{Example of four kinds of diagrams with arbitrary number of loops
arising in renormalization 
of four-fermion operators: in (a) and (b) no propagator
crosses the box or the axis; (c) and (d) exemplify the
rest of the diagrams. The rectangular drawn in dotted line in (b)
corresponds to operator structure $PP_{EU}$.}
\label{higher-order}
\end{figure} 

%------------------------------------------------------------
\subsection{Perturbative Matching And $\Re A_{0,2}$}
%------------------------------------------------------------

In calculating of $\Re A_0$ and $\Re A_2$, the main contributions
come from left-left operators. One-loop renormalization
factors for such (gauge-invariant) operators were computed by
Ishizuka and Shizawa~\cite{Ishizuka} (for current-current diagrams)
and by Patel and Sharpe~\cite{PatelSharpe} (for penguins). 
These factors are fairly small, so at the first glance the perturbation theory
seems to work well, in contrast to the case of left-right operators 
essential for estimating $\varepsilon '/\varepsilon$, as described below.
However, even in the case of $\Re A_0$ there is a certain
ambiguity due to mixing of $O_2$ with $O_6$ through
penguin diagrams. The matrix element of $O_6$ is rather large, so
it heavily influences $\langle O_2\rangle$ in spite of the small
mixing coefficient. The operator $O_6$ receives enormous renormalization 
corrections in the first order, as discussed below. Therefore, there
is an ambiguity as to whether the mixing should be evaluated
with renormalized or bare $O_6$. That is, the higher-order
diagrams (such as Fig.~\ref{higher-order}b and~\ref{higher-order}d) 
may be quite important here.  

In order to estimate the uncertainty of neglecting higher-order diagrams,
we evaluate the mixing with $O_6$ renormalized
by the partially non-perturbative procedure described below, and
compare with results obtained by evaluating mixing with bare $O_6$.
The first method amounts to resummation of those higher-order diagrams 
belonging only to type (b) in Fig.~\ref{higher-order}, while the second 
method ignores all higher-than-one-order corrections. 
Results quoted in the previous Section
were obtained by the first method, which is also close to using
tree-level non-diagonal matching. The second method would produce
values of $\Re A_0$ lower by about 20\%.
Thus we consider 20\% a reasonable estimate of the matching uncertainty.  

In calculating $\varepsilon '/\varepsilon$ the operator 
matching issue becomes a much more serious obstacle as explained below.

\begin{figure}[htb]
\begin{center}
\leavevmode
\centerline{\epsfysize=8cm \epsfbox{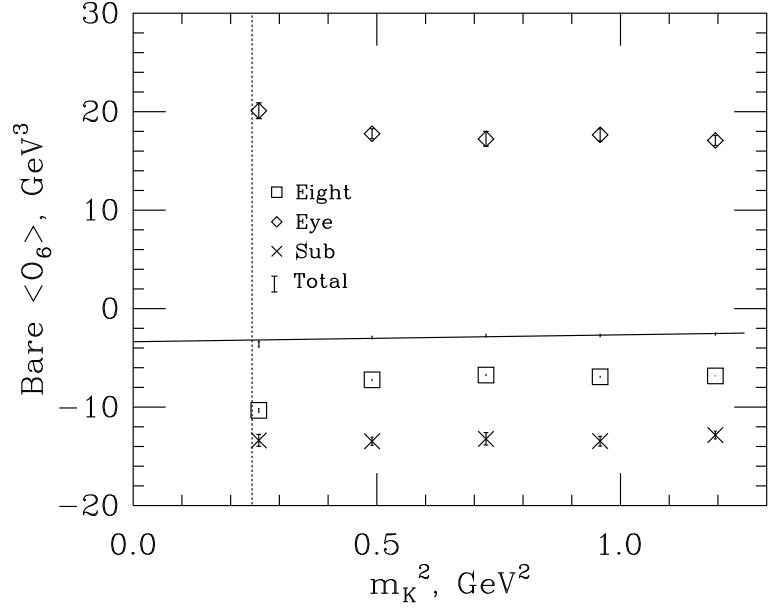}}
\end{center}
\caption{Three contributions to $\langle \pi^+|O_6|K^+\rangle$:
``eight'' (boxes), ``eye'' (diamonds) and ``subtraction'' (crosses).
These data represent bare operators for the dynamical ensemble.
The fit is done for the sum total of all contributions.}
\label{fig:O6}
\end{figure} 

%------------------------------------------------------------
\subsection{Problems With Perturbative Matching}
%------------------------------------------------------------

The value of $\varepsilon '/\varepsilon$ depends on a number of 
subtle cancellations between matrix elements. In particular, 
$O_6$ and $O_8$ are considered the most important operators
whose contributions have opposite signs and almost cancel. Furthermore,
the matrix elements of individual operators contain three main components 
(``eights'', ``eyes'',
and ``subtractions''), which again conspire to almost cancel each other
out (see Fig.~\ref{fig:O6}). 
%Finally, contractions with left-right operators 
%entering the expression for $\langle O_6\rangle$ compete with $(P-S)(P+S)$
%operators (see the Appendix), because the latter receive large 
%renormalization corrections and are made of comparable size with the former.  
Thus $\varepsilon '/\varepsilon$ is extremely sensitive to each of these 
components, and in particular to their matching. 

\begin{table}[tbh]
\begin{tabular}{l|c|c|c}
\hline\hline
Quark mass & 0.01 & 0.02 & 0.03 \\
\hline
Bare &
$-2.95 \pm 0.27 $ & 
$-2.61 \pm 0.11 $ & 
$-2.50 \pm 0.08 $ \\
$q^*=\pi /a$ &
$-0.69 \pm 0.10 $ & 
$-0.66 \pm 0.04 $ & 
$-0.66 \pm 0.03 $ \\
$q^* =1/a$ & 
$0.01  \pm 0.07 $ & 
$-0.07 \pm 0.03 $ & 
$-0.11 \pm 0.02 $ \\ 
Partially non-pert. &
$2.03 \pm  0.07 \pm 0.42$ & % \pm 0.62$ &
$0.74 \pm  0.03 \pm 0.37$ & % \pm 0.02$ &
$0.44 \pm  0.02 \pm 0.31$ \\ % \pm 0.03$ \\
\hline
\end{tabular}
\begin{tabular}{l|c|c}
\hline
Quark mass &  0.04 & 0.05 \\
\hline
Bare &
$-2.60 \pm 0.16 $ & 
$-2.94 \pm 0.27 $ \\
$q^*=\pi /a$ &
$-0.73 \pm 0.06 $ & 
$-0.83 \pm 0.09 $ \\
$q^* =1/a$ & 
$-0.19 \pm 0.04 $ & 
$-0.20 \pm 0.03 $ \\
Partially non-pert. &
$0.23 \pm  0.03 \pm 0.28$ & % \pm 0.02$ &
$0.25 \pm  0.05 \pm 0.36$ \\ % \pm 0.02$ \\
\hline\hline
\end{tabular}
\caption{$\langle \pi^+\pi^-|O_6|K^0\rangle$ for $Q_1$ ensemble in units of 
$(\mbox{GeV})^3$ 
with tree-level matching (bare); with one-loop perturbative matching using 
two values of $q^*$; and with matching obtained by the partially 
non-perturbative procedure. The errors are statistical, except for the 
last line, where the second error comes from
uncertainty in our determination of $Z_S$ and $Z_P$.
%and the third one is an estimate of higher-order diagrams. 
As mentioned in text, there is
an unknown uncertainty in the partially non-perturbative procedure.}
\label{tab:O6pert}
\end{table}

Consider fermion contractions with operators such as
\begin{eqnarray}
\label{eq:O6ops1}
(PP)_{EU} & = & (\overline{s} \gamma_5 \otimes \xi_5 u) 
(\overline{u} \gamma_5 \otimes \xi_5 d) \\
(SS)_{IU} & = & (\overline{s} \openone \otimes \openone d) 
(\overline{d} \openone \otimes \openone d) \\
\label{eq:O6ops3}
(PS)_{A2U} & = & (\overline{s} \gamma_5 \otimes \xi_5 d) 
(\overline{d} \openone \otimes \openone d) ,
\end{eqnarray}
which are main parts of, correspondingly, ``eight'', ``eye'' and 
``subtraction''
components of $O_6$ and $O_8$ (see the Appendix). The 
finite renormalization coefficients for these operators
have been  computed in Ref.~\cite{PatelSharpe}. 
The diagonal coefficients are very large, so the
corresponding one-loop corrections are in the 
neighborhood of $-100\%$ and strongly depend
on which $q^*$ is used (refer to Table~\ref{tab:O6pert}).
Thus perturbation theory fails in reliably 
matching the operators in Eqs.~(\ref{eq:O6ops1})---(\ref{eq:O6ops3}). 

The finite coefficients for other (subdominant)
operators, for example
$(PP)_{EF}$, $(SS)_{EU}$ and $(SS)_{EF}$,
are not known for formulation with gauge-invariant 
operators\footnote{Patel and Sharpe~\cite{PatelSharpe}
have computed corrections for gauge-noninvariant operators.
Operators in Eqs.~(\ref{eq:O6ops1})---(\ref{eq:O6ops3})
have zero distances, so the corrections are the same for
gauge invariant and non-invariant operators.
Renormalization of other operators
(those having non-zero distances) generally differs from the 
gauge-noninvariant operators.}.
For illustration purposes,
in Table~\ref{tab:O6pert} we have used coefficients for gauge
non-invariant operators computed in Ref.~\cite{PatelSharpe}, but 
strictly speaking this is not justified. 

To summarize, perturbative matching does not work and
some of the coefficients are even poorly known. A solution
would be to use a non-perturbative matching procedure, such as
described in Ref.~\cite{nonpert}. Before we proceed to this
procedure, let me discuss a simpler procedure for non-perturbative
calculation of matching coefficients for bilinears $Z_S$ and $Z_P$,
which is important in its own right since the perturbative 
matching error is the dominant source of errors in determination of
light quark masses via lattice QCD. 

%=====================================================
\section{Nonperturbative Matching Of The Scalar And Pseudoscalar 
Bilinears And Determination Of Light Quark Masses}
%=====================================================

\label{sec:bil}

\subsection{Framework}

The general method of nonperturbative operator matching has been 
suggested by Martinelli {\it et al.}~\cite{martinelli}. 
It has been successfully applied in several 
calculations~\cite{nonpert}. Here we follow one
variant of the method which allows only the determination
of $Z_S$ and $Z_P$ renormalization factors of the
scalar and the pseudoscalar operators, which is enough
for our purposes (see also Ref~\cite{IshizukaNonpert}).

The method is based on averaging the quark propagator
between external single quark states over gauge configurations
in a fixed (usually, Landau) gauge, considering 
off-shell momenta. The momenta should be large enough
($|p^2| \gg \Lambda_{QCD}^2$) so that it makes sense 
even to consider single quark states, and on the
other hand, the momenta should be sufficiently small compared to 
$1/a$ in order to avoid large discretization effects. Note that these 
criteria apply to any (perturbative or non-perturbative) 
method of operator matching.

The ensemble-averaged propagator is expected to have the form
\begin{equation}
\langle S(p) \rangle = \frac{Z_q(p)}{M(p)+\frac{i}{a}\sum_\mu 
(\gamma_\mu\otimes\openone) \sin (p_\mu a)}\;, 
\label{eq:S}
\end{equation}
where $Z_q$ is the renormalization of the fermion field. 
It can be shown that 
\begin{eqnarray}
\label{eq:Zs}
Z_S(p) = \frac{\partial M(p)}{\partial m}\\
\label{eq:Zp}
Z_P(p) = \frac{M(p)}{m}  ,
\end{eqnarray}
where $m$ is the bare quark mass and $Z_S$ and $Z_P$ are the
bilinear renormalization factors defined by
\begin{equation}
(S,P)^{\rm latt} = Z_{S,P} \;(S,P)^{\rm cont}. 
\end{equation}
It follows that $M(p)$ can be found as
\begin{equation}
M(p) = \frac{{\rm Tr}[S(p)^{-1}] Z_q(p)}{{\rm Tr}[\openone]},
\end{equation}
where the trace is taken with respect to both color and Dirac indices.
$Z_q$ can be obtained from
\begin{equation}
\label{eq:Zq}
Z_q(p)^{-1} = -ia\;\frac{\sum_\mu {\rm Tr}[S(p)^{-1}
(\gamma_\mu\otimes\openone )]\;
\sin(p_\mu a)}{\sum_\mu {\rm Tr}[\openone ]\;\sin^2 (p_\mu a)}.
\end{equation}

This procedure amounts to resummation of all perturbative
corrections to $Z_S$ and $Z_P$. In the continuum, it corresponds
to the so-called regularization-invariant (RI) scheme. In order
to obtain results in the $\overline{\rm MS}$ scheme, $Z_S$ and $Z_P$ 
need to be rescaled by an appropriate factor $Z_c$. This factor was 
computed at two-loop order in Ref.~\cite{RIfactor}. It is
of the order of 0.89 for the ensembles of interest, and the higher-order
perturbative corrections can be safely neglected. Thus $Z_{S,P}$
in $\overline{\rm MS}$ scheme are given by:
\begin{equation}
Z_{S,P}^{\overline{\rm MS}}(\mu ) = U_{\overline{\rm MS}}(\mu ,q^*)
Z_c(q^*)Z_{S,P}^{\rm RI}(q^*),
\end{equation}  
where $U_{\overline{\rm MS}}$ is the three-loop evolution factor.

\subsection{Procedure}

In applying the general method described above to the case
of staggered fermions, care needs to be taken in considering
the propagator. The complication here is that staggered quark
fields ``live'' on the coarse lattice ($2a$) and there are
several different but equivalent ways to write down the action.
Here I carefully describe the procedure and ideas behind our
computations.

Recall that there are two equivalent ways to work with 
staggered
fermions~\cite{montvay,kluberg}. The first one is based on fermion
fields $\chi (x,A)$ defined in the coordinate space 
(as in Section~\ref{sec:stag}). 
%The spin-flavor identification is achieved as
%\begin{equation}
%Q^{\alpha a}(x) = \frac{1}{8} \sum_{A\in \{0,1\}^4} 
%\label{eq:sf}
%\Gamma_A^{\alpha a} \chi (x,A)
%\end{equation}
%After transformation to momentum space in hypercube position $x$, 
%the (free) fermionic action can be written as
%\begin{eqnarray}
%\displaystyle
%\label{action-coordinate}
%S_F [\chi,\overline{\tilde{\chi}}] & = & 
%\frac{1}{V}\sum_p\sum_{A,B}
%\overline{\chi}(p,B)\;[\;m\;\openone_{B,A} 
%+ \frac{1}{2a}\sum_\mu (\;i\; \sin (2p_\mu a)\gam \mu_{B,A} \nonumber \\
%& & + (1-\cos (2p_\mu a)) \overline{(\gamma_5\otimes\xi_5\xi_\mu )}_{B,A})]\;
%\tilde{\chi} (p,A).
%\end{eqnarray}
%The sum over momentum is performed over the range $p\in [-\pi/2a,\pi/2a[^4$.

Another way is to start with fields $\psi (k)$ defined completely
in momentum space. As is well known, there are 16 poles in the
fermion propagator at $p_\mu = (0,\pi/a)$. These poles are used to
identify fermion spin and flavor. Thus it is convenient to 
divide the Brillouin zone into 16 regions, labeled by
$H \in \{0,1\}^4$. 
The momentum $k$ can be broken down into a 
physically meaningful part $p\in [-\pi/2a,\pi/2a[^4$ and
the offset corresponding to the given region as follows: 
$k=p+\frac{\pi}{a} H$. The (free) fermionic action can be written as
\begin{equation}
S_F = \frac{1}{V}\sum_p\sum_{F,H}  \overline{\psi}(p,F)
[\;m\;\openone_{F,H} + \frac{i}{a} \sum_\mu  \sin (p_\mu a) \ \ggam\mu_{F,H}]
\psi (p,F).
\end{equation}
The matrices $\ssfno S F$ are related to $\sfno S F$ by a unitary
transformation:
$$  \ssf S F A B = 
  \sum_{CD} \frac{1}{16} (-)^{A\cdot C}\ \sf S F C D \ (-)^{D\cdot B} \ ,
$$
where $A\cdot C \equiv \sum_\mu A_\mu C_\mu$.
The fields $\psi (k)$ are Fourier transforms of the fields
$\chi (x,A)$
($\psi (k) = \sum_{x,A} \exp (-ik(x+aA))\chi (x,A)$).
Consider fields $\phi (p,A)$ defined by
\begin{equation}
\phi (p,A) \equiv \sum_x \exp (-ip(x+aA))\chi (x,A)\ .
\end{equation}  
They are related to $\psi$ by
\begin{equation}
\psi (p,H) = \sum_A (-)^{A\cdot H} \phi (p,A)\ .
\end{equation}
The (free) action in terms of the new fields is given by
\begin{equation}
\label{action-momentum}
S_F = \frac{16}{V}\sum_p\sum_{A,B}
\overline{\phi}(p,B)[\;m\;\openone_{B,A} + 
\frac{i}{a}\sum_\mu \; \sin (p_\mu a)\gam \mu_{B,A}]\;\phi(p,A).
\end{equation}
Actions in coordinate and momentum space are completely equivalent
~\cite{kluberg,montvay,daniel}. 
It is with the latter (momentum space) form of action that we choose to 
work here.

%Note that $\tilde{\chi}$ differs from $\chi$ defined earlier by the
%factor $\exp {(iapA)}$. We work in the the second
%formalism (i.e., with $\tilde{\chi}$).

In practice, we use 16 
$\delta $-function sources at the corners of the hypercube containing 
the origin ($x=(0,A), A\in (0,1)^4$).
Propagator $G(y,B;0,A)$ is computed for each space-time point,
where $y$ denotes a hypercube and $B$ is a hypercube corner.
We perform Fourier transform in $y$ to obtain
$G(p,B,A)$, where $p\in [-\pi/2a,\pi/2a[^4$. In addition, 
in order to work with fields $\phi (p,A)$,
$G(p,B,A)$ is multiplied by the phase $\exp{[-iap(B-A)]}$ to obtain
$S(p,B,A)$. 
The propagator is 
averaged over the gauge ensemble (denoted $\langle\rangle$ below),
and afterwards the inverse is taken.
% 
%the one based on single-component field 
%$\chi (p,A)$ defined
%for all momenta $p\in [-\pi/2a,\pi/2a]^4$ and hypercube corners 
%in coordinate space $A$;
%and the one using staggered quark field $Q^{\alpha a}(p)$ with
%internal spin-flavor indices, again defined for all momenta 
%$p\in [-\pi/2a,\pi/2a]^4$. 
%
%Recall that in momentum space the action can be expressed
%equivalently in terms of three fields: the Brillouin zone field 
%$\Phi_A(p)$ (here $A$ denotes an appropriate zone),
%the single-component hypercube-based field 
%and the staggered
%quark field $Q^{\alpha a}(p)$. In terms of the staggered quark field,
%the propagator has the form of Eq.~(\ref{eq:S}). Equivalently, 
%the inverse propagator in terms of single-component hypercube-based field
%(which we have computed) is 
%
The inverse propagator then has the form (up to various factors that 
can be absorbed into $Z_q$)
\begin{eqnarray}
\langle S(p,B,A)\rangle^{-1} & = & (\sum_\mu \frac{i}{a} \sin (ap_\mu ) \;
\gam \mu _{B,A}
%\frac{1}{4}{\rm Tr} [\Gamma_B^\dag \gamma_\mu \Gamma_A] 
+ M(p)\iden_{B,A} ) 
%{\rm Tr} [\openone])
\;/\;Z_q
\label{eq:prop1}
\nonumber \\
& = & (\sum_\mu \frac{i}{a}\sin (ap_\mu ) \;\eta_\mu (A)
\;\delta_{(A,B+\hat{\mu})} + M(p)\;\delta_{A,B})\;/\;Z_q\ .
\end{eqnarray}
In $\delta_{(A,B+\hat{\mu})}$ the summation is
understood modulo 2. Thus $Z_q$ can be obtained as
\begin{equation}
Z_q^{-1} = -ia\;\frac{\sum_A\sum_\mu \langle S(p,A,A+\hat{\mu})\rangle^{-1}
\eta_\mu (A)\;\sin(p_\mu a)}{\sum_\mu \sin^2 (p_\mu a)},
\end{equation}
and
\begin{equation}
M(p) = \sum_A \langle S(p,A,A)\rangle^{-1} Z_q .
\end{equation}
Here $\langle S(p,A,B)\rangle^{-1}$ is the inverse of the matrix 
$\langle S(p)\rangle_{AB}$ in both
color and hypercube corner index space. In practice, the calculations
are made easier by the fact that the propagator is proportional
to the identity matrix in color space when one works in Landau gauge.

A number of checks have been performed to confirm the expected form
of the propagator in Eq.~(\ref{eq:S}). In particular, the traces of
the averaged inverse propagator with
all spin-flavor structures other than $(\gamma_\mu\otimes\openone )$
and $(\openone\otimes\openone)$ were found to be consistent with zero
within given statistical errors, and the momentum
dependence of the $(\gamma_\mu\otimes \openone)$ trace was found 
to be extremely close to $\;\sin (ap_\mu)$. 

\begin{figure}[htb]
\begin{center}
\leavevmode
\centerline{\epsfysize=8cm \epsfbox{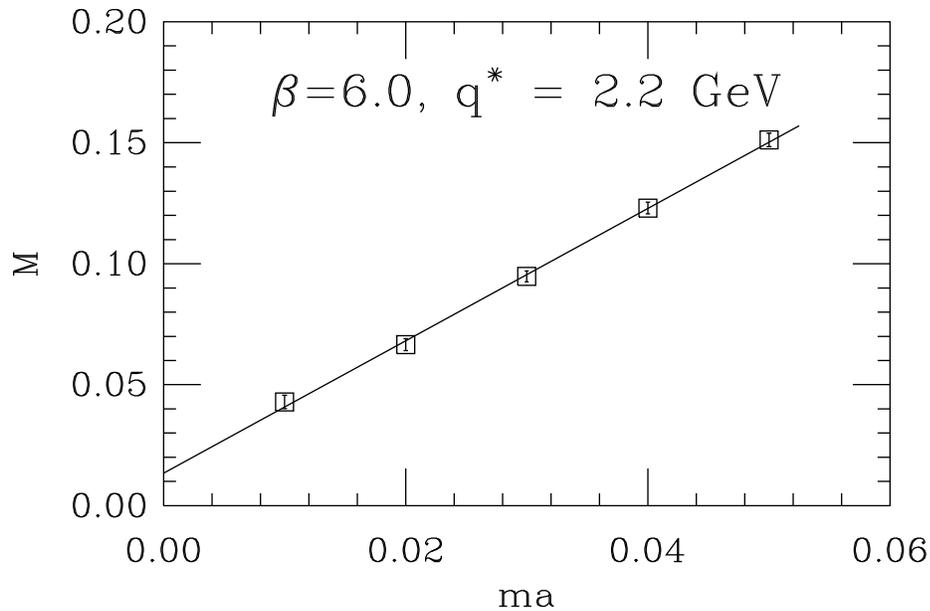}}
\end{center}
\caption{$M(p)$ vs. quark mass for one momentum with 
$p^2 = (2.2 \mbox{ GeV})^2$ at $\beta=6.0$.}
\label{Mp}
\end{figure} 

\begin{figure}[p]
\begin{center}
\leavevmode
\centerline{\epsfxsize=5.5in \epsfbox{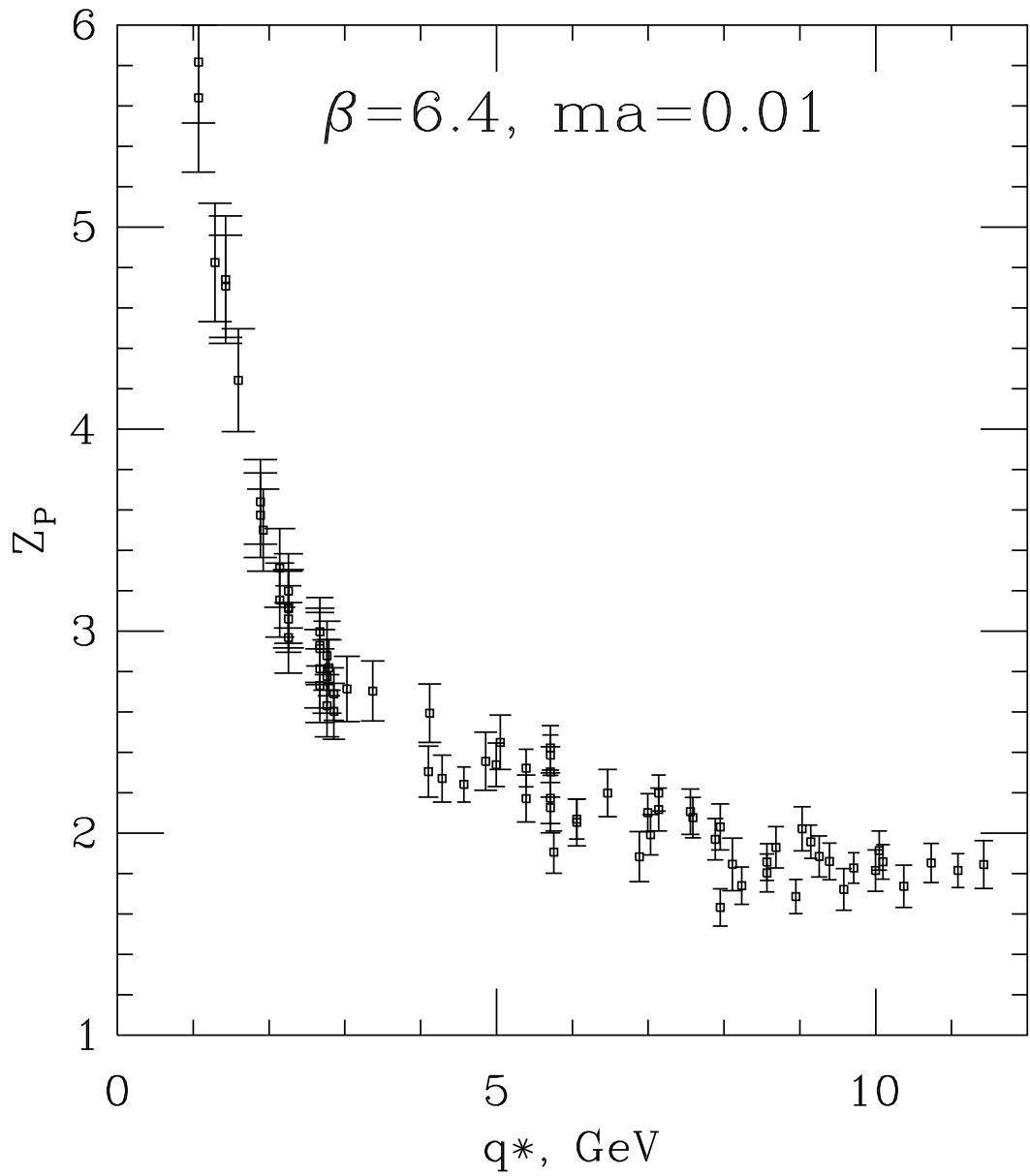}}
\end{center}
\caption{$Z_P$ at $ma=0.01$ vs. momentum at $\beta=6.4$.}
\label{fig:Zp}
\end{figure} 

%-------------------------------------------------------------
\subsection{Results For $Z_S$ And $Z_P$}
%-------------------------------------------------------------

\label{sec:Zs}

We have computed $Z_S$ and $Z_P$ on three quenched and one
dynamical ensembles (see Table~\ref{tab:param3} for details).

\begin{table}[tbh]
\begin{tabular}{ccccccc}
\hline\hline
$N_f$ & $\beta $ & Size & $a^{-1}$, GeV & L, fm & Number of & 
Quark masses\\  & & & & & configurations & considered \\
\hline
2 & 5.7 & $16^3\times 32$ & 2.0  & 1.6 & 49 & 0.02, 0.03 \\
0 & 6.0 & $16^3\times 32$ & 2.07 & 1.5 & 70 & 0.01, 0.02, 0.03, 0.10 \\
0 & 6.2 & $24^3\times 48$ & 2.77 & 1.7 & 69 & 0.01, 0.015, 0.02, 0.03 \\
0 & 6.4 & $32^3\times 64$ & 3.64 & 1.8 & 57  & 0.0100, 0.0125, 0.0150 \\
\hline\hline
\end{tabular}
\caption{Parameters of ensembles used for calculation of $Z_S$, $Z_P$ and 
light quark masses.}
\label{tab:param3}
\end{table}

Plotted in Fig.~\ref{Mp} is $M(p)$ at a given $p$ with 
$q^*\equiv\sqrt{p^2}=2.2$ GeV, as a function of quark mass
at $\beta=6.0$.
The dependence is linear, which is also the case for the rest of the
momenta. From Eq.~(\ref{eq:Zs}) it is clear that
$Z_S$ is the slope of this function. The fact that the intersect
is non-zero signals the presence of non-perturbative pseudoscalar 
vertex contribution, arising from the capability of the pseudoscalar
density operator to create Goldstone bosons from the vacuum.  
This contribution 
is a consequence of spontaneous chiral symmetry breakdown,
and corresponds to the dynamical generation of a (constituent) 
light quark mass, of magnitude about 300--400 MeV, off-shell and 
gauge-dependent (for further discussion see 
Refs.~\cite{IshizukaNonpert,Pittori} and references therein). 
This effect (measured by the intersect of $M(p)$ as a function of $m$) behaves
as $\sim\frac{1}{p^2}$ for large enough $p^2$ ($p^2\gg m_\pi^2$). 

Since the intersect is non-zero, $Z_P$ is not identical to $Z_S$.
This is in contrast to $Z_S=Z_P$ in perturbation theory, where 
Goldstone boson pole is entirely omitted.
The ratio of $Z_P/Z_S$ (Fig.~\ref{Zps}) is almost constant 
and close to 1 for larger momenta, while it grows rapidly
at smaller momenta. This shows that the Goldstone boson pole
contribution is very large. It is 
proportional to $1/m$ (again, see Eq.~(\ref{eq:Zp})).
Thus $Z_P$ (Fig.~\ref{fig:Zp}) diverges in the limit of small 
momenta and quark masses.

$Z_S$ is finite and independent of quark mass (Fig.~\ref{Zs}).
The errors were obtained by combining line fitting and 
jackknife. In Fig.~\ref{Zs} I compare the non-perturbative results 
for $Z_S$ with those 
obtained in improved perturbation theory (using $g_{\overline{MS}}$).

\begin{figure}[p]
\begin{center}
\leavevmode
\centerline{\epsfxsize=5.5in \epsfbox{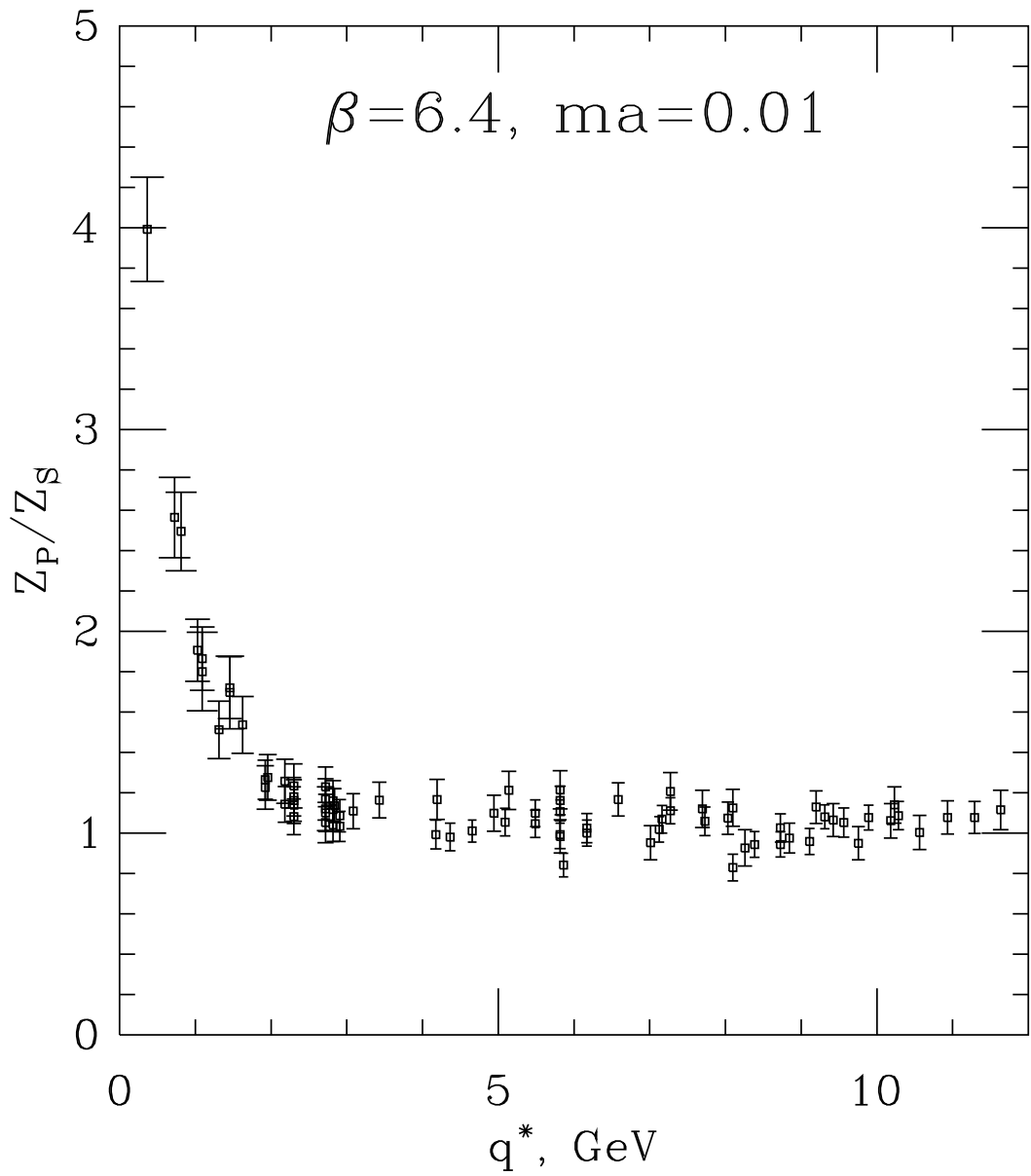}}
\end{center}
\caption{Ratio $Z_P/Z_S$ for $\beta=6.4$ vs. momentum.}
\label{Zps}
\end{figure} 

\begin{figure}[p]
\begin{center}
\leavevmode
\centerline{\epsfxsize=5.5in \epsfbox{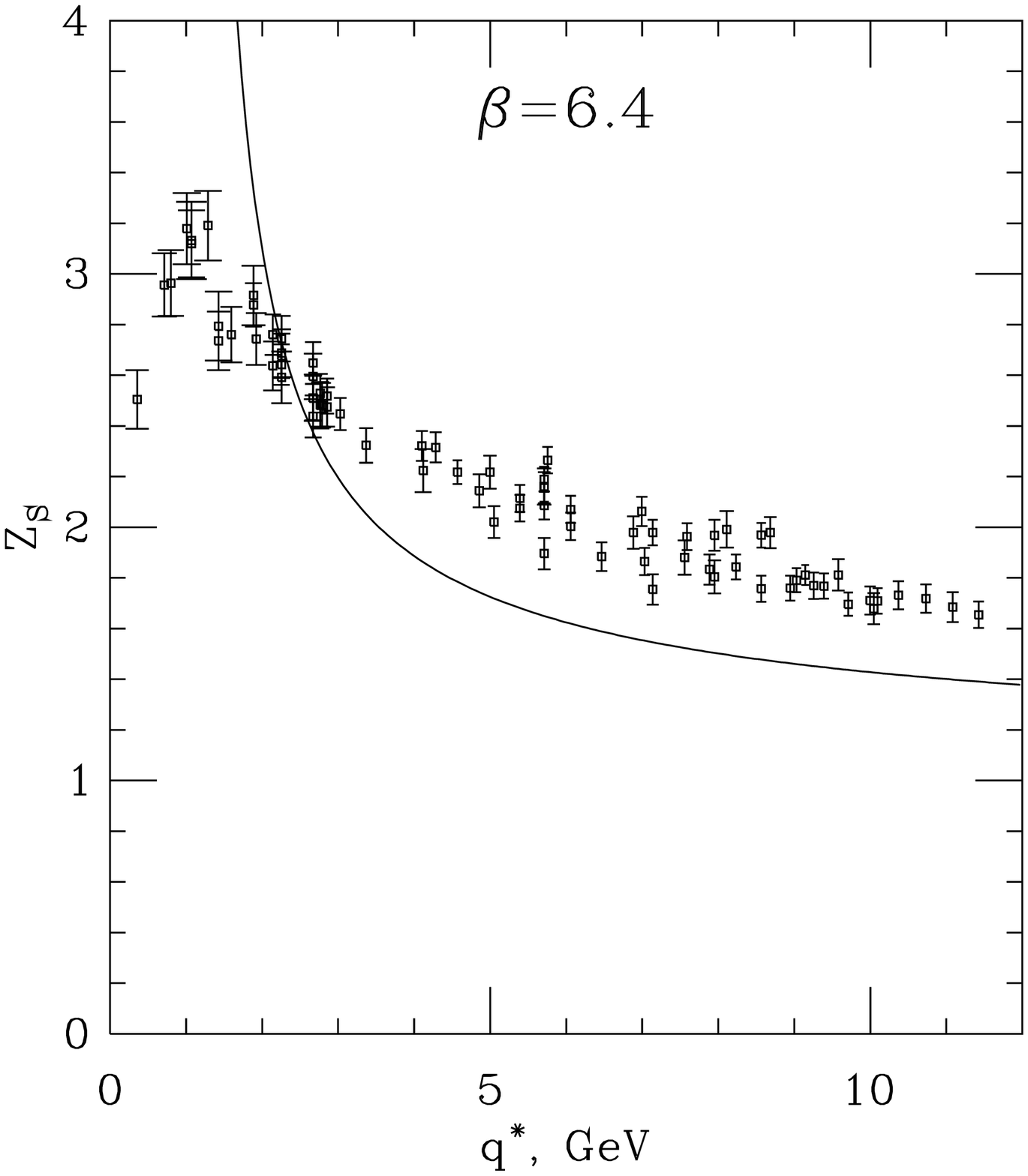}}
\end{center}
\caption{$Z_S$ at $\beta=6.4$ vs. momentum. Solid curve shows the
$Z_S$ value calculated in perturbation theory with $g_{\overline{\rm MS}}$
coupling.}
\label{Zs}
\end{figure} 

We have checked the discrete rotation invariance
by comparing results for several momenta related by rotation
by $90^\circ$ around one or several of the axes. Another 
interesting check is to compare results obtained with two 
momenta which have the same $p^2$ but different component 
structure, for example: (1,1,1,1) and (0,0,0,2). This allows one 
to check for continuous rotation
invariance. As expected, this invariance is slightly broken
by the lattice cutoff. In the above figures, all kinds of 
momenta component structures are present. Sometimes, 
two (or more) momenta have the same $p^2$. By looking at the 
differences in the plotted function between these two momenta 
it is possible to obtain an indication of the discretization effects.
Indeed, these effects grow larger for larger momenta, 
as expected. In addition, they decrease for any
given momentum as $\beta$ grows (i.e., closer to the continuum limit).
These effects are discussed more in the next subsection.

%------------------
\subsection{Results For Light Quark Masses}
%------------------------------------

\label{sec:QM}

Values of light quark masses can be obtained from lattice calculations
with staggered fermions as follows. By computing masses of mesons
at a variety of bare quark mass values, the values of
bare quark masses are obtained by extrapolation to 
the corresponding physical meson masses.
For example, the strange quark mass corresponds to the lattice
meson mass being $\sqrt{2}$ times the physical kaon mass (this is 
because the kaon on the lattice is composed of two equal-mass quarks, 
while in the real world the kaon is composed of the strange quark and
the comparatively massless down-quark; and $m_K^2 \propto (m_s+m_d)$).
The bare quark mass in physical units is obtained by dividing the
mass in lattice units by $a$:
\begin{equation}
m = \frac{(ma)_{M_{\rm phys}}}{a}\ .
\end{equation}
This procedure gives the masses of the quarks in lattice
regularization scheme. But their values need to be matched to 
continuum just the same as for the operators, and herein lies
a well-known problem, since the perturbative operator matching
does not work very well in this case.

To overcome this problem, we use the bilinear renormalization constant 
$Z_S$, computed non-perturbatively as described in the previous
subsection. Using
$$Z_S = Z_m^{-1},$$
we compute the quark masses in $\overline{\rm MS}$ NDR scheme 
as follows:
\begin{equation}
m^{\rm cont}(\overline{\rm MS} \;{\rm NDR}, \mu) = 
U(\mu ,q^*) Z_c(q^*) Z_S^{\rm RI}(q^*) m^{\rm latt}.  
\end{equation}

\begin{figure}[p]
\begin{center}
\leavevmode
\centerline{\epsfxsize=5.5in \epsfbox{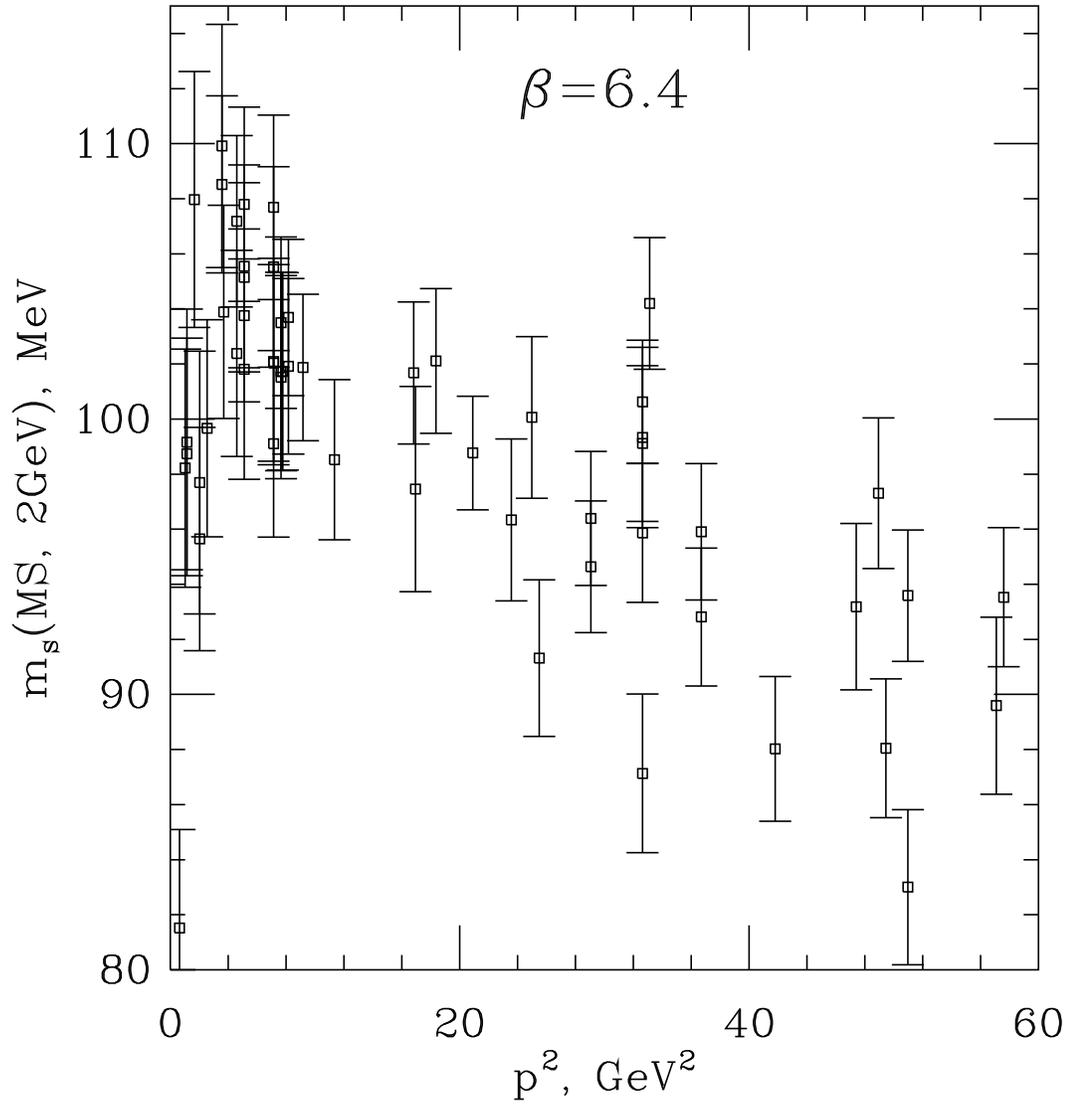}}
\end{center}
\caption{Strange quark mass at 2 GeV ($\beta=6.4$) vs. the matching 
momentum scale $p^2=q^{*2}$.}
\label{ms}
\end{figure} 

\begin{figure}[p]
\begin{center}
\leavevmode
\centerline{\epsfxsize=5.5in \epsfbox{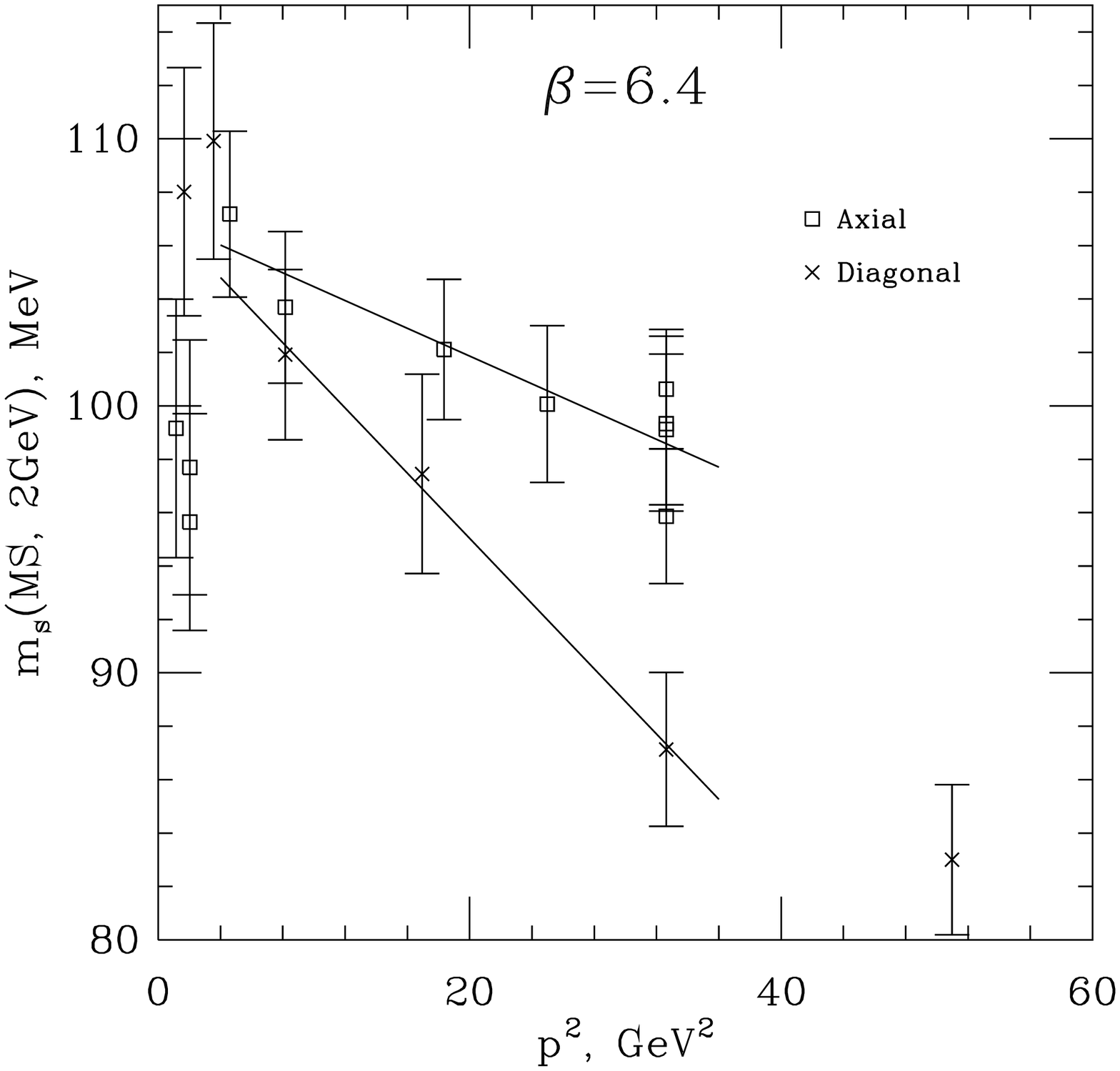}}
\end{center}
\caption{Strange quark mass at 2 GeV ($\beta=6.4$) vs. the matching 
momentum scale $p^2=q^{*2}$ for the axial and diagonal momentum
families.}
\label{fam}
\end{figure} 

In Fig.~\ref{ms} I show the results for $m_s$ at the scale of 2 GeV
in $\overline{\rm MS}$ NDR scheme obtained from a variety of 
matching momenta at $\beta=6.4$.
Ideally, one expects to see a plateau in the middle of the
plot, where $p^2$ is sufficiently above the hadronization scale,
and sufficiently low so that $p^2a^2\ll 1$. However, the data
show (1) sizable dependence on the momentum component structure
(a spread along the vertical axis) and (2)
a noticeable downward trend. Both are evidently due to the
discretization effects, and are expected to be proportional
to $p^2a^2$, which is not unreasonable according to the data.
In order to study it more, in Fig.~\ref{fam} I have plotted 
two families of momenta each having unique component structure but 
different overall magnitude.
These families are: (1) axial, with momenta
(0,0,0,1), (0,0,0,2), (0,0,0,4) etc.; and
(2) diagonal, with momenta (1,1,1,1), (2,2,2,2) etc.
One expects these  families to be the opposite extreme cases, with
the rest of the momenta being in between them in certain sense. 
%The fact that the linear  extrapolation to $p^2= 0$ 
%gives very close results for these two families confirms the
%hypothesis that both the slope and the spread in data in Fig.~\ref{ms}
%are mostly due to discretization effects which are linear in $p^2$. 
Linear extrapolation to $p^2=0$ gives very close results for these
two families. Therefore, such an extrapolation is
the most natural way to obtain reliable estimates of the quark mass 
at any given $\beta$, since a large part of
the discretization effects is removed. 

\begin{figure}[htb]
\begin{center}
\leavevmode
\centerline{\epsfysize=10cm \epsfbox{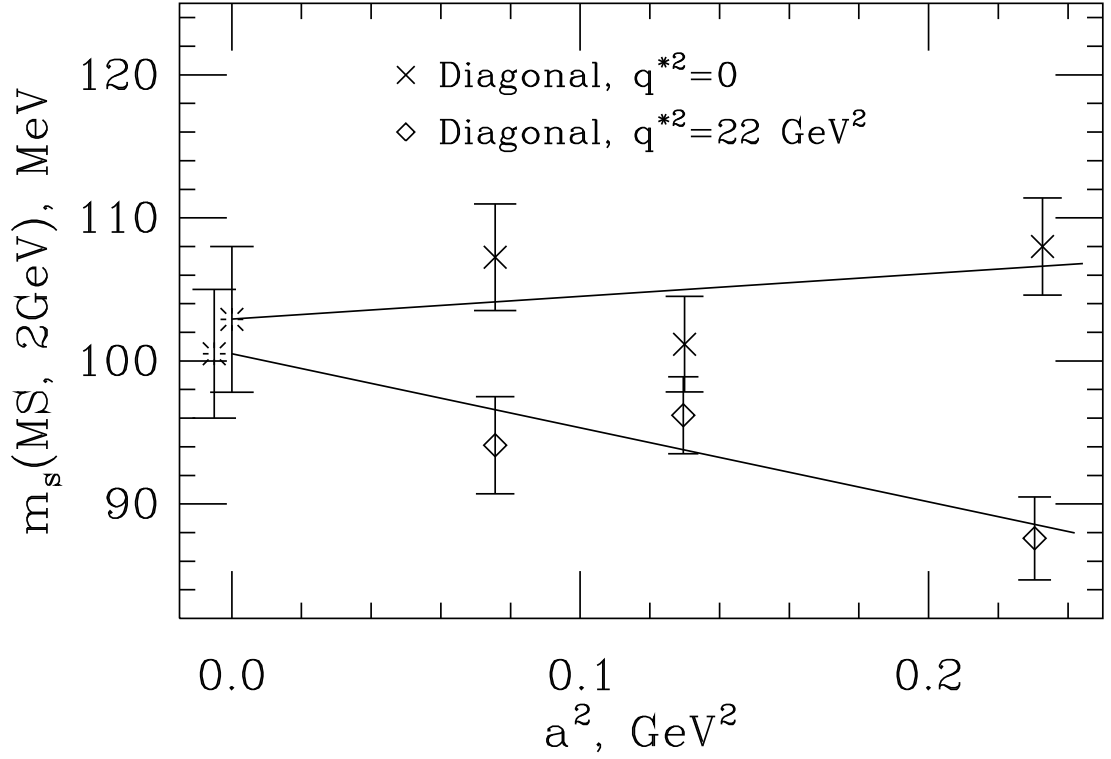}}
\end{center}
\caption{Strange quark mass at 2 GeV, a continuum extrapolation
in quenched QCD. Continuum limit results are shown with bursts
at $a^2=0$.}
\label{msvsa2}
\end{figure} 

One can also study the discretization effects by performing
continuum extrapolation. This is done in Fig.~\ref{msvsa2}
for ensembles with $\beta=6.4$, 6.2 and 6.0.
%Repeating similar procedure for the case of $\beta =6.2$ and 6.0,
%we obtain data sufficient to make a continuum extrapolation.
We are using the diagonal family for the
extrapolation, since arguably it has minimal discretization corrections.  
Two sets of data are plotted in Fig.~\ref{msvsa2}: one corresponding
to using the matching scale $p^2 = 22\ {\rm GeV}^2$, and the other 
one obtained at each $\beta$ by extrapolating to $p^2=0$,
as discussed above. The results are consistent, which was
expected since the discretization errors vanish in continuum limit.
This adds to the strength of the hypothesis that the slope 
and spread in data in Fig.~\ref{ms} is due to discretization effects. 

%The extrapolation is done separately
%for the two families, with very close results.
As our best estimate we quote the value of the strange quark mass
in quenched QCD\footnote{In obtaining this result we have used 
kaon mass to fix the bare strange quark mass, 
as opposed to another frequently-used alternative of using the $\phi$
meson mass}:
\begin{equation}
m_s(\overline{\rm MS} \;{\rm NDR}, 2\; {\rm  GeV}, a=0) = 102.9 \pm 5.1 \; {\rm MeV}.
\end{equation}
This was obtained by extrapolating to $p^2=0$ at each $\beta$.
The mass $\overline{m}\equiv (m_u+m_d)/2$ can be obtained by dividing
the above value by 26, according to chiral perturbation theory.
Our results agree with those of the JLQCD group~\cite{IshizukaNonpert},
which were obtained with a slightly different nonperturbative method.
These results are also consistent with those obtained with $O(a)$
improved Wilson fermions~\cite{WilsonQuarkMasses}.
To close the discussion of light quark masses, I report the result for 
dynamical $N_f=2$ ensemble at $\beta=5.7$:
\begin{equation}
m_s(\overline{\rm MS} \;{\rm NDR}, 2 \;{\rm  GeV}) = 105.1 \pm 16.4 \; {\rm MeV}.
\end{equation}
At current level of precision, the dynamical result is consistent
with the quenched ones. 

%=====================================================
\section{Partially Nonperturbative Matching For Basis Operators.}
%=====================================================

\label{sec:ansatz}

Ideally, one would like to perform a matching procedure
for four-fermion operators, not only for bilinears. 
It is quite conceivable that such a procedure will be done
in the nearest future. As a temporary solution, however, we have adopted 
a partially nonperturbative operator matching procedure, which makes 
use of bilinear renormalization coefficients $Z_P$ and $Z_S$
computed in the previous section. An estimate of the
renormalization of four-fermion operators can be obtained as follows. 

\begin{figure}[htb]
\begin{center}
\leavevmode
\centerline{\epsfysize=8cm \epsfbox{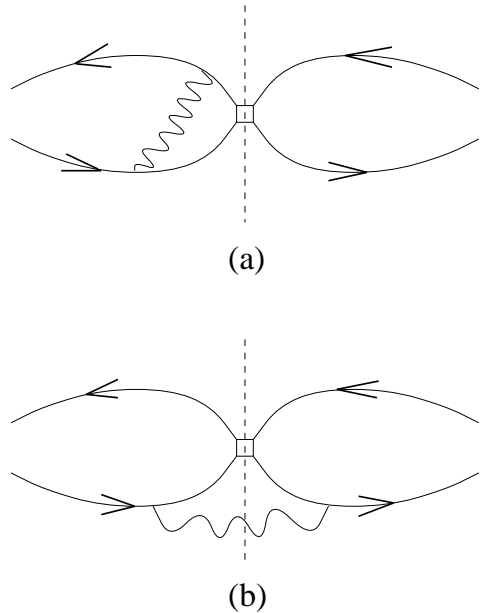}}
\end{center}
\caption{Example of one loop diagrams arising in 
renormalization of four-fermion operators: in type (a) no propagator
crosses the axis, and type (b) includes the rest of the diagrams.}
\label{Opp}
\end{figure} 

Consider
renormalization of the pseudoscalar--pseudoscalar operator in
Eq.~(\ref{eq:O6ops1}). 
At one-loop level, the diagonal renormalization coefficient 
$C_{PP}$ (involving diagrams shown in Fig.~\ref{Opp}) 
is almost equal to twice the pseudoscalar bilinear correction $C_P$. 
%$C_{PP}$ is almost equal to $2C_P$, where $C_P$ is the 
%renormalization coefficient for the pseudoscalar bilinear operator.
This suggests that, at least at one-loop level,
the renormalization of $(PP)_{EU}$ comes mostly from diagrams
in which no gluon propagator crosses the vertical axis of the diagram
(for example, diagram $(a)$ in Fig.~\ref{Opp}), and very little
from the rest of the diagrams
%from the gluon exchange between quarks of different bilinears
(such as diagram $(b)$ in Fig.~\ref{Opp}). In other words, the
renormalization of $(PP)_{EU}$ would be identical to 
the renormalization of product of two pseudoscalar bilinears,
were it not for the diagrams of type $(b)$, which give a subdominant
contribution. Mathematically, 
%
%Bilinear renormalization factor $Z_P$ has been computed.
%We can write:
%$$
%C_{PP} = 2*C_P + \widetilde{C_{PP}},
%$$
$$
(PP)_{EU}^{\mathrm{cont}} = (PP)_{EU}^{\mathrm{latt}}\; Z_{PP} + ... \, ,
$$
with
\begin{equation}
Z_{PP} = Z_P^2 (1 + \frac{g^2}{16\pi^2} \widetilde{C_{PP}} + O(g^4))\, ,
\label{eq:Zpp}
\end{equation}
\begin{equation}
Z_P = 1 + \frac{g^2}{16\pi^2} C_P + O(g^4)\, ,
\label{eq:Zp-pert}
\end{equation}
and dots indicate mixing with other operators (non-diagonal part).
The factor $\widetilde{C_{PP}} \equiv C_{PP} - 2 C_P$ contains
diagrams of type $(b)$ in Fig.~\ref{Opp} and is quite small.

In order to proceed, it may be reasonable to {\it assume} that the same 
holds at all orders in perturbation
theory; namely the diagrams of type $(c)$ in Fig.~\ref{higher-order} 
give subdominant contribution compared to type $(a)$ of the same
Figure. This assumption should be verified
separately by performing non-perturbative renormalization procedure
for four-fermion operators. If this ansatz is true, we can substitute
the non-perturbative value of $Z_P$ into Eq.~(\ref{eq:Zpp}) instead
of using the perturbative expression from Eq.~(\ref{eq:Zp-pert}).
Thus a partially nonperturbative estimate of $(PP)_U^{\mathrm cont}$
is obtained. This procedure is quite similar to the tadpole
improvement idea: the bulk of diagonal renormalization is
calculated non-perturbatively, while the rest is reliably computed
in perturbation theory.  
Analogously we obtain diagonal renormalization
of operators $(SS)_{IU}$ and $(PS)_{A2U}$ by using
$Z_{SS} = Z_S^2(1+\frac{g^2}{16\pi^2} \widetilde{C_{SS}} + O(g^4))$ and 
$Z_{PS} = Z_S Z_P(1+\frac{g^2}{16\pi^2} \widetilde{C_{PS}} + O(g^4))$.
(Note that $Z_P \neq Z_S$, even though they are equal in perturbation
theory.) We match operators at the scale $q^*=1/a$ and use the
continuum two-loop anomalous dimension to evolve to $\mu =2$ GeV.

Unfortunately, the above procedure does not solve completely the problem 
of operator renormalization, since it deals only with diagonal 
renormalization of the zero-distance operators in
Eqs.~(\ref{eq:O6ops1})---(\ref{eq:O6ops3}). Even though these operators
are dominant in contributing to $\varepsilon '/\varepsilon$, other
operators (such as $(SS)_{EU}$ and $(PP)_{EF}$)
can be important due to mixing with the dominant ones.
The mixing coefficients for these operators are not known 
even in perturbation theory. For a reasonable estimate we use
the coefficients
obtained for gauge non-invariant operator mixing~\cite{PatelSharpe}.
%For a reasonable
%estimate, we vary them from zero to twice their value
%The resulting variation in $\varepsilon '/\varepsilon$ is 
%reasonably small, even though the variations
%of subdominant operators' matrix elements are huge.

Secondly, since renormalization of operators $(PP)_{EU}$, $(SS)_{IU}$ 
and $(PS)_{A2U}$ is dramatic\footnote{For example, at $m_q=0.01$ and 
$\mu =2$ GeV for $Q_1$ ensemble we obtain $Z_{PP} = 0.055 \pm 0.007$, 
$Z_{PS} = 0.088 \pm 0.007$ and $Z_{SS} = 0.142 \pm 0.010$.}, their 
influence on other operators 
through non-diagonal mixing is ambiguous at one-loop order, 
even if the mixing coefficients are known.
The ambiguity is due to higher
order diagrams (for example, those shown in Fig.~\ref{higher-order}). 
In order to partially resum them
we use operators $(PP)_{EU}$, $(SS)_{IU}$ and $(PS)_{A2U}$ 
multiplied by factors $Z_P^2$, $Z_S^2$ and $Z_P Z_S$, correspondingly,
whenever they appear in non-diagonal mixing with other operators.
This is equivalent to evaluating the diagrams of type ($a$) and ($b$)
in Fig.~\ref{higher-order} at all orders, but ignoring the rest
of the diagrams (such as diagrams ($c$) and ($d$) in Fig~\ref{higher-order})
at all orders higher than first.
A completely analogous procedure was used for mixing of $O_6$ with $O_2$ 
through penguins when evaluating $\Re A_0$. 
To estimate a possible error in this procedure
we compare with a simpler one, whereby bare operators
are used in non-diagonal corrections (i.e. we apply strictly one-loop 
renormalization).
The difference in $\varepsilon '/\varepsilon$ between these two approaches
is of the same order or even less than the error due to uncertainties in 
determination
of $Z_P$ and $Z_S$ (see Tables~\ref{tab:epsp1} and~\ref{tab:epsp2}). 

\begin{table}[tbh]
\begin{tabular}{l|ccc}
\hline\hline
Quark & 0.01 & 0.02 & 0.03  \\
mass & & &  \\
\hline
M1 (p.r.) &
$-61.2  \pm 2.8   \pm 10.6 $ & % 6.1, was 8.8  
$-27.4   \pm 0.9   \pm 8.9 $ & %4.2, was 5.5  
$-16.8 \pm 0.5 \pm 8.0 $ \\ %4.0, was 4.8
M1 (1-loop.) &
$-52.3 \pm 2.2 \pm 10 $ &
$-22.0 \pm 0.8 \pm 8.3 $ &
$-12.2 \pm 0.5 \pm 6.9 $ \\
M2 (p.r.) &
$-38.6 \pm 2.1 \pm 6.0 $ &   %  6.8, was 4.4  
$-18.7 \pm 0.3 \pm 7.0 $ &  % 0.1, was 3.7
$-11.7 \pm 0.2 \pm 6.0 $ \\ % 1.3, was 3.4 
M2 (1-loop) &
$-45.4 \pm 3.5 \pm 8.6 $ & 
$-18.8 \pm 0.4 \pm 7.0 $ &
$-10.3 \pm 0.3 \pm 6.0 $ \\
M3 (p.r.) &     
$-97   \pm 14  \pm 13 $ &   % 45, was 13 
$-81   \pm 4   \pm 23 $ &    % 7, was 2?  
$-79   \pm 2   \pm 27 $ \\  % 5, was 26
M3 (1-loop) &
$-142 \pm 28 \pm 29 $ &
$-88 \pm 5 \pm 35 $ &
$-75 \pm 2 \pm 39 $ \\
\end{tabular}
\begin{tabular}{l|ccc}
\hline
Quark & 0.04 & 0.05 &\\
mass & & & \\
\hline
M1 (p.r.) &
$-8.0 \pm 0.9 \pm 7.2  $ &        %3.9, was 4.7  
$-4.4  \pm 0.9 \pm 7.2 $ & \\   %3.2, was 3.9 
M1 (1-loop) &
$-4.2 \pm 1.1 \pm 6.5 $ &
$-1.2 \pm 1.0 \pm 6.6 $ \\
M2 (p.r.) &
$-6.1  \pm 0.5 \pm 5.3  $ &   %2.4, was 3.5
$-3.1  \pm 0.5 \pm 4.9  $ & \\  %2.1, was 2.7
M2 (1-loop) &
$-3.7 \pm 0.8 \pm 5.8 $ &
$-0.9 \pm 0.8 \pm 5.2 $ \\
M3 (p.r.) &
$-81   \pm 5   \pm 38  $ &   %18, was 37
$-74   \pm 4   \pm 38  $& \\ %19, was 33
M3 (1-loop) &
$-64 \pm 4 \pm 52 $ &
$-55 \pm 5 \pm 51$ \\
\hline\hline
\end{tabular}
\caption{$\varepsilon '/\varepsilon$ in units of $10^{-4}$ for $Q_1$
ensemble ($\beta=6.0$), computed in three ways mentioned in text.
In all cases, partially-nonperturbative matching have been used to obtain 
the results. }
\label{tab:epsp1}
\end{table}

\begin{table}[tbh]
\begin{tabular}{l|ccc}
\hline\hline
Quark mass & 0.005 & 0.010 & 0.015 \\
\hline
M1 (p.r.) &
$-68.1 \pm 6.9 \pm 36.0$ & 
$-33.6 \pm 2.9 \pm 23.9$ & 
$-24.9 \pm 1.8 \pm 22.0$ \\

%$  -109\pm   15  \pm   84 $ &     % \pm 21 $ &  
%$  -64 \pm   7   \pm   63 $ \\    % \pm 25 $ \\
M1 (1-loop) &
$-60.9 \pm 6.9 \pm 31.2$ &
$-29.3 \pm 2.9 \pm 21.0$ &
$21.4 \pm 1.9 \pm 19.4$ \\

% $ -95 \pm 16 \pm 74 $&
% $-52 \pm 8 \pm 56$ \\
M2 (p.r.) &
$ -43.9 \pm 9.5 \pm 16.5$ &
$-33.1 \pm 3.8 \pm 18.3$ &
$-22.1 \pm 1.2 \pm 16.4$ \\

%$  -39 \pm  9   \pm     22 $ &    % \pm 5$ &  
%$  -22 \pm  2   \pm     18 $ \\   % \pm 4$ \\
M2 (1-loop) &
$ -53.6 \pm 16.9 \pm 25.0$ &
$-37.9 \pm 6.0 \pm 25.3$ &
$-23.0 \pm 1.5 \pm 20.0 $ \\

% $ -48 \pm 18 \pm 33$ &
% $-23 \pm 3 \pm 23 $ \\
M3 (p.r.) &
$-63.3 \pm 35.1 \pm 15.5$ &
$-103.9 \pm 31.7 \pm 45.2$ &
$-82.4 \pm 12.9 \pm 51.0$ \\

% $-30.0 \pm 17.0 \pm 23.2 $ &      % \pm 5.8$ &
% $-21.1 \pm 2.3 \pm   20.7 $ \\  % \pm 8.2$ \\
M3 (1-loop) &
$-98.3 \pm 72.9 \pm 46.2$ &
$-138.1 \pm 53.1 \pm 101.5$ &
$-92.4 \pm 16.3 \pm 86.3$ \\
% $-48 \pm 42 \pm 39$ &
% $-24 \pm 7 \pm 28$ \\ 
\end{tabular}
\begin{tabular}{l|ccc}
\hline
Quark mass & 0.020 & 0.030 & \\
\hline
M1 (p.r.) &
$-14.8 \pm 1.0 \pm 24.8$ &
$-10.3 \pm 0.6 \pm 19.6$ & \\

% $  -36 \pm   4   \pm   60 $ &      % \pm 11 $ &  
% $  -19 \pm   2   \pm   38 $ & \\     % \pm 6  $ \\ 
M1 (1-loop) &
$-11.8 \pm 1.1 \pm 21.9$ &
$-7.9 \pm 0.7 \pm 17.3$ & \\

% $-27 \pm 5 \pm 54 $ &
% $-14 \pm 2 \pm 34 $ & \\
M2 (p.r.) &
$-14.2 \pm 0.4 \pm 20.3$ &
$-9.5 \pm 0.3 \pm 16.1$ & \\

% $  -13.7 \pm  0.6 \pm   20 $ &     %\pm 4$ &  
% $  -9.5 \pm  0.6 \pm    17 $ & \\    %\pm 3$ \\ 
M2 (1-loop) &
$ -13.6 \pm 0.6 \pm 24.3$ &
$-8.5 \pm 0.5 \pm 18.0$ & \\

% $-12.4 \pm 1.1 \pm 23.3 $ &
% $-8.1 \pm 0.9 \pm 18.5 $ & \\ 
M3 (p.r.) &
$-74.9 \pm 5.7 \pm 88.0$ &
$-80.4 \pm 3.9 \pm 92.3$ & \\

% $-15.9 \pm 1.8  \pm   26.5 $ &    % \pm 4.9$ &
% $-10.8 \pm 1.1 \pm   21.6 $ & \\    % \pm 3.4$\\
M3 (1-loop) &
$ -72.6 \pm 5.7 \pm 142.0$ &
$  -73.5 \pm 4.0 \pm 131.0$ & \\
% $-14.7 \pm 1.4 \pm 31.1 $ &
% $-9.6 \pm 0.8 \pm 18.4 $ & \\
\hline\hline
\end{tabular}
\caption{$\varepsilon '/\varepsilon$ results for $Q_3$ ensemble 
($\beta=6.2$).}
\label{tab:epsp2}
\end{table}

%=====================================================
\section{Estimates Of $\varepsilon '/\varepsilon$}
%=====================================================

\label{sec:epsp_res}

Within the procedure outlined in the previous section we have found that 
$\langle O_6\rangle$ has a different sign from the expected one
due to a large renormalization factor.
This translates into a negative or very slightly positive value of 
$\varepsilon '/\varepsilon$ (Tables~\ref{tab:epsp1} and~\ref{tab:epsp2}
and Fig.~\ref{epsp}). 

Tables~\ref{tab:epsp1} and~\ref{tab:epsp2} contain our results 
for \epsp obtained in the above-described procedure. 
In all perturbative corrections
we have used one-loop non-diagonal coefficients computed for 
gauge-noninvariant operators, which are assumed to be of the same order 
as those for gauge-invariant operators.
The first quoted error is statistical (obtained by combining the individual 
errors in matrix elements by jackknife).
The second error is the diagonal operator matching error due to uncertainty 
in the determination of $Z_P$ and $Z_S$. 
In order to estimate the non-diagonal operator matching error we compare
two renormalization procedures: using strictly one-loop 
non-diagonal corrections (denoted ``(1-loop)''), and resumming part of 
higher-order corrections in non-diagonal mixing by using 
non-perturbative renormalization factors $Z_P$ and $Z_S$ (as explained in 
Section~\ref{sec:ansatz}). The latter method is denoted ``(p.r.)''. 

Finite volume and quenching effects were found small
compared to noise. In addition to the quoted errors, there
are uncertainties due to an unknown degree of validity of 
our ansatz in Sec.~\ref{sec:ansatz}, and due to unknown
non-diagonal coefficients in the mixing matrix.
The former error is uncontrolled at this point, since it 
is difficult to rigorously check our assumption made 
in Sec.~\ref{sec:ansatz}.  
The latter error is likely to be of the same order as the spread
in $\varepsilon '/\varepsilon$ between two different
approaches to higher-order corrections (strictly one-loop and partial
resummation), studied in Tables~\ref{tab:epsp1} and~\ref{tab:epsp2}. 

Some other parameters used in obtaining
these results are: $\Im \lambda_t = 1.5\cdot 10^{-4}$, $m_t=170$ GeV,
$m_b=4.5$ GeV, $m_c=1.3$ GeV, $\Omega_{\eta +\eta '} = 0.25$,
$\alpha^{(n_f=0)}_{\overline{\mathrm MS}} (2\quad {\mathrm GeV}) = 0.195$
(the latter is based on setting the lattice scale by $\rho$ meson mass).
Short distance coefficients were obtained by two-loop running
using the anomalous dimension and threshold matrices computed by
Buras {\it et al.}~\cite{buras_an}. 

Note that there are several choices to make in using our data to estimate
$\varepsilon '/\varepsilon$. One can use the experimental values of 
$\Re A_0$ and $\Re A_2$ in Eq.~(\ref{eq:epsp}) (M1), or one can use the 
values obtained on the lattice (M3).
One can also adopt an intermediate strategy (M2) of using the experimental
amplitude ratio $\omega$ and computed $\Re A_0$. When the higher-order
chiral corrections are taken into account and the continuum limit is taken
(so that $\omega = 22$),
these three methods should converge. The method M2 is preferred,
since in this method the overall error due to final state interactions
cancels between real and imaginary parts of $A_0$ amplitude, while 
the relative size of $\Pi_0$ and $\Pi_2$ contributions is given
by the physical $\omega$. 

\begin{figure}[htbp]
\begin{center}
\leavevmode
\centerline{\epsfysize=8cm \epsfbox{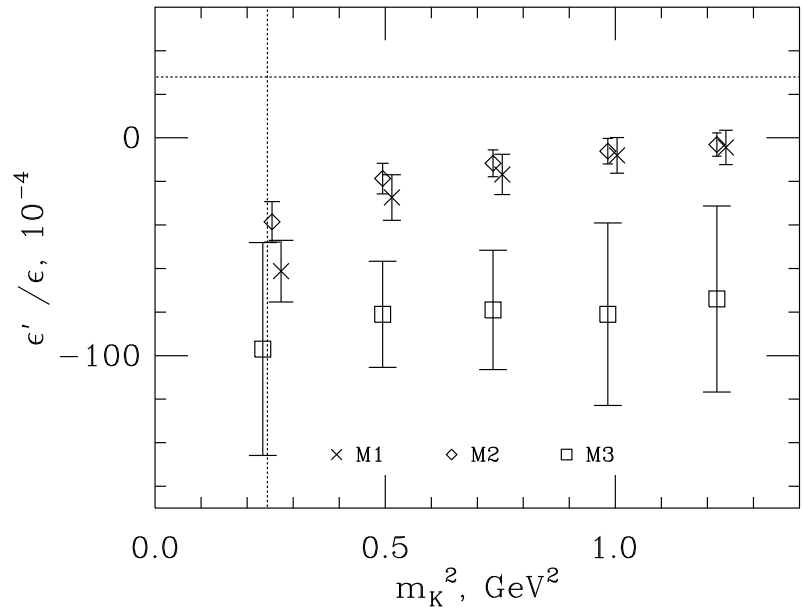}}
\end{center}
\caption{A rough estimate of $\varepsilon '/\varepsilon$ 
for $Q_1$ ($\beta=6.0$) ensemble using the partially-nonperturbative procedure
described in text. The Fermilab result's central
value is shown with a horizontal line. The points are displaced
horizontally for convenience.}
\label{epsp}
\end{figure} 

Fig.~\ref{epsp} presents the data obtained on ensemble $Q_1$, 
with error bars showing all three errors quoted in Tables~\ref{tab:epsp1}
and~\ref{tab:epsp2}, combined in quadrature.

\begin{figure}[htb]
\begin{center}
\leavevmode
\centerline{\epsfysize=8cm \epsfbox{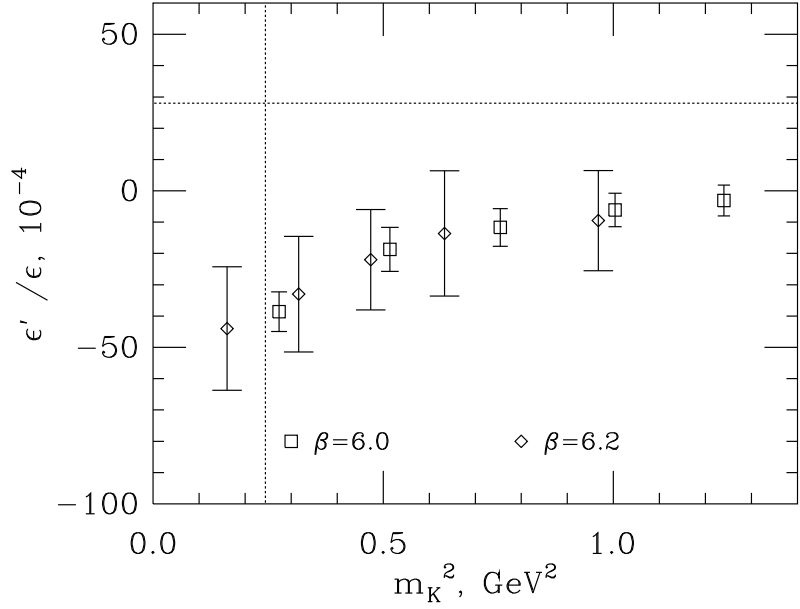}}
\end{center}
\caption{A study of cutoff dependence of \epsp. Plotted are data
obtained on $\beta=6.0$ and $\beta=6.2$ ensembles for M2 method.
The error bars show only the statistical error 
in matrix elements and in $Z_S$ and $Z_P$ constants. 
Systematic errors are significant but common to both ensembles. }
%The lowest mass point for $\beta=6.2$ is displaced
%horizontally for convenience.}
\label{epsp2}
\end{figure} 

Fig.~\ref{epsp2} contains comparison of results for \epsp on
$\beta=6.0$ and $\beta=6.2$ ensembles. With current errors,
the cutoff dependence
of \epsp does not appear to be significant. The errors shown are purely
statistical.

%=====================================================
\section{Summary}
%=====================================================

\label{epsp.sum}

I have shown that perturbative operator matching conventionally
employed in lattice calculations breaks down in the case 
of estimating \epsp. A partially nonperturbative procedure is
proposed as a temporary resolution. Light quark masses and
(pseudo)scalar bilinear renormalization constants have been
computed nonperturbatively. 

Our best result (at kaon mass, $\beta=6.0$, M2 method)
is 
\begin{equation}
\Re (\varepsilon '/\varepsilon ) = (-38.6 \pm 9.3) \cdot 10^{-4}\ .
\end{equation}
The quoted error is a combination of the statistical and several 
known systematic errors. The reader should keep in mind that 
additional systematic errors present are very difficult to estimate.
These errors are due to higher-order chiral terms (of the order of
40 \%) and due to neglecting certain types of higher-order
diagrams in partially nonperturbative operator matching. 

Taken at face value, 
our present estimate of \epsp would contradict the present
experimental results, which favor a positive value. 
Therefore, the minimal Standard Model's
description of direct CP violation would be proved to
be invalid. At this point, it is too early to draw
strong conclusions, since potentially large systematic uncertainties
are present in our estimates.

A recent study of \epsp with domain wall fermions by the
RIKEN-BNL-Columbia group~\cite{blum} has also produced
negative values for \epsp. While this consistency is encouraging
(in particular, it shows that the assumptions made in
our operator matching procedure do not change the
qualitative picture), the reader should keep in mind that
the errors due to using lowest-order chiral perturbation
theory are common to both ours and RIKEN-BNL-Columbia
calculations.

\chapter{Conclusions}
\label{conclusion}

In this thesis I report the techniques and results
of work directed at calculating hadronic matrix
elements of $\Delta S=1$  basis operators in Lattice QCD.
Matrix elements of all basis operators have been computed 
with good statistical precision. Combined with lowest-order 
chiral perturbation theory, they allow one to make a theoretical
prediction for $\Delta I=1/2$ rule and direct CP violation
parameter \epsp .

Based on the minimal Standard Model, we predict a substantial 
enhancement of $\Delta I=1/2$ transitions with respect to 
$\Delta I=3/2$ ones, consistent with experimental observations,
although the exact amount 
of enhancement is subject to large systematic uncertainties
due to higher-order chiral terms.
Thus our calculations confirm the currently dominant view
that most of the enhancement comes from strong interactions.

A theoretical prediction for \epsp is also made.
Evaluated within the minimal Standard Model framework, 
\epsp is found to have a negative value, in apparent
conflict with experiments, although our prediction at present
suffers from large systematic error.
In the future, it is possible to improve the accuracy of
this prediction by performing fully nonperturbative operator
matching as well as including higher orders in the chiral expansion.
These improvements will come at a relatively easy computational
cost, since the main results of our calculation 
(matrix elements $\langle \pi^+|O_i|K^+\rangle$ and
$\langle 0|O_i|K^0\rangle$ in
the lattice regularization scheme, which 
require most of the computational power) can be used 
along with above-mentioned improvements without
repeating the calculations. If the sign of \epsp stays negative
after these improvements, this would be the first failure of
the minimal Standard Model to describe the real-world data,
and so would have profound consequences in particle physics.

In addition, we have computed light quark masses using nonperturbative
renormalization.

%
% If you have appendices in your dissertation, you will need the
% following, else keep it commented. The following appendices are in
% files called ``app1.tex'', and ``app2.tex'', and they
% look just like any chapter.

\appendix
%===================================================================
\chapter{Explicit expressions for matrix elements in terms of
fermion contractions.}
%===================================================================
\label{sec:app}

Operators in Eqs.~(\ref{eq:ops1})--\ref{eq:ops10}) can be decomposed
into $I=0$ and $I=2$ parts, which contribute, correspondingly, 
to $\Delta I=1/2$ and $\Delta I=3/2$ transitions. Here we give the 
expressions for these parts for completeness, since $\Re A_0$, 
$\Re A_2$ and $\varepsilon '/\varepsilon$ are directly expressible
in terms of their matrix elements. The $I=0$ parts are given
as follows:
\begin{eqnarray}
O_1^{(0)} & = & \frac{2}{3} 
(\overline{s}\gamma_\mu (1-\gamma_5)d)(\overline{u}\gamma^\mu (1-\gamma_5)u)
-\frac{1}{3}(\overline{s}\gamma_\mu (1-\gamma_5)u)(\overline{u}\gamma^\mu 
(1-\gamma_5)d)  \nonumber \\
& + & \frac{1}{3}(\overline{s}\gamma_\mu (1-\gamma_5)d)
(\overline{d}\gamma^\mu (1-\gamma_5)d) \\
O_2^{(0)} & = & \frac{2}{3} 
(\overline{s}\gamma_\mu (1-\gamma_5)u)(\overline{u}\gamma^\mu (1-\gamma_5)d)
-\frac{1}{3}(\overline{s}\gamma_\mu (1-\gamma_5)d)(\overline{u}\gamma^\mu 
(1-\gamma_5)u) \nonumber \\
& + & \frac{1}{3}(\overline{s}\gamma_\mu (1-\gamma_5)d)
(\overline{d}\gamma^\mu (1-\gamma_5)d) \\
O_3^{(0)} & = & 
(\overline{s}\gamma_\mu (1-\gamma_5)d) \sum_{q=u,d,s}(\overline{q}\gamma^\mu (1-\gamma_5)q) \\
O_4^{(0)} & = & 
(\overline{s}_\alpha\gamma_\mu (1-\gamma_5)d_\beta ) \sum_{q=u,d,s}(\overline{q}_\beta
\gamma^\mu (1-\gamma_5)q_\alpha ) \\
O_5^{(0)} & = & 
(\overline{s}\gamma_\mu (1-\gamma_5)d) \sum_{q=u,d,s}
(\overline{q}\gamma^\mu (1+\gamma_5)q) \\
O_6^{(0)} & = & 
(\overline{s}_\alpha\gamma_\mu (1-\gamma_5)d_\beta ) \sum_{q=u,d,s}
(\overline{q}_\beta \gamma^\mu (1+\gamma_5)q_\alpha ) \\
O_7^{(0)} & = & \frac{1}{2} [
(\overline{s}\gamma_\mu (1-\gamma_5)d)(\overline{u}\gamma^\mu (1+\gamma_5)u)
-(\overline{s}\gamma_\mu (1-\gamma_5)u)(\overline{u}\gamma^\mu (1+\gamma_5)d)
\nonumber \\ & - & (\overline{s}\gamma_\mu (1-\gamma_5)d)
(\overline{s}\gamma^\mu (1+\gamma_5)s)
] \\
O_8^{(0)} & = & \frac{1}{2} [
(\overline{s}_\alpha\gamma_\mu (1-\gamma_5)d_\beta )
(\overline{u}_\beta\gamma^\mu (1+\gamma_5)u_\alpha )
-(\overline{s}_\alpha\gamma_\mu (1-\gamma_5)u_\beta )
(\overline{u}_\beta\gamma^\mu (1+\gamma_5)d_\alpha ) \nonumber \\
& - & (\overline{s}_\alpha\gamma_\mu (1-\gamma_5)d_\beta )
(\overline{s}_\beta\gamma^\mu (1+\gamma_5)s_\alpha )
] \\
O_9^{(0)} & = & \frac{1}{2} [
(\overline{s}\gamma_\mu (1-\gamma_5)d)(\overline{u}\gamma^\mu (1-\gamma_5)u)
-(\overline{s}\gamma_\mu (1-\gamma_5)u)(\overline{u}\gamma^\mu (1-\gamma_5)d)
\nonumber \\ & - & (\overline{s}\gamma_\mu (1-\gamma_5)d)
(\overline{s}\gamma^\mu (1-\gamma_5)s)
] \\
O_{10}^{(0)} & = & \frac{1}{2} [
(\overline{s}\gamma_\mu (1-\gamma_5)u)(\overline{u}\gamma^\mu (1-\gamma_5)d)
-(\overline{s}\gamma_\mu (1-\gamma_5)d)(\overline{u}\gamma^\mu (1-\gamma_5)u)
\nonumber \\ & - & (\overline{s}\gamma_\mu (1-\gamma_5)d)
(\overline{s}\gamma^\mu (1-\gamma_5)s)
] 
\end{eqnarray}

Expressions for $I=2$ parts are as follows:

\begin{eqnarray}
O_1^{(2)} & = & O_2^{(2)} = \frac{2}{3}O_9^{(2)} = \frac{2}{3}O_{10}^{(2)}
= \frac{1}{3} [ 
(\overline{s}\gamma_\mu (1-\gamma_5)u)(\overline{u}\gamma^\mu (1-\gamma_5)d) 
\nonumber \\
& & +(\overline{s}\gamma_\mu (1-\gamma_5)d)(\overline{u}\gamma^\mu 
(1-\gamma_5)u) 
- (\overline{s}\gamma_\mu (1-\gamma_5)d)
(\overline{d}\gamma^\mu (1-\gamma_5)d)] \\
O_7^{(2)} & = & \frac{1}{2} [
(\overline{s}\gamma_\mu (1-\gamma_5)u)(\overline{u}\gamma^\mu (1+\gamma_5)d)
+(\overline{s}\gamma_\mu (1-\gamma_5)d)(\overline{u}\gamma^\mu 
(1+\gamma_5)u) \nonumber \\
& & 
-  (\overline{s}\gamma_\mu (1-\gamma_5)d)
(\overline{d}\gamma^\mu (1+\gamma_5)d)] \\
O_8^{(2)} & = & \frac{1}{2} [
(\overline{s}_\alpha\gamma_\mu (1-\gamma_5)u_\beta )(\overline{u}_\beta
\gamma^\mu (1+\gamma_5)d_\alpha ) 
+(\overline{s}_\alpha\gamma_\mu (1-\gamma_5)d_\beta )(\overline{u}_\beta
\gamma^\mu (1+\gamma_5)u_\alpha ) \nonumber \\
& &
-  (\overline{s}_\alpha\gamma_\mu (1-\gamma_5)d_\beta )
(\overline{d}_\beta\gamma^\mu (1+\gamma_5)d_\alpha )] \\
O_3^{(2)} & = & O_4^{(2)} = O_5^{(2)} = O_6^{(2)} = 0 
\end{eqnarray}
(Whenever the color indices are not shown, they are contracted
within each bilinear, i.e. there are two color traces.)

As mentioned in Sec.~\ref{sec:diag}, in order to compute
matrix elements of $I=0$ operators one needs to evaluate 
three types of diagrams: ``eight'' (Fig.~\ref{diagrams}a),
``eye'' (Fig.~\ref{diagrams}b) and ``annihilation''
(Fig.~\ref{diagrams}c). In the previous Appendix section we
have given detailed expressions for computation
of these contractions, given the spin-flavor structure.
Here we assign this structure to all contractions required for each
operator, i.e. we express each matrix element in terms of
contractions which were ``built'' in the previous section. 

Let us introduce some notation. The matrix element of the above
operators have three components:
\begin{equation}
\langle\pi^+\pi^-|O_i|K^0\rangle = (E_i + I_i - S\,(\,2\,m\alpha_i )\,  )\,
\frac{m_K^2-m_\pi^2}{(p_\pi \cdot p_K)f},
\end{equation}
where $m$ is the common quark mass for $s$, $d$ and $u$, and
\begin{equation}
\label{eq:alpha}
\alpha_i = \frac{A_i}{P}.
\end{equation}
Here $E_i$ and $I_i$ stand for ``eight and ``eye'' contractions
of the $\langle\pi^+|O_i|K^+\rangle$ matrix element, 
$A_i \sim \langle 0|O_i|K^0\rangle \,/\,(m_d-m_s)$ is the
``annihilation'' diagram, 
%(or more precisely, its
%derivative with respect to the mass difference $(m_d-m_s)$),
$S =\langle \pi^+|\overline{s}d|K^+\rangle$
is the ``subtraction'' diagram, and 
$P =\langle 0|\overline{s}\gamma_5 d|K^0\rangle$ is the two-point function.
We compute $\alpha_i$ by averaging the ratio in the right-hand side of
Eq.~(\ref{eq:alpha}) over a suitable time range. 

Detailed expressions for $E_i$, $I_i$ and $A_i$ 
are given below in terms of the basic contractions on the lattice.
We label basic contractions by two letters, each representing a bilinear.
For example, $PP$ stands for contraction of the operator with 
spin structure $(\gamma_5)(\gamma_5)$, $SS$ is for $(\openone )(\openone )$,
$VV$ stands for $(\gamma_\mu )(\gamma^\mu )$, and $AA$ is for
$(\gamma_\mu \gamma_5)(\gamma^\mu\gamma_5)$. The staggered flavor
is determined by the type of contraction, as explained in the
previous Appendix section. Basic contractions
are also labeled by their subscript.
The first letter indicates whether it is an ``eight'', ``eye'' or
``annihilation'' contraction, and the second is ``U'' for two, or
``F'' for one color trace.
For example: $PP_{EU}$ stands for the ``eight'' contraction
of the operator with spin-flavor structure 
$(\gamma_5\otimes\xi_5)(\gamma_5\otimes\xi_5)$ with two color traces; 
$VA_{A1F}$ stands for the ``annihilation'' contraction
of the first type, in which the derivative is taken with respect to
quark mass on the external leg (see the previous Appendix section),
the spin-flavor structure is $(\gamma_\mu\otimes\xi_5)
(\gamma^\mu\gamma_5\otimes\openone )$,
and one color trace is taken. What follows are the full 
expressions\footnote{Signs of operators $O_7$ and $O_8$ have been changed
in order to be consistent with the sign convention of Buras 
{\it et al.}~\cite{buras}.}.

``Eight'' parts:
\begin{eqnarray}
E_1^{(0)} & = &\frac{2}{3}(VV_{EF} + AA_{EF}) - \frac{1}{3}(VV_{EU}+AA_{EU}) \\
E_2^{(0)} & = &\frac{2}{3}(VV_{EU} + AA_{EU}) - \frac{1}{3}(VV_{EF}+AA_{EF}) \\
E_3^{(0)} & = & VV_{EF}+AA_{EF} \\
E_4^{(0)} & = & VV_{EU}+AA_{EU} \\
E_5^{(0)} & = & 2(PP_{EF}-SS_{EF}) \\
E_6^{(0)} & = & 2(PP_{EU}-SS_{EU}) \\
E_7^{(0)} & = & SS_{EF} - PP_{EF} +\frac{1}{2}(VV_{EU}- AA_{EU}) \\
E_8^{(0)} & = & SS_{EU} - PP_{EU} +\frac{1}{2}(VV_{EF}- AA_{EF}) \\
E_9^{(0)} & = & -E_{10}^{(0)}=\frac{1}{2} (VV_{EF}+AA_{EF}-VV_{EU}-AA_{EU}) \\
E_1^{(2)} & = & E_2^{(2)} = \frac{2}{3}E_9^{(2)} = \frac{2}{3}E_{10}^{(2)}
= \frac{1}{3} (VV_{EU}+AA_{EU}+VV_{EF}+AA_{EF}) \\
E_3^{(2)} & = &  E_4^{(2)} = E_5^{(2)} = E_6^{(2)}  = 0 \\
E_7^{(2)} & = & \frac{1}{2} (AA_{EU} - VV_{EU}) + SS_{EF}-PP_{EF} \\
E_8^{(2)} & = & \frac{1}{2} (AA_{EF} - VV_{EF}) + SS_{EU}-PP_{EU} 
\end{eqnarray}

``Eye'' parts:
\begin{eqnarray}
I_1^{(0)} & = & VV_{IU}+AA_{IU} \\
I_2^{(0)} & = & VV_{IF}+AA_{IF} \\
I_3^{(0)} & = & 3(VV_{IU}+AA_{IU}) + 2(VV_{IF}+AA_{IF}) \\
I_4^{(0)} & = & 3(VV_{IF}+AA_{IF}) + 2(VV_{IU}+AA_{IU}) \\
I_5^{(0)} & = & 3(VV_{IU}-AA_{IU}) + 4(PP_{IF} - SS_{IF}) \\
I_6^{(0)} & = & 3(VV_{IF}-AA_{IF}) + 4(PP_{IU} - SS_{IU}) \\
I_7^{(0)} & = & 2(PP_{IF} - SS_{IF}) \\
I_8^{(0)} & = & 2(PP_{IU} - SS_{IU}) \\
I_9^{(0)} & = & VV_{IF} + AA_{IF} \\
I_{10}^{(0)} & = & VV_{IU} + AA_{IU} 
\end{eqnarray}

``Annihilation'' parts are obtained by inserting the 
derivative with respect to \mbox{$(m_d-m_s)$} into every propagator involving 
the strange quark:
\begin{eqnarray}
A_1^{(0)} & = & -(VA_{A1U} + AV_{A1U}) \\
A_2^{(0)} & = & -(VA_{A1F} + AV_{A1F}) \\
A_3^{(0)} & = & -3(VA_{A1U} + AV_{A1U}) - (VA_{A2U} + AV_{A2U}) \nonumber \\
& & -2 (VA_{A1F} + AV_{A1F}) - (VA_{A2F} +AV_{A2F})  \\
A_4^{(0)} & = & -3(VA_{A1F} + AV_{A1F}) - (VA_{A2F} + AV_{A2F}) \nonumber \\
& & -2 (VA_{A1U} + AV_{A1U}) - (VA_{A2U} +AV_{A2U})  \\
A_5^{(0)} & = & 3 (VA_{A1U} - AV_{A1U}) + (VA_{A2U} - AV_{A2U})
+ 2(PS_{A2F} - SP_{A2F}) \\
A_6^{(0)} & = & 3 (VA_{A1F} - AV_{A1F}) + (VA_{A2F} - AV_{A2F})
+ 2(PS_{A2U} - SP_{A2U}) \\
A_7^{(0)} & = & \frac{1}{2}(VA_{A2U} - AV_{A2U}) + (PS_{A2F} - SP_{A2F}) \\
A_8^{(0)} & = & \frac{1}{2}(VA_{A2F} - AV_{A2F}) + (PS_{A2U} - SP_{A2U}) \\
A_9^{(0)} & = & VA_{A1F} + AV_{A1F} + \frac{1}{2} (VA_{A2U}+AV_{A2U}+VA_{A2F}+AV_{A2F}) \\ 
A_{10}^{(0)} & = & VA_{A1U} + AV_{A1U} + \frac{1}{2} (VA_{A2F}+AV_{A2F}+VA_{A2U}+AV_{A2U}) 
\end{eqnarray}
Of course, ``eye'' and ``annihilation'' contractions are not present 
in $I=2$ operators.

%The raw results of our calculations on the $Q1$ ensemble are given
%in Table~\ref{tab:results} for the purpose of reference. 

%
% The all important bibliography file at the end of your document!! Use
% the bibstyle you (your department) like in the \bibliographystyle{}
% statement and list the name of your bibliography database file in
% the \bibliography{} statement.  In this example, ``bibfile.bib'' is
% the name of the database.  See the LaTeX manual appendix B for details
% about the bibliography database and BibTeX.
%

%\bibliographystyle{plain}
%\bibliography{bib}

\end{document}